\documentclass[twocolumn]{aastex631}

\pdfoutput=1 
\usepackage{xspace}
\usepackage[encapsulated]{CJK}  
\usepackage{gensymb}

\graphicspath{{Figures/}}

\newcommand{\coto}{$^{12}{\rm CO}~(2-1)$\xspace}
\newcommand{\halpha}{H$\alpha$\xspace}
\newcommand{\hii}{\ion{H}{2}\xspace}

\newcommand{\hagasratio}{$I_{\rm H{\alpha}}/\Sigma_{\rm H{\sc I}+H_2}$\xspace}
\newcommand{\newhagasratio}{\ensuremath{I_{\rm H{\alpha}}/\Sigma_{\rm H{\sc I}+H_2}}}

\newcommand{\rpah}{$R_{\rm PAH}$\xspace}
\newcommand{\rpahdiff}{$R_{\rm PAH, Diffuse}$\xspace}
\newcommand{\rpahneb}{$R_{\rm PAH,Neb}$\xspace}
\newcommand{\avrpahneb}{$\langle R_{\rm PAH,Neb}\rangle$\xspace}
\newcommand{\avrpahdiff}{$\langle R_{\rm PAH,Diffuse}\rangle$\xspace}

\newcommand{\qpah}{$q_{\rm PAH}$\xspace}
\newcommand{\mjysr}{MJy~sr$^{-1}$\xspace}

\newcommand{\fmol}{$f_{\rm{Mol}}$}

\begin{document}

\title{The Fraction of Dust Mass in the Form of PAHs on 10--50~pc Scales in Nearby Galaxies}
\correspondingauthor{Jessica Sutter}
\email{sutterjs@whitman.edu}


\newcommand{\Whitman}{Whitman College, 345 Boyer Avenue, Walla Walla, WA 99362, USA}

\newcommand{\Ox}{Sub-department of Astrophysics, Department of Physics, University of Oxford, Keble Road, Oxford OX1 3RH, UK}

\newcommand{\UGent}{Sterrenkundig Observatorium, Universiteit Gent, Krijgslaan 281 S9, B-9000 Gent, Belgium}

\newcommand{\STScI}{Space Telescope Science Institute, 3700 San Martin Drive, Baltimore, MD 21218, USA}

\newcommand{\MPIA}{Max-Planck-Institut f\"{u}r Astronomie, K\"{o}nigstuhl 17, D-69117, Heidelberg, Germany}

\newcommand{\AURA}{AURA for the European Space Agency (ESA), Space Telescope Science Institute, 3700 San Martin Drive, Baltimore, MD 21218, USA}

\newcommand{\STScIESA}{AURA for the European Space Agency (ESA), Space Telescope Science Institute, 3700 San Martin Drive, Baltimore, MD 21218, USA}

\newcommand{\UCSD}{Department of Astronomy \& Astrophysics, University of California, San Diego, 9500 Gilman Dr., La Jolla, CA 92093, USA}

\newcommand{\JHU}{Department of Physics and Astronomy, The Johns Hopkins University, Baltimore, MD 21218, USA}

\newcommand{\OSU}{Department of Astronomy, The Ohio State University, 140 West 18th Avenue, Columbus, OH 43210, USA}

\newcommand{\CCAPP}{Center for Cosmology and Astroparticle Physics (CCAPP), 191 West Woodruff Avenue, Columbus, OH 43210, USA}

\newcommand{\ARI}{Astronomisches Rechen-Institut, Zentrum f\"{u}r Astronomie der Universit\"{a}t Heidelberg, M\"{o}nchhofstr. 12-14, D-69120 Heidelberg, Germany}

\newcommand{\UConn}{Department of Physics, University of Connecticut, 196A Auditorium Road, Storrs, CT 06269, USA}

\newcommand{\UHawaii}{Institute for Astronomy, University of Hawaii, 2680 Woodlawn Drive, Honolulu, HI 96822, USA}

\newcommand{\UniCA}{Universit\'{e} C\^{o}te d'Azur, Observatoire de la C\^{o}te d'Azur, CNRS, Laboratoire Lagrange, 06000, Nice, France}

\newcommand{\UAlberta}{Dept. of Physics, University of Alberta, 4-183 CCIS, Edmonton, Alberta, T6G 2E1, Canada}

\newcommand{\Arcetri}{INAF — Osservatorio Astrofisico di Arcetri, Largo E. Fermi 5, I-50125, Florence, Italy}

\newcommand{\UWyoming}{Department of Physics and Astronomy, University of Wyoming, Laramie, WY 82071, USA}

\newcommand{\LJMU}{Astrophysics Research Institute, Liverpool John Moores University, 146 Brownlow Hill, Liverpool L3 5RF, UK}

\newcommand{\ITA}{Universit\"{a}t Heidelberg, Zentrum f\"{u}r Astronomie, Institut f\"{u}r Theoretische Astrophysik, Albert-Ueberle-Str 2, D-69120 Heidelberg, Germany}

\newcommand{\CfA}{Center for Astrophysics $\mid$ Harvard \& Smithsonian, 60 Garden St., 02138 Cambridge, MA, USA}

\newcommand{\MPE}{Max-Planck-Institut f\"{u}r Extraterrestrische Physik (MPE), Giessenbachstr. 1, D-85748 Garching, Germany}

\newcommand{\Surrey}{Department of Physics, University of Surrey, Guildford GU2 7XH, UK}

\newcommand{\ESO}{European Southern Observatory, Karl-Schwarzschild Stra{\ss}e 2, D-85748 Garching bei M\"{u}nchen, Germany}

\newcommand{\IWR}{Universit\"{a}t Heidelberg, Interdisziplin\"{a}res Zentrum f\"{u}r Wissenschaftliches Rechnen, Im Neuenheimer Feld 205, D-69120 Heidelberg, Germany}

\newcommand{\ulyon}{Univ Lyon, Univ Lyon1, ENS de Lyon, CNRS, Centre de Recherche Astrophysique de Lyon UMR5574, F-69230 Saint-Genis-Laval France}

\newcommand{\COOL}{Cosmic Origins Of Life (COOL) Research DAO, coolresearch.io}

\newcommand{\OAN}{Observatorio Astron{\'o}mico Nacional (IGN), C/ Alfonso XII 3, E-28014 Madrid, Spain}

\newcommand{\UBonn}{Argelander-Institut f\"{u}r Astronomie, Universit\"{a}t Bonn, Auf dem H\"{u}gel 71, 53121 Bonn, Germany}

\newcommand{\kipac}{Kavli Institute for Particle Astrophysics \& Cosmology (KIPAC), Stanford University, CA 94305, USA}

\newcommand{\Umanc}{Jodrell Bank Centre for Astrophysics, Department of Physics and Astronomy, University of Manchester, Oxford Road, Manchester M13 9PL, UK}

\newcommand{\NRAO}{National Radio Astronomy Observatory, 520 Edgemont Road, Charlottesville, VA 22903, USA}

\newcommand{\ANU}{Research School of Astronomy and Astrophysics, Australian National University, Canberra, ACT 2611, Australia}

\newcommand{\AThreeD}{ARC Centre of Excellence for All Sky Astrophysics in 3 Dimensions (ASTRO 3D), Australia}

\newcommand{\IAC}{Instituto de Astrof\'isica de Canarias, C/ V\'ia L\'actea s/n, E-38205, La Laguna, Spain}

\newcommand{\ULL}{Departamento de Astrof\'isica, Universidad de La Laguna, Av. del Astrof\'isico Francisco S\'anchez s/n, E-38206, La Laguna, Spain}

\newcommand{\Princeton}{Department of Astrophysical Sciences, Princeton University, 4 Ivy Lane, Princeton, NJ 08544, USA}

\newcommand{\IRAM}{IRAM, 300 rue de la Piscine, 38400 Saint Martin d'H\'{e}res, France}

\newcommand{\LERMA}{LERMA, Observatoire de Paris, PSL Research University, CNRS, Sorbonne Universit\'{e}s, 75014 Paris}

\newcommand{\YB}{Centro de Desarrollos Tecnol\'ogicos, Observatorio de Yebes (IGN), 19141 Yebes, Guadalajara, Spain}

\newcommand{\uwa}{International Centre for Radio Astronomy Research, University of Western Australia, 7 Fairway, Crawley, 6009, WA, Australia}



\author[0000-0002-9183-8102]{Jessica Sutter}
\affiliation{\Whitman}
\affiliation{\UCSD}

\author[0000-0002-4378-8534]{Karin Sandstrom}
\affiliation{\UCSD}

\author[0000-0002-5235-5589]{J\'{e}r\'{e}my Chastenet}
\affiliation{\UGent}

\author[0000-0002-2545-1700]{Adam~K.~Leroy}
\affiliation{\OSU}
\affiliation{\CCAPP}

\author[0000-0001-9605-780X]{Eric W. Koch}
\affiliation{\CfA}

\author[0000-0002-0012-2142]{Thomas~G.~Williams}
\affiliation{\Ox}

\author[0000-0001-8241-7704]{Ryan Chown}
\affiliation{\OSU}







\author[0000-0002-2545-5752]{Francesco~Belfiore}
\affiliation{\Arcetri}

\author[0000-0003-0166-9745]{Frank Bigiel}
\affiliation{\UBonn}

\author[0000-0003-0946-6176]{M\'ed\'eric Boquien}
\affiliation{\UniCA}

\author[0000-0001-5301-1326]{Yixian Cao}
\affiliation{Max-Planck-Institut f\"ur Extraterrestrische Physik (MPE), Giessenbachstr. 1, D-85748 Garching, Germany}

\author[0000-0002-5635-5180]{M\'{e}lanie Chevance}
\affiliation{\ITA}
\affiliation{\COOL}

\author[0000-0002-5782-9093]{Daniel~A.~Dale}
\affiliation{\UWyoming}

\author[0000-0002-4755-118X]{Oleg~V.~Egorov}
\affiliation{\ARI}

\author[0000-0001-6708-1317]{Simon~C.~O.~Glover}
\affiliation{\ITA}

\author[0000-0002-9768-0246]{Brent~Groves}
\affiliation{\uwa}

\author[0000-0002-0560-3172]{Ralf S. Klessen}
\affiliation{\ITA}
\affiliation{\IWR}

\author[0000-0001-6551-3091]{Kathryn Kreckel}
\affiliation{\ARI}

\author[0000-0003-3917-6460]{Kirsten~L.~Larson}
\affiliation{\STScIESA}

\author[0000-0002-0119-1115]{Elias K. Oakes}
\affiliation{\UConn}

\author[0000-0003-2721-487X]{Debosmita Pathak}
\affiliation{\OSU}

\author[0000-0002-9190-9986]{Lise Ramambason}
\affiliation{\ITA}

\author[0000-0002-5204-2259]{Erik Rosolowsky}
\affiliation{\UAlberta}

\author[0000-0002-7365-5791]{Elizabeth J. Watkins}
\affiliation{\Umanc}

\begin{abstract}
Polycyclic aromatic hydrocarbons (PAHs) are a ubiquitous component of the interstellar medium (ISM) in $z \sim 0$ massive, star-forming galaxies and play key roles in ISM energy balance, chemistry, and shielding.  Wide field of view, high resolution mid-infrared (MIR) images from JWST provides the ability to map the fraction of dust in the form of PAHs and the properties of these key dust grains at 10$-$50 pc resolution in galaxies outside the Local Group.  We use MIR JWST photometric observations of a sample of 19 nearby galaxies from the ``Physics at High Angular Resolution in Nearby GalaxieS'' (PHANGS) survey to investigate the variations of the PAH fraction. By comparison to lower resolution far-IR mapping, we show that a combination of the MIRI filters (\rpah = [F770W+F1130W]/F2100W) traces the fraction of dust by mass in the form of PAHs (i.e., the PAH fraction, or \qpah). Mapping \rpah across the 19 PHANGS galaxies, we find that the PAH fraction steeply decreases in \hii regions, revealing the destruction of these small grains in regions of ionized gas. Outside \hii regions, we find \rpah is constant across the PHANGS sample with an average value of 3.43$\pm$0.98, which, for an illuminating radiation field of intensity 2--5 times that of the radiation field in the solar neighborhood, corresponds to \qpah values of 3--6\%.

\end{abstract}



\section{Introduction} \label{sec:intro}
Dust is a key component of the interstellar medium (ISM) of galaxies, and plays important roles regulating the thermal and ionization balance within the ISM.  Dust also absorbs and scatters ultraviolet and optical light, reprocessing on average an estimated $\sim$30\% of the starlight produced in galaxies and re-emitting it in the infrared \citep{Draine2003}. The composition, properties, and distribution of the dust set the ISM heating efficiency and determine attenuation, making the dust a key component of any ISM model. Of particular interest is the breakdown between larger carbonaceous and silicate dust grains and the smallest carbonaceous grains, the polycyclic aromatic hydrocarbons (PAHs). PAHs are widely considered\footnote{A variety of alternative explanations for the carrier of the MIR bands can be found in the literature, including a number of materials with primarily hydrocarbon composition \citep[see e.g.][]{Jones2013}.} to be responsible for broad emission features that dominate the near- and mid-infrared (NIR and MIR, respectively) spectra of galaxies, including prominent emission features that are produced by the stretching and bending vibrational modes of the C--C (6.2\micron, 7.7\micron) and C--H (3.3\micron, 11.3\micron) bonds in PAHs \citep{Allamandola1989}.  PAH emission has been detected from a wide variety of objects, including \ion{H}{2} regions, planetary nebulae, young stellar objects, asymptotic giant branch stars, and in the ISM of dwarf, spiral, elliptical, and ultra-luminous infrared galaxies \citep[see][and references within]{Li2020}.


\begin{figure*}[t]
\centering
\includegraphics[width= \textwidth]{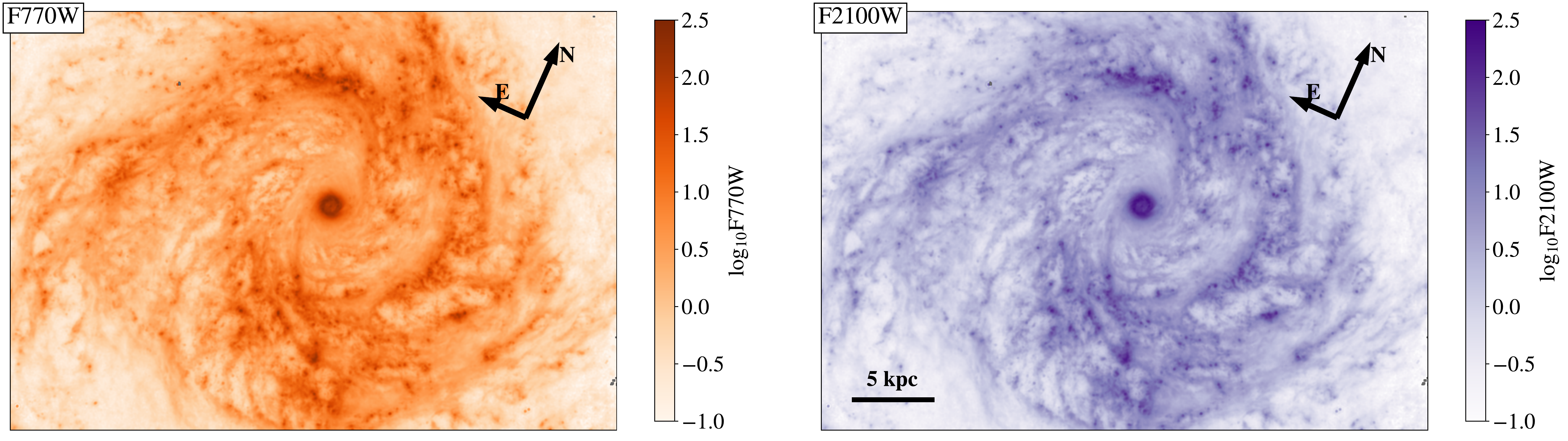}
\caption{NGC~4303 in F770W (left) and F2100W (right).  F770W and F2100W intensities are both shown in log scale to capture the full dynamic range, and F770W has been corrected to remove starlight emission (see Section~\ref{sec:starsub} for details).  The \rpah map of NGC~4303 is shown in Figure~\ref{fig:rpah_4303_big}. The observed emission in F770W and F2100W over large parts of the galaxy look very similar, indicating that \rpah does not change substantially in the diffuse neutral gas. The ratio map shown in Figure~\ref{fig:rpah_4303_big}  reveals that in the bright, compact \hii regions, \rpah drops steeply.  }
\label{fig:miri_4303}
\end{figure*}

A key aspect of understanding the roles of PAHs in the ISM and using their emission as a tracer of galaxy properties is quantifying their abundance relative to the total dust grain population \citep{Draine2007, Jones2015}. Since PAHs are typically stochastically heated  \citep{Sellgren1983,DraineLi2001,LiDraine2001} and therefore respond linearly to the radiation field intensity, the fraction of the dust mass in the form of PAHs can be inferred from comparisons of the MIR PAH emission to the far-IR emission from dust in thermal equilibrium with the radiation field \citep[e.g.\ ][]{Draine&Li2007}, given assumptions about the dust grain populations and radiation field strength and spectrum. 
Dust models in the literature make a variety of assumptions about the specific characteristics of the PAHs (or alternative small carbonaceous grain components), other small grains that contribute to the MIR emission, and the radiation field heating those grains  \citep{Draine&Li2007,Compiegne2011,Jones2013,Jones2017,Galliano2021,Hensley2023}. Because all of these choices impact the MIR spectral energy distribution, different models will produce different measurements of the PAH fraction and its variation.  In the following, we adopt the \citet{Draine&Li2007} dust model to guide our interpretation of the MIR SED and PAH fraction. This model does not have a variable ``very small grain'' population and assumes a distribution of radiation field intensities heating the dust described by a delta function plus power-law component \citep{Dale2001,Draine&Li2007,Aniano2012}.  The fraction of the dust mass in the form of PAHs (with less than 10$^3$ carbon atoms) in the \citet{Draine&Li2007} model is denoted as \qpah.

Studies of local universe galaxies \citep{Weingartner2001b, LiDraine2001, Draine&Li2007, Aniano2020} including the Large and Small Magellanic Clouds \citep{LiDraine2002bSMC, Sandstrom2010, Paradis2011, Chastenet2019, Paradis2023, Dale2023b} have found \qpah varies between approximately 0--5\%.  The PAH fraction has been shown to decrease in low metallicity galaxies \citep{Engelbracht2005, Draine2007,Khramtsova2013, RemyRuyer2015, Galliano2018, Chastenet2019, Aniano2020} and to fall dramatically within \hii regions \citep{Cesarsky1996,Chastenet2019}.  Tracing \qpah across different environments allows for insights into where PAHs are being produced or destroyed and how they impact their surroundings.  \qpah measurements have, however, been limited in resolution and sample size due to the necessity of infrared observations covering both PAH emission and FIR emission from larger dust grains.  

An alternative approach to full MIR to FIR SED modeling is to trace the PAH fraction with ratios of PAH emission to small dust grain continuum emission in the MIR.  \citet{Engelbracht2005} used {\em Spitzer} IRAC 8 \micron\ and MIPS 24 \micron\ measurements to show strong metallicity trends in PAH-to-continuum ratios, which were then validated with MIR {\em Spitzer} spectroscopy \citep{Engelbracht2008}.  These MIR only measurements have greatly expanded our view of PAHs in the ISM, \citep[see][and references therein]{Li2020}, and can be translated to use WISE 12 and 22 \micron\ photometry for all nearby galaxies (though at lower resolution).  However, they remain limited by the sensitivity and resolution of \textit{Spitzer}, with a $6''$ resolution at 24\micron, corresponding to $\sim100$s of pc in nearby galaxies.  The sensitivity of \textit{Spitzer} also made it difficult to trace PAH emission from the diffuse ISM.  With the advent of the \textit{James Webb Space Telescope} (JWST), a new realm of environments for PAH studies are available.

Within its first year, JWST has already expanded our view of PAHs, detecting PAH emission at a distance of 100~pc from an actively accreting supermassive black hole \citep[NGC~7469,][]{Lai2022,Armus2023}, in an extended disk around merging galaxies \citep[VV 114,][]{Evans2022}, in the diffuse ISM of local star-forming galaxies \citep{Leroy2023a,Sandstrom2023}, tracing PAH emission up to $z=2$ \citep{Shivaei2024}, and has tracked the properties of PAHs in stellar associations \citep{Dale2023} and star-forming regions \citep{Egorov2023, Ujjwal2024}.  While JWST spectroscopy can provide precise details about the PAH population \citep[see e.g.][]{Chown2023}, the small field of view of both the MIRI IFU and NIRSpec make it observationally costly to study the distribution of PAHs in Local Universe galaxies using JWST spectroscopy.  Instead, MIRI and NIRCam photometric bands situated precisely on many of the strongest PAH features at $z \sim 0 $ can be used as indicators of PAH emission across the disks of nearby galaxies.  Following the work of \citet{Chastenet2023_RPAH} and \citet{Egorov2023}, this paper presents an analysis of a proposed photometrically derived tracer of the fraction of dust stored in PAHs using three of the MIRI filters: F770W, F1130W, and F2100W.

Tracing PAHs in the ISM of nearby galaxies with JWST will help address many outstanding questions surrounding these small dust grains.  Understanding the formation and destruction mechanisms of the PAHs across ISM phases and galaxy environments will provide context for using  PAH emission as a tracer of star-formation rate \citep[SFR, e.g.][]{Peeters2004,Calzetti2007, Shipley2016, Whitcomb2020,Belfiore2023} or gas column density \citep[e.g.][]{Cortzen2019, Whitcomb2023, Leroy2023b} and will help to clarify why \qpah decreases at low metallicity \citep[e.g.][]{Aniano2020}.  Previous work with \textit{Spitzer} has already provided clues into the complex life-cycle of the PAHs, showing suppression of \qpah in regions with warm dust measured by increased $\nu L_{\nu} (70\mu$m)/$L_{\rm{TIR}}$ values \citep{Aniano2020}, implying that PAHs are destroyed in regions surrounding young stars.  This was further confirmed in \citet{Chastenet2019} and \citet{Chastenet2023_RPAH}, both of which show decreasing PAH fractions in \ion{H}{2} regions in the LMC and initial work on four of the galaxies from the PHANGS sample, respectively.

Using the nineteen galaxies observed by JWST as part of the Physics at High Angular resolution in Nearby GalaxieS (PHANGS)--JWST Cycle 1 Treasury \citep{Lee2023, Williams2024}, along with additional multi-wavelength observations, we examine how galaxy environment, specific star formation rate (sSFR), and ISM conditions impact the PAH fraction at spatial scales of 10--100~pc.

This paper is organized as follows: Section~\ref{sec:samp} describes the observations used to complete this analysis.  Section~\ref{sec:RPAH_qpah} provides a justification for the proposed photometric tracer of the PAH fraction.  Section~\ref{sec:Analysis} uses this tracer to map the distribution of PAHs within our sample, and describes trends seen as a function of environment, gas phase, metallicity, and proximity to sites of active star formation. Section~\ref{sec:Discussion} situates the results of this paper in the broader astronomical context.  Finally, Section~\ref{sec:Conclusions} summarizes the conclusions drawn from this analysis.

\begin{figure*}[t]
    \centering
    \includegraphics[width=\textwidth,]{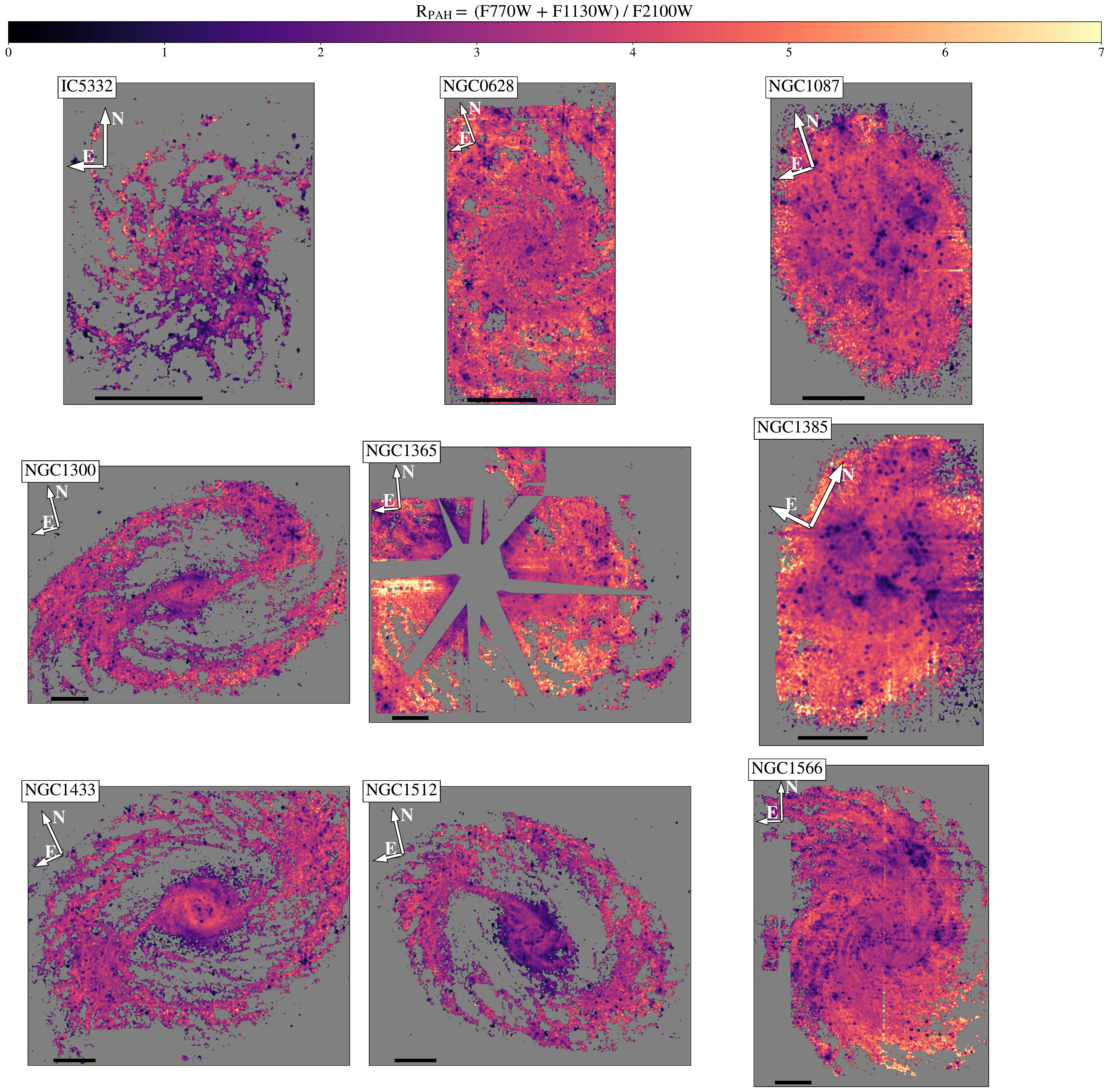}
    \caption{\rpah mapped in nine of the eighteen galaxies included in this work. See  Figure~\ref{fig:rpah2} and \ref{fig:rpah_4303_big} for the other targets. Pixels below the $3\sigma$ noise limit are cut, and shown in gray.  Additional by-eye masking is done on galaxies where a bright nucleus saturated in F2100W, creating diffraction spikes which are cut for our analysis.  Black bars in the lower-left corner of each image represent 5~kpc in each source.  Each galaxy has been rotated to better fill the space, arrows showing the NE directions are displayed in the upper left of each panel.  Maps have been smoothed to 0\farcs9 to increase the signal to noise.  The starlight subtraction method described in Section~\ref{sec:starsub} was performed on the F770W fluxes before producing these \rpah maps.}
    \label{fig:rpah1}
\end{figure*}

\begin{deluxetable*}{lccccccccc}[t]
    \tablecaption{Galaxy Sample}
        \tablehead{
        \colhead{Target} & \colhead{$\log_{10}$M$_{\star, \rm{g}}$}  & \colhead{$\log_{10}$M$_{\star, \rm{JWST}}$} & \colhead{$\log_{10}$SFR$_{\rm{JWST}}$} & \colhead{Distance } & \colhead{P.A.} & \colhead{\textit{i} } & \colhead{${\rm r_{25}}$ } & \colhead{HI Map}  & \colhead{MUSE Res.} \\
        \colhead{} & \colhead{(M$_{\odot}$)}  & \colhead{(M$_{\odot}$)} & \colhead{(M$_{\odot}$ yr$^{-1}$)} & \colhead{(Mpc)} & \colhead{(\degree)} & \colhead{(\degree)} & \colhead{($'$)} & \colhead{} & \colhead{($''$)} }
        \startdata
        IC~5332  &  9.68  &  9.20  &$-$1.30  & 9.01   & 74.4     & 26.9 & 3.0 & MeerKAT & 0.87\\
        NGC~0628 & 10.34  & 10.08  &$-$0.35  & 9.84   & 20.7     & 8.9  & 4.9 & VLA & 0.92 \\
        NGC~1087 & 9.94   &  9.94  & 0.26    & 15.85  & 359.1    & 42.9 & 1.5  & VLA    & 0.92 \\
        NGC~1300 & 10.62  & 10.62  &$-$0.08  & 18.99  & 278.0    & 31.8 & 3.0 & MeerKAT & 0.89 \\
        NGC~1365 & 11.00  & 10.92  & 0.35    & 19.57  & 201.1    & 55.4 & 6.01 & -- & 1.15 \\
        NGC~1385 & 9.98   & 9.98   & 0.44    &  17.22  &  181.3   &  44.0  &  1.7 & VLA & 0.77\\
        NGC~1433 & 10.87  & 10.83 & $-$0.25  & 18.63  &  199.7   &  28.9  &  3.1 & -- & 0.91 \\
        NGC~1512 & 10.72  & 10.71 & $-$0.19  &  18.83  &  261.9   &  42.5  &  4.2 & MeerKAT & 1.25 \\
        NGC~1566 & 10.79  & 10.78 & 0.50     & 17.69  &  214.7   &  29.5  &  3.6 & -- & 0.80 \\
        NGC~1672 & 10.73  & 10.73 & 0.79     &  19.40  &  134.3   &  43.6  &  3.1 & -- &  0.96\\
        NGC~2835 & 10.00  &  9.87 & $-$0.27  & 12.22  &  1.0     &  41.3  &  3.2 & VLA & 1.15 \\
        NGC~3351 & 10.37  & 10.37 & $-$0.24  & 9.96   &  193.2   &  45.1  &  3.6 & VLA & 1.05 \\
        NGC~3627 & 10.84  & 10.84 & 0.71     & 11.32  &  173.1   &  57.3  &  5.1 & VLA & 1.05 \\
	NGC~4254 & 10.42  & 10.42 & 0.52     & 13.10  &  68.1    &  34.4  &  2.5 & VLA & 0.89 \\ 
        NGC~4303 & 10.51  & 10.51 & 0.69     & 16.99  &  312.4   &  23.5  &  3.4 & VLA & 0.78 \\
        NGC~4321 & 10.75  & 10.75 & 0.48     & 15.21  &  156.2   &  38.5  &  3.0 & VLA & 1.16\\
        NGC~4535 & 10.54  & 10.47 &$-$0.01   & 15.77  &  179.7   &  44.7  &  4.1 & MeerKAT & 0.56 \\
        NGC~5068 & 9.41   &  9.27 & $-$0.74  & 5.20    &  342.4   &  35.7  &  3.7 & VLA & 1.04\\
        NGC~7496 & 10.00  & 10.00 & $-$0.29  & 18.72  & 193.7    & 35.9   & 1.7 & MeerKAT & 0.89 
    \enddata
    \tablecomments{Global stellar masses (M$_{\star, \rm{g}}$) from \citet{Leroy2021b} as well as the stellar mass (M$_{\star, \rm{JWST}}$) and SFR (SFR$_{\rm{JWST}}$) measured within the area of the galaxy mapped by JWST.  SFR measurements are made using the extinction corrected \halpha measurements from \citet{Belfiore2023}.  Inclinations ($i$) and Position Angles (P.A.) are both listed in degrees, and were obtained from \citet[][]{Leroy2012}. ${\rm r_{25}}$, the radius of the 25th magnitude isophotal contour in the B band, is listed in arc-minutes from the HyperLeda database \citep[][]{Makarov2014}. Distances are from \citet[][]{Anand2021}.  The HI map column lists the origin of the HI data, and the MUSE Res.\ column displays the final `copt' resolution of the MUSE data (see Section~\ref{sec:ancil_data} for details).  Additional information can be found in Table~1 of the survey paper \citep[][]{Lee2023}. }
    \label{tab:samp}
    \end{deluxetable*}

\section{Data and Methods} \label{sec:samp}

The galaxies included in this study are all part of the PHANGS--JWST Cycle 1 sample \citep{Lee2023, Williams2024}.  This survey includes observations of 19 galaxies in four NIRCam and four MIRI bands. Information about the 19 galaxies can be found in Table~\ref{tab:samp}.  The 19 galaxies included in the PHANGS--JWST sample are all nearby ($D \lesssim 20$~Mpc) and all have relatively low inclinations ($i\leq 60\degree$).  All are star-forming galaxies, with metallicities spanning  $12+\log_{10}$(O/H) = 8.4--8.7 \citep{Groves2023} and stellar masses between $\log_{10}$(M$_{\star}$/M$_{\odot}$) = 9.5--11.1 \citep{Leroy2021b}.  The galaxies included in this sample include a range of morphologies and a subset that host AGN.  

\subsection{JWST MIRI \& NIRCam}
\label{sec:jwst}

This work uses data from three of the MIRI bands: F770W, F1130W, and F2100W  \citep{Rieke2015} as well as the NIRCam F200W. The MIRI F770W and F1130W filters capture the emission from two of the strongest PAH features, the 7.7~\micron\ and 11.3~\micron\ features.  The F2100W filter is dominated by dust continuum emission.  As an example, the F770W and F2100W maps of NGC~4303 are shown in Figure~\ref{fig:miri_4303}.  The NIRCam F200W filter primarily covers stellar emission and is therefore used to remove starlight from the F770W data (see Section~\ref{sec:starsub}).

Each target was covered by small 1- to 4-pointing mosaics, chosen to maximize the overlap with the available optical and millimeter spectroscopy and cover the majority ($\geq 70\%$) of star formation activity traced by MIR emission in each galaxy (defined by a contour of 0.5 \mjysr\ at WISE 12 \micron).  Details of the data processing are described in \citet{Williams2024}.  Briefly, JWST data are reduced using the most up-to-date calibration reference data system (CRDS) context (jwst\_1201), and JWST calibration pipeline version v1.13.4, and the PHANGS-JWST data version 1.1.0.  In addition to the calibration pipeline, the background level in each filter is matched to existing wide-field \textit{Spitzer} or WISE data that extend off of the galaxy, as discussed in \citet{Leroy2023b}.  The JWST data are finally all convolved to a common Gaussian resolution of 0\farcs90, a resolution slightly larger than the F2100W beam, to improve signal-to-noise at F2100W and match the resolution of existing multiwavelength data \citep[see discussion in][]{Williams2024}.  This corresponds to only a moderate decrease in resolution (native resolution of 0\farcs67 at F2100W, the largest of this filter set) but greatly improves the signal-to-noise, as the noise in these filters is pixel based rather than resolution element based.  The convolution was implemented using the method described in \citet{Aniano2011}, using the WebbPSF\footnote{\url{https://stsci.app.box.com/v/jwst-simulated-psf-library}} models to generate kernels\footnote{\url{https://github.com/francbelf/jwst_kernels}}.  The pixel scale is left at the original pixel scale of 0\farcs11 which over samples the PSF, but this does not affect the analysis presented in this paper.

A 3$\sigma$ signal-to-noise cut is performed on each MIRI filter used in this work.  We determine the typical noise by finding empty sky regions in a subset of the maps that were large enough to include areas off the galaxy.  This was possible for: NGC~1087, NGC~1385, NGC~1433, NGC~1512, NGC~1566, and NGC~7496.  The 1$\sigma$ level was determined as the average of the standard deviations of the sky regions found in these six maps. All background regions showed similar standard deviations and median values. With this method, our 3$\sigma$ limits are 0.09~MJy/sr in F770W, 0.13 MJy/sr in F1130W, and 0.29 MJy/sr in F2100W. These values are in good agreement with the characterization of noise levels presented in \citet{Williams2024} and match well with the expected values from convolving the pipeline-generated error maps. Additionally, several of the nuclei in our sample saturated in F2100W, producing diffraction spikes across the maps.  By-eye masking is done to exclude regions where these spikes contaminate the maps.  In convolving to lower resolution, we also convolve the spike masks and use them to conservatively exclude any contaminated pixels by removing all pixels where the convolved mask has a value greater than 0.9.

\subsection{z0MGs Dust Maps}
\label{sec:z0mgs}
In order to assess the combination of JWST bands as tracers of dust properties, we compare the JWST data to the results of MIR to FIR spectral energy distribution fitting to the \citet{Draine&Li2007} dust models produced as part of the ``$z=0$ Multiwavelength Galaxy Synthesis'' program \citep[z0MGs,][Chastenet et al, in prep]{Leroy2019z0mgs, Chastenet2021}. The results of the \citet{Draine&Li2007} model fitting include maps of $q_{\rm{PAH}}$, which we use to calibrate our JWST photometric measurement of the PAH fraction, \rpah.  These fit results were produced using a combination of WISE and \textit{Herschel} Space Observatory data and the \citet{Draine&Li2007} dust models \citep[using the DustBFF code,][]{Gordon2014}.  

\subsection{Ancillary Data}
\label{sec:ancil_data}

In addition to the infrared data, we use ancillary data products to determine the gas and stellar properties of the galaxies in our sample.  These include \coto moment 0 maps from PHANGS--ALMA \citep{Leroy2021a, Leroy2021b}, \halpha maps from PHANGS--MUSE \citep{Emsellem2022}, and HI maps from several VLA projects including THINGS \citep{Walter2008} or MeerKAT \citep{Sun2022}, depending on data availability.  MeerKAT HI maps are used in instances where both VLA and MeerKAT data are available.

The sources of HI maps are listed in Table~\ref{tab:samp}.  These maps have resolutions ranging from 11\arcsec\ to 15\arcsec, and are left at their native resolutions for our analysis.  While the typical HI  resolution is much larger than the other maps used in this work, the smoothness and flatness of the atomic gas profile observed in the Milky Way and other galaxies suggests that convolution to matched resolutions will not greatly impact our comparisons \citep{Schruba2011, Bigiel2012, Leroy2013, Wong2013}.  The HI intensities are converted to atomic gas densities including helium and assuming optically thin HI gas, using the equation $\Sigma_{\rm{HI}}[\rm{M}_{\odot}\rm{pc}^{-2}] = 0.020 ~I_{\rm{HI}}$[K~km~s$^{ -1 }$].

The \coto moment 0 maps were observed with ALMA using the 12m, 7m, and total power array, producing maps with $\sim$1\arcsec~resolution \citep{Leroy2021b}.  The \coto intensities were converted to a molecular gas surface density using a constant $\alpha_{\rm CO, 1-0}$ of 4.35~M$_{\odot}$~pc$^{-2}$ (K~km~s$^{-1}$)$^{-1}$ and a \coto/ $^{12}$CO($1-0$) line ratio $R_{21} = 0.65$ \citep{denBrok2022, Leroy2021b}.  For this work we require only a coarse estimate of the molecular gas surface density and so neglect conversion factor variations due to metallicity effects, excitation variations, and opacity variations. We use the broad moment 0 masks, which maximize the \coto emission included in the moment 0 maps and are further described in \citet{Leroy2021b}.  

\halpha maps were obtained from the PHANGS--MUSE data \citep{Emsellem2022} for all 19 galaxies.  For this work, we use the convolved-optimized (``copt'') resolution maps.  The copt resolution provides a uniform PSF for all MUSE data for a single galaxy using the broadest PSF for all observations. Resolutions for each MUSE map are listed in Table~\ref{tab:samp}.  The \halpha line was fit using the \texttt{pPXF} tool to fit the stellar continuum (including H$\alpha$ absorption) and a Gaussian profile for each spaxel.  Further discussion of the MUSE data products can be found in \citet{Emsellem2022}. In general, the MUSE ``copt'' resolutions are very similar to or slightly larger than the 0\farcs9 resolution of the convolved F2100W data.  We do not do any additional steps of resolution matching, since we primarily rely on the nebular catalog for our analysis, as described below.

\subsection{PHANGS Data Products}
\label{sec:data_prods}
In addition to the maps described above, we rely on several of the PHANGS higher order data products to distinguish between environments and compare to physical properties in individual regions.  These data products are described below.

\hii regions are identified based on the catalogs compiled by \cite{Santoro2022} and further described in \cite{Groves2023}.  These nebular region catalogs were produced using the \halpha maps from the PHANGS--MUSE survey (described above) and the \texttt{HIIPhot} algorithm for segmentation described in \citet{Thilker2000}.  Below we distinguish between \rpah\ inside (\rpahneb) and outside (\rpahdiff) of these nebular regions. When we do so, we include all nebular regions in the catalogs, which are mostly \hii\ regions but also include some other nebular regions like supernova remnants. 

Metallicity maps for each galaxy were produced in \citet{Williams2022}.  These maps were created with the MUSE data described above \citep{Emsellem2022} and the \citet{Pilyugin2016} S calibration, which relies on the relative strengths of the H$\beta$, [OIII]$\lambda 5007$, [NII]$\lambda 6584$, and [SII]$\lambda 6717+6731$ emission lines.  All lines were corrected for dust extinction using the Balmer decrement before the metallicity was calculated.  The \citet{Williams2022} metallicity maps interpolate between \hii\ region metallicities using Gaussian Process Regression, yielding a fully sampled metallicity map for each galaxy. 

We also use the environmental maps from \citet{Querejeta2021} to isolate morphological environments within each galaxy.  These maps were created with the \textit{Spitzer} 3.6\micron\ images and primarily distinguish between centers, bars, spiral arms, inter-arm disks, and disks without strong spiral arms.

Finally, we use the extinction-corrected \halpha star formation rate (SFR) and stellar mass (M$_{\star}$) maps produced by \citet{Belfiore2023}.  These maps were produced using the integral field spectroscopy from the PHANGS--MUSE survey \citep{Emsellem2022}.  The \halpha measurements are corrected for attenuation using an $E(B-V)$ determined based on the measured Balmer decrement across the maps.  Stellar masses were determined using Voronoi binning of the MUSE data cubes and full-spectral fitting  using the \texttt{pPXF} tool \citep{Cappellari2004ppxf}, as described in \citet{Emsellem2022}, Section 5.2.4.   

\renewcommand{\thefigure}{\arabic{figure} (Cont.)}
\addtocounter{figure}{-1}

\begin{figure*}[t!]
    \centering
    \includegraphics[width=\textwidth,]{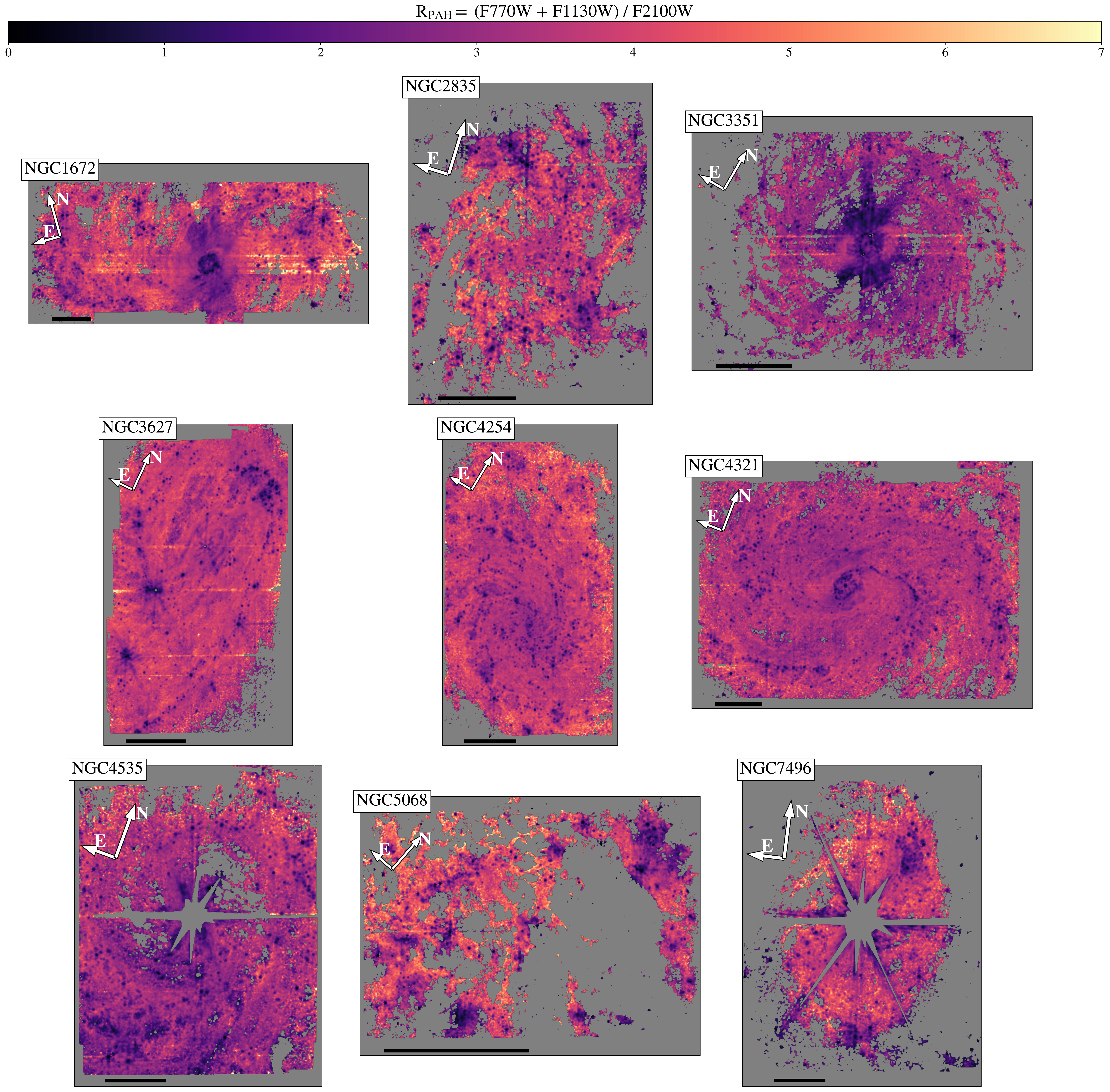}
    \caption{}
    \label{fig:rpah2}
\end{figure*}
\renewcommand{\thefigure}{\arabic{figure}}

\subsection{Methods}

\begin{figure*}[t]
    \includegraphics[width=\textwidth,]{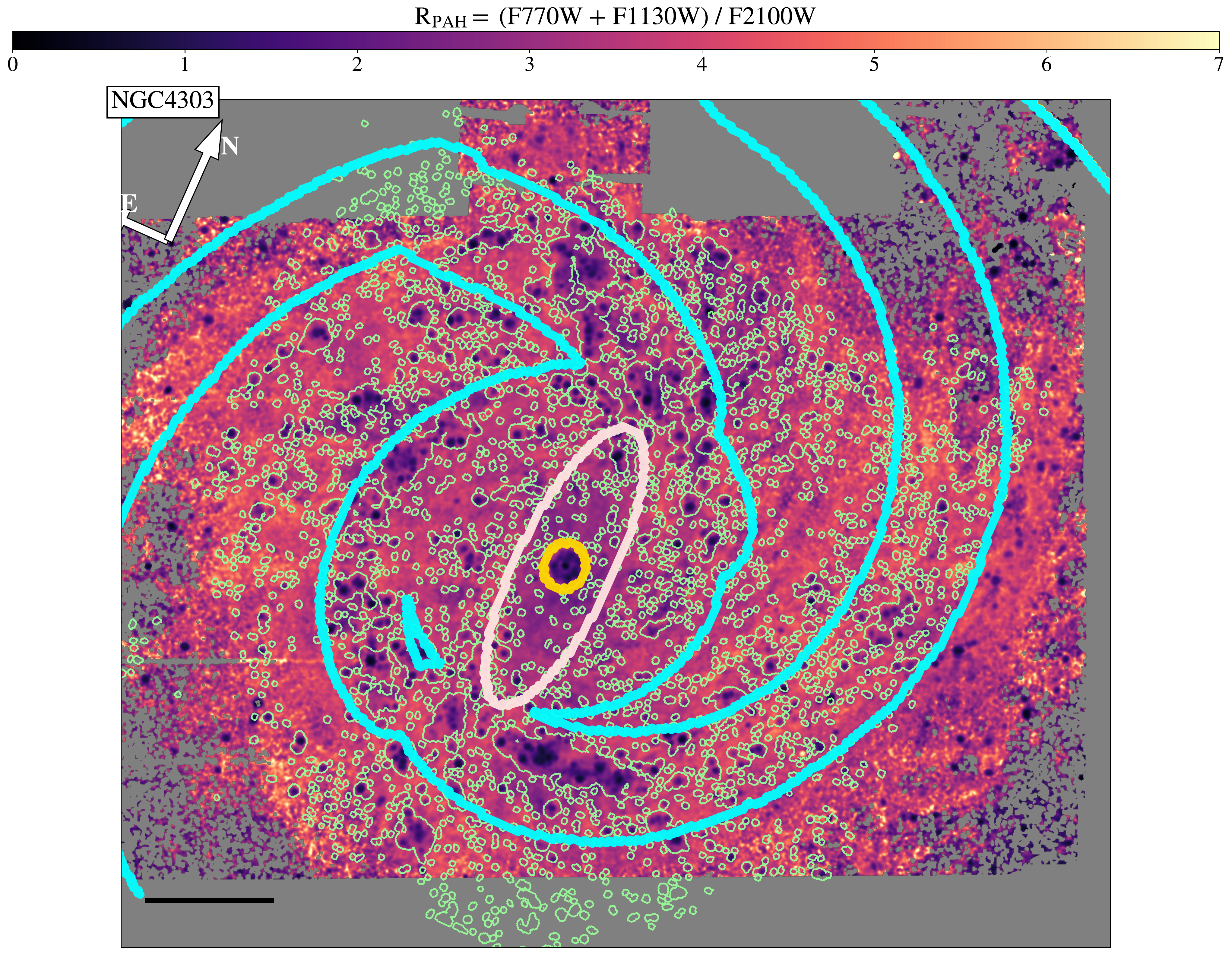}
   \caption{\rpah mapped in NGC~4303, with the F770W star subtraction described in Section~\ref{sec:starsub} completed before incorperation of the F770W into the \rpah map.  Yellow, pink, and blue contours show the location of the nucleus, bar, and spiral arms, respectively, as defined by the maps from \citet{Querejeta2021}.  Light green contours show the locations of the nebular regions from the PHANGS--MUSE nebular catalogs \citep{Santoro2022, Groves2023}.  The black bar on the bottom left represents a scale of 5~kpc.  Maps have been smoothed to 0\farcs9 to increase the signal to noise.}
    \label{fig:rpah_4303_big}
\end{figure*}

To test trends in the PAH distribution, we examine the MIRI data from each galaxy on three scales: pixel-by-pixel, in circular regions with radii of 1~kpc spaced using a hexagonal grid to evenly sample the full MIRI maps, and the average of each band across the full maps.  The pixel-by-pixel trends leverage the resolution of JWST and are the finest statistical sample available with this dataset. Because we are also interested in how large-scale environment affects \rpah\ (e.g. $\Sigma_{\star}$, $\Sigma_{\rm{SFR}}$, sSFR) we also use larger (1 kpc) regions.  Within each 1 kpc region, we record the values of \rpah in the nebular regions (\rpahneb) and also diffuse gas (\rpahdiff) and also note that this cannot be done pixel-by-pixel, as each pixel can only be in either a nebular region or a diffuse region.  Finally, in using the full maps, we determine how integrated galaxy properties might enhance or decrease the PAH fraction.  These integrated diagnostics also provide the most direct comparison to higher redshift measurements of PAHs, which will primarily yield galaxy--integrated PAH data.

\begin{deluxetable*}{lcccccccc}[t]
 \tablecaption{Average \rpah\ Values Across Environments}
    \tablehead{
         \colhead{Target} & \colhead{$\langle R_{\rm{PAH}}\rangle$} & \colhead{$\langle R_{\rm{PAH, Diff.}}\rangle$} & \colhead{$\langle R_{\rm{PAH, Neb.}}\rangle$} & \colhead{$\langle R_{\rm{PAH, Nuc.}}\rangle$} & \colhead{$\langle R_{\rm{PAH, Bar}}\rangle$} & \colhead{$\langle R_{\rm{PAH, Arms}}\rangle$} & \colhead{$\langle R_{\rm{PAH, IA}}\rangle$} & \colhead{$\langle R_{\rm{PAH, Disk}} \rangle$}}

    \startdata
        IC5332 &  2.53$_{1.42}^{3.62}$ &  2.54$_{1.43}^{3.64}$ &  2.57$_{1.47}^{3.60}$ &  3.26$_{2.28}^{3.89}$ &    ... &    ... &     ... &  2.51$_{1.42}^{3.61}$ \\
        NGC0628 &  3.68$_{2.79}^{4.45}$ &  3.72$_{2.84}^{4.50}$ &  3.56$_{2.66}^{4.28}$ &  3.41$_{2.88}^{3.84}$ &    ... &  3.70$_{2.85}^{4.43}$ &   3.72$_{2.72}^{4.55}$ &    ... \\
        NGC1087 &  3.47$_{2.45}^{4.16}$ &  3.55$_{2.53}^{4.23}$ &  3.25$_{2.34}^{3.92}$ &    ... &  3.21$_{2.43}^{3.66}$ &    ... &     ... &  3.51$_{2.55}^{4.18}$ \\
        NGC1300 &  2.99$_{1.99}^{3.83}$ &  2.97$_{1.96}^{3.83}$ &  3.10$_{2.21}^{3.82}$ &  3.14$_{2.34}^{3.64}$ &  2.84$_{1.88}^{3.57}$ &  3.26$_{2.40}^{4.05}$ &   2.78$_{1.79}^{3.75}$ &    ... \\
        NGC1365 &  3.85$_{2.70}^{4.95}$ &  3.91$_{2.78}^{5.01}$ &  3.60$_{2.41}^{4.63}$ &  2.79$_{2.49}^{3.00}$ &  3.77$_{2.84}^{4.71}$ &  3.73$_{2.77}^{4.48}$ &   3.92$_{2.64}^{5.07}$ &    ... \\
        NGC1385 &  3.59$_{2.56}^{4.56}$ &  3.76$_{2.80}^{4.69}$ &  3.14$_{2.17}^{4.12}$ &    ... &    ... &  3.74$_{2.62}^{4.70}$ &   3.55$_{2.54}^{4.49}$ &    ... \\
        NGC1433 &  2.89$_{1.80}^{3.77}$ &  2.85$_{1.76}^{3.76}$ &  3.08$_{2.17}^{3.83}$ &  3.52$_{2.73}^{4.13}$ &  2.74$_{1.62}^{3.62}$ &    ... &     ... &  2.91$_{1.84}^{3.79}$ \\
        NGC1512 &  2.81$_{1.75}^{3.73}$ &  2.76$_{1.70}^{3.71}$ &  3.05$_{2.17}^{3.81}$ &  2.72$_{2.06}^{3.13}$ &  2.29$_{1.37}^{3.22}$ &  3.07$_{2.13}^{3.85}$ &   2.85$_{1.77}^{3.85}$ &    ... \\
        NGC1566 &  3.41$_{2.45}^{4.19}$ &  3.44$_{2.53}^{4.20}$ &  3.27$_{2.31}^{3.97}$ &  2.48$_{1.65}^{3.51}$ &  3.22$_{2.60}^{3.62}$ &  3.43$_{2.47}^{4.15}$ &   3.44$_{2.44}^{4.28}$ &    ... \\
        NCG1672 &  3.57$_{2.56}^{4.40}$ &  3.65$_{2.68}^{4.47}$ &  3.31$_{2.26}^{4.14}$ &  1.77$_{1.11}^{2.49}$ &  3.43$_{2.56}^{4.20}$ &  3.70$_{2.89}^{4.45}$ &   3.64$_{2.57}^{4.50}$ &    ... \\
        NGC2835 &  3.37$_{2.22}^{4.29}$ &  3.49$_{2.34}^{4.38}$ &  3.14$_{2.07}^{4.05}$ &  3.52$_{3.11}^{3.95}$ &  3.48$_{2.88}^{3.98}$ &  3.33$_{2.15}^{4.23}$ &   3.37$_{2.20}^{4.37}$ &    ... \\
        NGC3351 &  2.77$_{1.84}^{3.47}$ &  2.79$_{1.89}^{3.48}$ &  2.84$_{1.96}^{3.48}$ &  1.73$_{1.22}^{2.24}$ &  2.35$_{1.33}^{3.23}$ &    ... &     ... &  2.84$_{2.02}^{3.52}$ \\
        NGC3627 &  3.60$_{2.96}^{4.06}$ &  3.68$_{3.21}^{4.08}$ &  3.39$_{2.53}^{3.90}$ &  3.11$_{2.50}^{3.57}$ &  3.39$_{2.76}^{3.77}$ &  3.58$_{2.83}^{4.01}$ &   3.65$_{3.05}^{4.11}$ &    ... \\
        NGC4254 &  3.63$_{2.96}^{4.26}$ &  3.73$_{3.18}^{4.36}$ &  3.34$_{2.55}^{3.95}$ &  2.88$_{2.29}^{3.11}$ &    ... &  3.61$_{2.94}^{4.28}$ &   3.64$_{3.02}^{4.25}$ &    ... \\
        NGC4303 &  3.53$_{2.73}^{4.11}$ &  3.64$_{3.03}^{4.15}$ &  3.36$_{2.50}^{3.93}$ &  2.29$_{1.33}^{3.14}$ &  3.03$_{2.65}^{3.42}$ &  3.54$_{2.73}^{4.07}$ &   3.57$_{2.78}^{4.16}$ &    ... \\
        NGC4321 &  3.28$_{2.69}^{3.82}$ &  3.32$_{2.78}^{3.85}$ &  3.11$_{2.35}^{3.64}$ &  2.61$_{1.84}^{3.31}$ &  3.21$_{2.75}^{3.58}$ &  3.26$_{2.70}^{3.75}$ &   3.33$_{2.69}^{3.92}$ &    ... \\
        NCG4535 &  2.99$_{2.18}^{3.76}$ &  2.95$_{2.15}^{3.69}$ &  2.91$_{2.13}^{3.55}$ &    ... &  2.64$_{1.89}^{3.15}$ &  3.20$_{2.52}^{3.83}$ &   2.96$_{2.12}^{3.78}$ &    ... \\
        NGC5068 &  3.68$_{2.28}^{4.74}$ &  4.00$_{2.68}^{4.95}$ &  3.21$_{1.95}^{4.31}$ &  2.67$_{2.00}^{3.39}$ &  3.86$_{2.75}^{4.59}$ &    ... &     ... &  3.66$_{2.24}^{4.76}$ \\
        NGC7496 &  3.29$_{1.86}^{4.24}$ &  3.33$_{1.95}^{4.29}$ &  3.30$_{2.11}^{4.13}$ &    ... &  3.31$_{2.44}^{3.86}$ &    ... &     ... &  3.29$_{1.84}^{4.28}$ \\
    \enddata
    \tablecomments{List of the average \rpah values for various environments in each galaxy.  A `...' is listed when the specified environment does not exist in the galaxy or was not included in the JWST maps.  The upper and lower values for each represent the 84$^{\rm{th}}$ and 16$^{\rm{th}}$ percentile for each environment, respectively.}
    \label{tab:rpah}
    \end{deluxetable*}

\section{Defining R$_{\rm PAH}$} \label{sec:RPAH_qpah}

Work by \citet{Engelbracht2005, Wu2006, Engelbracht2008} showed the utility of MIR photometric tracers of PAH fraction with {\em Spitzer} data.
In order to leverage the increased resolution of JWST to efficiently map the PAH fraction in a wide range of environments, we propose to update previous \textit{Spitzer}-based photometric tracers of \qpah with MIRI photometric bands. 
Previous studies with \textit{Spitzer} have shown 8~\micron\ / 24~\micron\ from the IRAC4 and MIPS1 bands respectively, is a good indicator of the PAH fraction \citep{Engelbracht2005,Smith2007, Engelbracht2008,Marble2010, Croxall2012}.  With the narrower MIRI filters, we can improve upon this indicator by including both the 7.7\micron\ and 11.3\micron\ PAH features (F770W and F1130W) in place of IRAC4 and replacing the MIPS1 24\micron\ emission with F2100W.  Just like the PAH emission features, the 21\micron\ continuum is often produced by stochastic heating events, though in regions of high radiation field intensity it can include a contribution from small grains in equilibrium with the radiation field. 
In the case where all three filters are dominated by stochastically heated dust emission, almost any dependence of \rpah\ on the interstellar radiation field will be removed \citep[][]{Draine&Li2007,Chastenet2023_RPAH}.  This is further explored in Appendix~\ref{app:models}, where we use the \citet{Draine2021} dust models to explore variations in predicted \rpah for different ISRFs.  Although it is possible that changes in the ISRF could slightly shift the measured \rpah, on the 10--50~kpc scales we are measuring this should not be the dominant effect. As we discuss below, current evidence suggests that almost all of our data will lie in or near this regime where all three bands are dominated by stochastic heating.

Based on the work done in \cite{Chastenet2023_RPAH} and \citet{Egorov2023}, we define our photometric tracer \rpah as:
$$
R_{\rm{PAH}} = \frac{\rm{F770W_{ss}} + \rm{F1130W}}{\rm{F2100W}},
$$
where F1130W and F2100W are the surface brightness in the MIRI bands measured in MJy/sr and F770W$_{\rm{ss}}$ is the surface brightness in the F770W band with a starlight subtraction (see Sec~\ref{sec:starsub}).  As the F770W and F1130W bands include two of the strongest PAH features (the 7.7~$\mu$m and 11.3~$\mu$m features, respectively) they are used as a proxy for total PAH emission.  We normalize by the F2100W band, which includes a significant contribution from non-PAH, small dust grain continuum emission.  \rpah is thus the ratio of PAH emission to other small grain emission, providing a high spatial resolution estimation of the PAH fraction without costly spectroscopy.   An example of a map of \rpah in NGC~4303 is shown in Figure~\ref{fig:rpah_4303_big}, with environments (yellow=nucleus, pink=bar, and cyan=spiral arms) and nebular regions (light green) indicated with colored contours.

We will use \rpah\ as a tracer of the PAH fraction. That interpretation relies on several assumptions: 
(1) the F770W and F1130W bands are dominated by PAH emission where PAHs are present, (2) that the combination of the 7.7~$\mu$m and 11.3~$\mu$m feature strengths is a good indicator of the total PAH emission,  (3) that the F2100W band includes a contribution from non-PAH dust continuum sufficient to give leverage on the PAH-to-continuum ratio, and (4) the MIR emission spectrum is primarily altered by the PAH fraction, or that any other effects altering the ratio (e.g. increases in the radiation field intensity) are covariant with changes in the PAH fraction and do not undermine the correlation between \rpah\ and \qpah.  

These assumptions all have clear evidence in their support. (1) Observations and dust emission models show that when PAHs are present and illuminated by typical interstellar radiation field intensities and spectra their emission dominates the F770W and F1130W filters \citep[see spectral decomposition of MIR spectra from][]{Whitcomb2023b}.  In cases where there is low ISM column relative to stellar mass surface density (e.g. central bulges of galaxies) the F770W filter can have a significant contribution from starlight which we will correct for (see further discussion in Section~\ref{sec:starsub}). 
(2) Models of PAH emission show that in a typical interstellar radiation field, the 7.7~$\mu$m emission is produced primarily by charged PAHs and the 11.3~$\mu$m feature by neutral PAHs, and the sum of both provides a robust tracer of PAH emission regardless of ionization state of the PAHs \citep{Smith2007, Draine&Li2007, Draine2021}. (3) The ratio of (F770W+F1130W)/F2100W traces the PAH fraction in data and in dust emission models in such a way that even with varying radiation field intensity and \qpah\ the F2100W provides sufficient leverage on dust continuum emission in comparison to PAH emission \citep[see Section~\ref{sec:rpah_qpah} for more detailed analysis;][]{Draine2021, Hensley2023}.  Finally, (4) Over the range of radiation field intensities relevant for the $10-50$ pc resolution of our observations, the \citet{Draine&Li2007} models of dust emission do not show substantial changes to \rpah\ related to radiation field intensity.  We will discuss this point in more detail in Appendix~\ref{app:radiation}. Changing the radiation field spectrum can alter the relationship between \qpah\ and \rpah, in the sense that harder spectra produce more PAH emission for a given PAH surface density \citep{Draine2021}.  This means that in regions with harder radiation fields, our \rpah\ tracer may suggest higher \qpah.  Since our primary observation is a decrease in \rpah\ towards regions of recent star formation, this effect would strengthen our conclusions.  We also investigate \rpah\ diagnostics using only F770W/F2100W and only F1130W/F2100W. These are described in Appendix~\ref{app:rpah_alt}, where we find that using only one PAH tracing band can provide a viable alternative tracer of PAH fraction, although both alternatives have the potential to confuse changes in the PAH fraction with changes in the size, ionization state, and heating of the PAHs \citep{Draine2021}.

\begin{figure*}[t]
    \centering
    \includegraphics[width=0.85\textwidth]{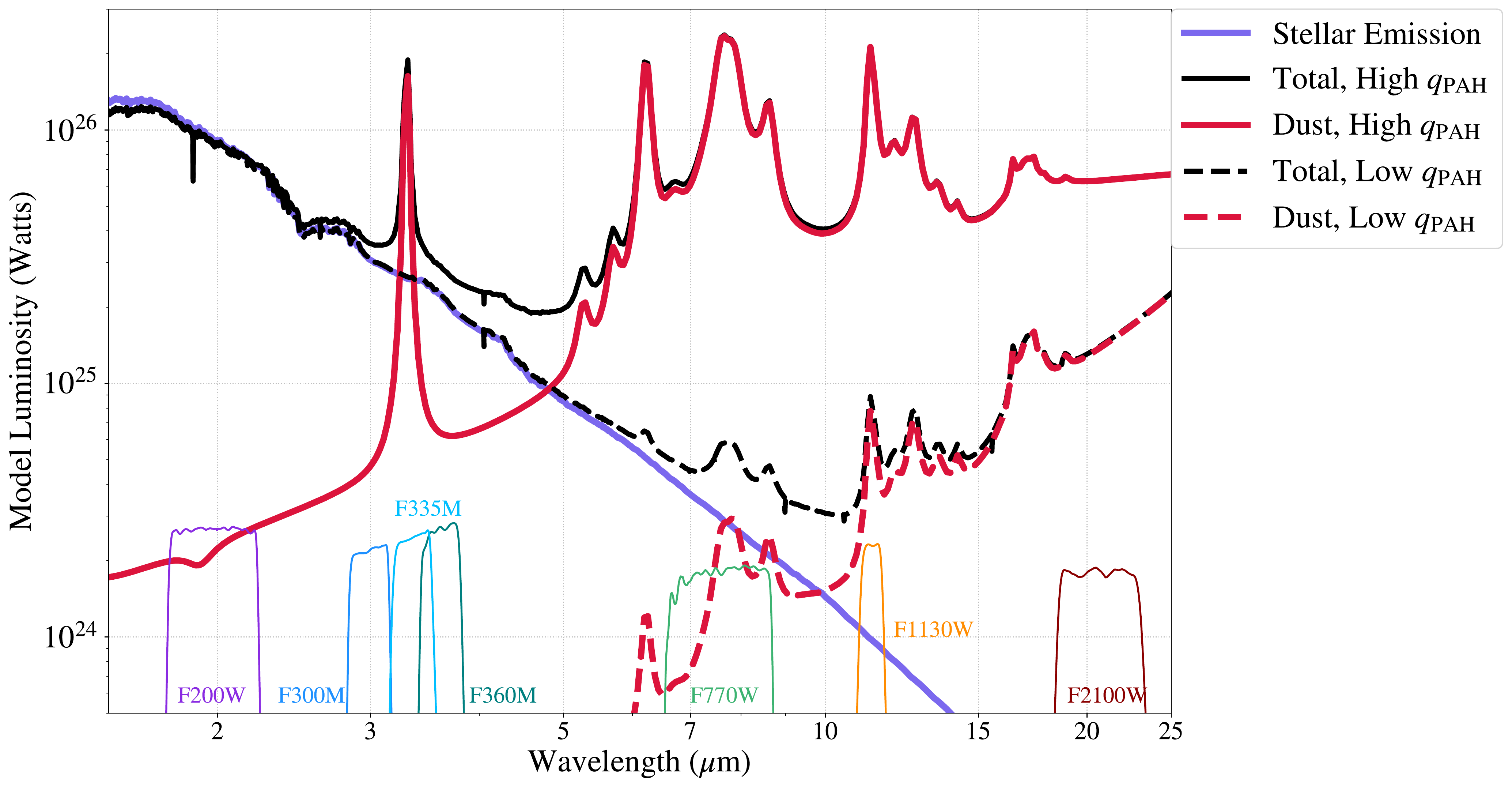}
    \caption{Two models of mid-infrared emission compiled using CIGALE.  Solid (high PAH content) and dashed (low PAH content) red lines represent the dust emission, while the solid and dashed black lines represent the total luminosity.  The same stellar radiation field was used in both models, and is shown as a solid blue line.  The filter profiles for a subset of the JWST filters are shown along the x-axis, for reference.}
    \label{fig:models}
\end{figure*}

Maps of \rpah for the complete sample are shown in Figures~\ref{fig:rpah1} and \ref{fig:rpah_4303_big}.  A 5~kpc scale bar has been placed in the lower-left of each image, and all images have the same linear scaling for \rpah.  In instances where an active galactic nucleus (AGN) saturated the center (NGC~1365, NGC~4535, and NGC~7496), masks have been placed to remove the affected pixels.  Galaxies have been rotated for ease of display.  The average \rpah values for each galaxy as well as for isolated environments are listed in Table~\ref{tab:rpah}.

\subsection{Starlight Subtraction}
\label{sec:starsub}

In rare cases in the PHANGS sample, where there is low dust column relative to the stellar surface density (e.g.\ in some galactic bulges) and/or a very low PAH fraction, the F770W band can have a significant contribution from starlight. F1130W is at a long enough wavelength where the contribution from starlight will always be minimal.  To demonstrate this, we show two SED models produced using the Code for Investigating GALaxy Emission \citep[CIGALE,][]{Boquien2019} in Figure~\ref{fig:models}.  In one model, the PAH fraction is set to a relatively high value of \qpah~=~6.63\% (solid lines), while in the other model the PAH fraction is set to a low value of \qpah~=~0.47\% (dashed lines).  The red and black lines represent the dust and total emission, respectively.  Both models assume the same stellar population, with the unattenuated starlight from this population shown as the solid purple line.  In the case of the low-PAH fraction model (or in the case of a model with the same \qpah but a far lower dust column density), it is clear that the F770W MIRI filter will have a substantial contribution from starlight. 

\begin{figure}
    \centering
    \includegraphics[width=0.48\textwidth]{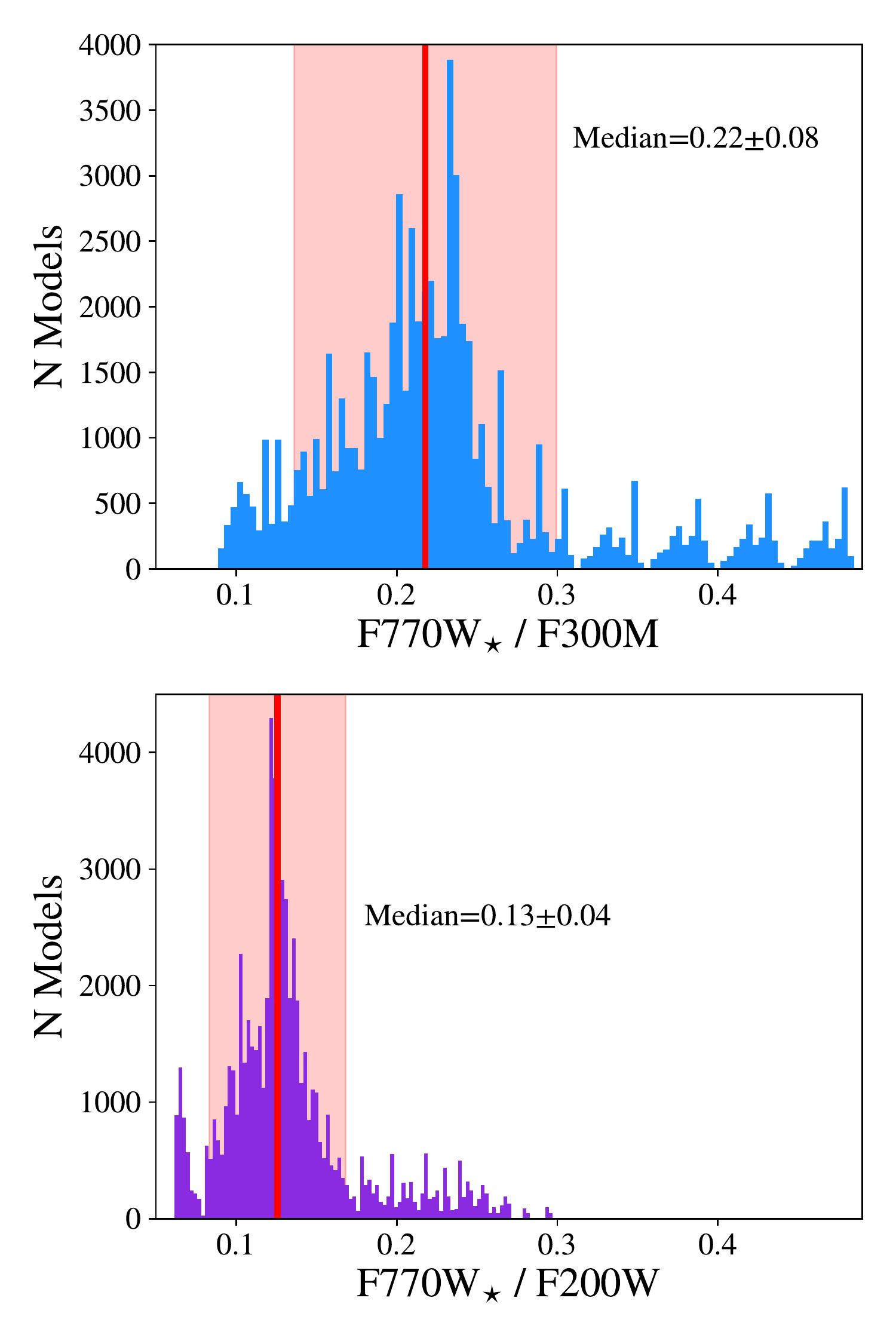}
    \caption{Using CIGALE to produce a range of stellar emission models, we determine the expected F770W$_{\star}$/F300M and F770W$_{\star}$/F200W ratios assuming only stellar emission in each band.  The resulting histograms are displayed, along with the medians as a solid red line, with the standard deviation displayed as a shaded area surrounding the median line.}
    \label{fig:star_ratios}
\end{figure}

In order to remove contributions from starlight from the F770W photometry, we scale the NIRCam F200W maps to predict the amount of starlight in the F770W band, following similar procedure to what was done by \citet{Helou2004} using {\em Spitzer} 3.6 \micron\ \citep[see also][]{Draine2007,Dale2009}. We use a ratio of F770W$_{\rm{starlight}}$/F200W = 0.13$\pm$0.04, determined from a range of reasonable stellar emission models with ages spanning 0.5~Gyr--13~Gyr, a later burst of star-formation, and the \citet{Chabrier2003} initial mass function.  In general, the results are fairly insensitive to the details of the stellar populations \citep[see also Appendix B of][for a detailed investigation]{Ciesla2014}. These starlight models are also assumed to experience attenuation following \citet{Calzetti2000}.  Dust emission is set using the \citet{Draine2014} models, with the available range of \qpah.  CIGALE does not couple the attenuation curve and \qpah\ assumption.  It assumes an energy balance, so all light attenuated in the optical and UV is re--emitted in the infrared. CIGALE does not adjust the 2175$\AA$ feature when \qpah changes, but for energy balance to be maintained, the integral of the infrared SED will impact the level of attenuation applied to the starlight.  The stellar emission models are produced using CIGALE \citep{Boquien2019}, with all additional properties held constant using the default CIGALE settings.  Synthetic photometry is performed on each stellar emission model to determine the relative fluxes in the F770W and the attenuated stellar emission model to determine the flux in the F200W bands.  We use the F200W band to correct for starlight instead of the F300M due to the higher signal-to-noise of the F200W measurements and the narrower range of F770W/F200W ratios produced by different stellar emission models, as shown in Figure~\ref{fig:star_ratios}.  

One potential issue with using the F200W band to correct for starlight is the possible presence of Paschen~$\alpha$ emission in this filter.  To confirm the small influence of the Paschen~$\alpha$ line, we once again use the grid of CIGALE models produced to estimate the contribution of Paschen~$\alpha$ emission to the F200W band.  As CIGALE also models the nebular emission, we are able to estimate the fraction of the F200W flux from Paschen~$\alpha$ emission across all of our models.  Within the range of models we test, we find the maximum contribution  Paschen~$\alpha$ is 2\%, with a median contribution of 0.04\%, confirming that this emission line will not significantly affect our starlight correction using the F200W filter.  It should be noted that the models tested here do not include a wide range of nebular gas emission properties, which is likely limiting the range of contributions from Paschen $\alpha$ to the F200W filter in these models.   We still advocate for the use of the F200W filter as an indicator of starlight in F770W for the galaxies included in our sample due to the higher signal to noise in F200W compared to F300M. 

As an additional test, star subtractions in F770W using both F200W and F300M were compared.  We find that the average differences between these measurements are $\leq0.01$~MJy/Sr, below the SNR threshold for the F770W data.  This suggests that the choice of starlight filter will not impact our overall results.  There are some instances where this may no longer be the case, specifically when the interstellar radiation field is dominated by an older stellar population.  In the top panel of Figure~\ref{fig:star_ratios}, several repeating peaks can be seen in the F770W$_\star$/F300M plot.  These peaks represent models with increasingly older stellar populations, which shifts the amount of modeled starlight in the F300M band relative to the F770W band.  Because the median does not reflect these older stellar populations, we find that the nuclei of our galaxies, which will be dominated by an older stellar population, starlight subtractions using F300M remove much less flux than the F200W starlight subtraction.  This is a further reason for our choice of the F200W band for the starlight subtraction, which is less dependant on the age of the stellar population.

\subsection{R$_{\rm PAH}$ as a Proxy for $q_{\rm PAH}$}
\label{sec:rpah_qpah}

\subsubsection{Comparison to Infrared SED Modeling Results}

\begin{figure*}[t]
    \centering
    \includegraphics[width=0.9\textwidth, ]{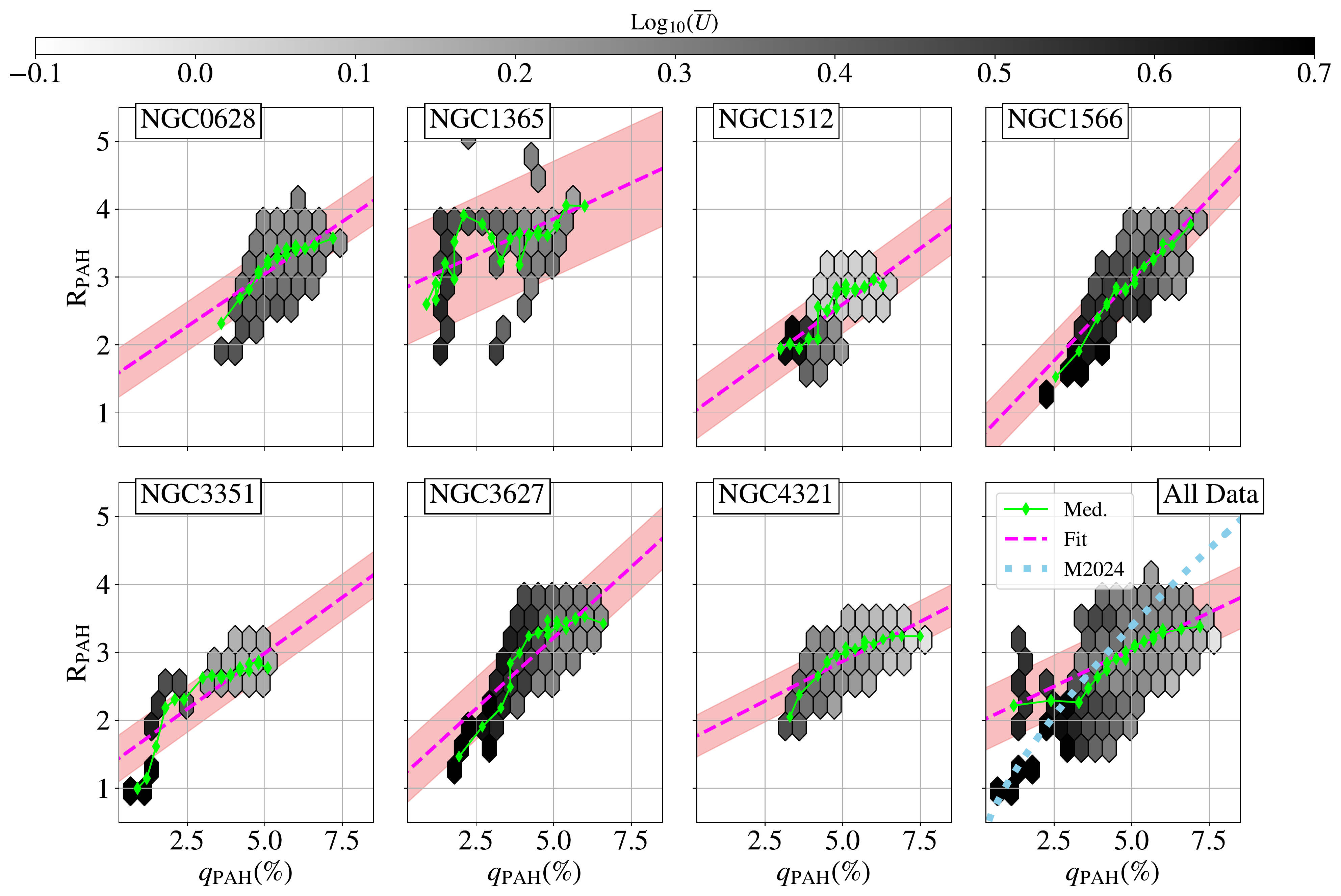}
    \caption{Binned data showing \rpah vs \qpah for the sub-sample of galaxies with available \qpah maps from the z0MGs survey \citep[][Chastenet et al. in prep]{Chastenet2021}.  Bins are color-coded by the average radiation field heating the dust, represented by the average $\log_{10}(\overline{U}) $ value.  Only bins with significant numbers of measurements are plotted.   Binned medians are shown as solid lines, while lines of best fit are displayed in each panel as dashed pink lines, with the 1$\sigma$ scatter shaded on either side.  The final panel shows the binned medians for all galaxies together, and also includes the predicted relationship between \rpah and \qpah from \citet{Matsumoto2024} [M2024] for dust grains heated by a 3~Gyr stellar population as a blue dotted line.}
    \label{fig:Rpahqpahbins}
\end{figure*}

To determine the effectiveness of using \rpah as a proxy for the fraction of the interstellar dust that is PAHs, we compare our measurements of \rpah to \qpah in a selection of galaxies where mid- to far-IR SED modeling has been carried out using the \citet{Draine&Li2007} models.  \qpah maps were created as part of z0MGs \citep{Leroy2019z0mgs}, as described in Section~\ref{sec:z0mgs} and \citet{Chastenet2021} and Chastenet et al. (in prep).  As the dust modeling uses far-IR photometry from \textit{Herschel} PACS and SPIRE, the maps of \qpah are much lower resolution (18\arcsec, the resolution of the SPIRE 250\micron\ data) than the MIRI data, and are only available for a subset of the galaxies in our sample: NGC~0628, NGC~1365, NGC~1512, NGC~1566, NGC~3351, NGC~3627, and NGC~4321.  These galaxies cover a wide range in morphological type and physical properties, and therefore can be considered representative of the full sample.  

In order to compare our measurements of \rpah to the modeled \qpah maps, we first smooth the MIRI data to the SPIRE resolution using smoothing kernel made by convolving the WebbPSF models with a 16\arcsec\ Gaussian, using the same method as described in Section~\ref{sec:jwst}.  In NGC~1365, which was masked to remove saturated pixels, the mask was also smoothed to the SPIRE resolution and all pixels with a convolved--mask value greater than 0.9 were discarded to remove all pixels impacted by the large diffraction spikes after convolution.  The comparison between these \rpah and \qpah measurements for each galaxy are shown in Figure~\ref{fig:Rpahqpahbins}. Bins are color-coded by the dust model estimates of $\log_{10}(\overline{U})$, the dust mass surface density weighted average intensity of the radiation field heating the dust \citep[see e.g.][for details]{Draine&Li2007, Aniano2012, Chastenet2019}.  Only bins with at least 8 measurements are included in the individual galaxy plots, while only bins with 20 or more measurements are included in the final panel with all seven sources.  Each panel also includes solid lines representing the binned medians and dashed lines showing the best-fit linear relationship between \rpah and \qpah.  The slopes and intercepts for each linear relationship are listed in Table~\ref{tab_qpahfits}.   The relative similarity between the slopes and interecepts of the five galaxies shows that the linear relationship between \rpah and \qpah is relatively consistent regardless of galaxy properties.  The one exception to this seems to be NGC~1365, where the JWST data had to undergo masking to remove many saturated pixels from the bright nucleus.  As we have tried to preserve as much usable data as possible, the chosen cuts may not be adequately removing pixels impacted by the saturation in the nucleus, leading to erroneous high or low values of F2100W, which could be partially responsible for the increased scatter here.  Additionally, cutting all the pixels from the nucleus, where \rpah tends to be lower and more uniform, will increase the scatter we see in the binned histograms shown in Figure~\ref{fig:Rpahqpahbins}.  Further discussion of the correlation between infrared photometry ratios and \qpah can be found in Appendix~\ref{app:radiation}.

\begin{deluxetable}{lcccc}[t]
    \tablecaption{Slopes of \rpah\ vs. \qpah\ Relationships.}
    \tablehead{
    \colhead{Target} & \colhead{Slope} & \colhead{Intercept} & \colhead{RMS} &
    \colhead{$R^2$} }
    \startdata
        NGC~0628 & 0.308 & 1.51 & 0.35 & 0.35 \\
        NGC~1365 & 0.228 & 2.73 & 0.85 & 0.15\\
        NGC~1512 & 0.332 & 0.96 & 0.42 & 0.40\\
        NGC~1566 & 0.473 & 0.60 & 0.42 & 0.60\\
        NGC~3351 & 0.329 & 1.33 & 0.34 & 0.55\\
        NGC~3627 & 0.421 & 1.11 &  0.45 & 0.62 \\
        NGC~4321 & 0.235 & 1.68 & 0.35 & 0.44\\
        All  & 0.330 & 1.30 & 0.42 & 0.49 \\
    \enddata

    \tablecomments{Slopes and intercepts found assuming a linear relationship between \rpah and \qpah, in the form $R_{\rm{PAH}} = m \times q_{\rm{PAH}} +b$. The RMS scatter and coefficient of determination ($R^2$, with values closer to 1 showing a better fit) are listed to demonstrate significance of relationships.}
    \label{tab_qpahfits}

\end{deluxetable}

\subsubsection{Expectation from Dust Models}
We also explore the \rpah-\qpah\ relation expected based on models.  We use the models of \citet{Draine&Li2007} and \citet{Hensley2023} to determine \rpah\ for models with a range of radiation field intensities $U$, \qpah, and size distributions.  Here $U$ is the radiation field intensity in units of the \citet{MMP1983} Solar neighborhood radiation field ($U_{\rm MMP}$).  We use the \cite{Draine&Li2007} over the \citet{Draine2021} models as these include \qpah as a model parameter, while the \citet{Draine2021} models do not.  Synthetic photometry for the F770W, F1130W, and F2100W filters was then performed on a range of the available \citet{Draine&Li2007} and \citet{Hensley2023} models, to track how \rpah varies with \qpah in these widely used models of PAH emission.  As these models \textit{only} include dust emission, we do not replicate our star-subtraction for the model fluxes.  

\begin{figure*}[t]
    \centering
    \includegraphics[width=0.95\textwidth,]{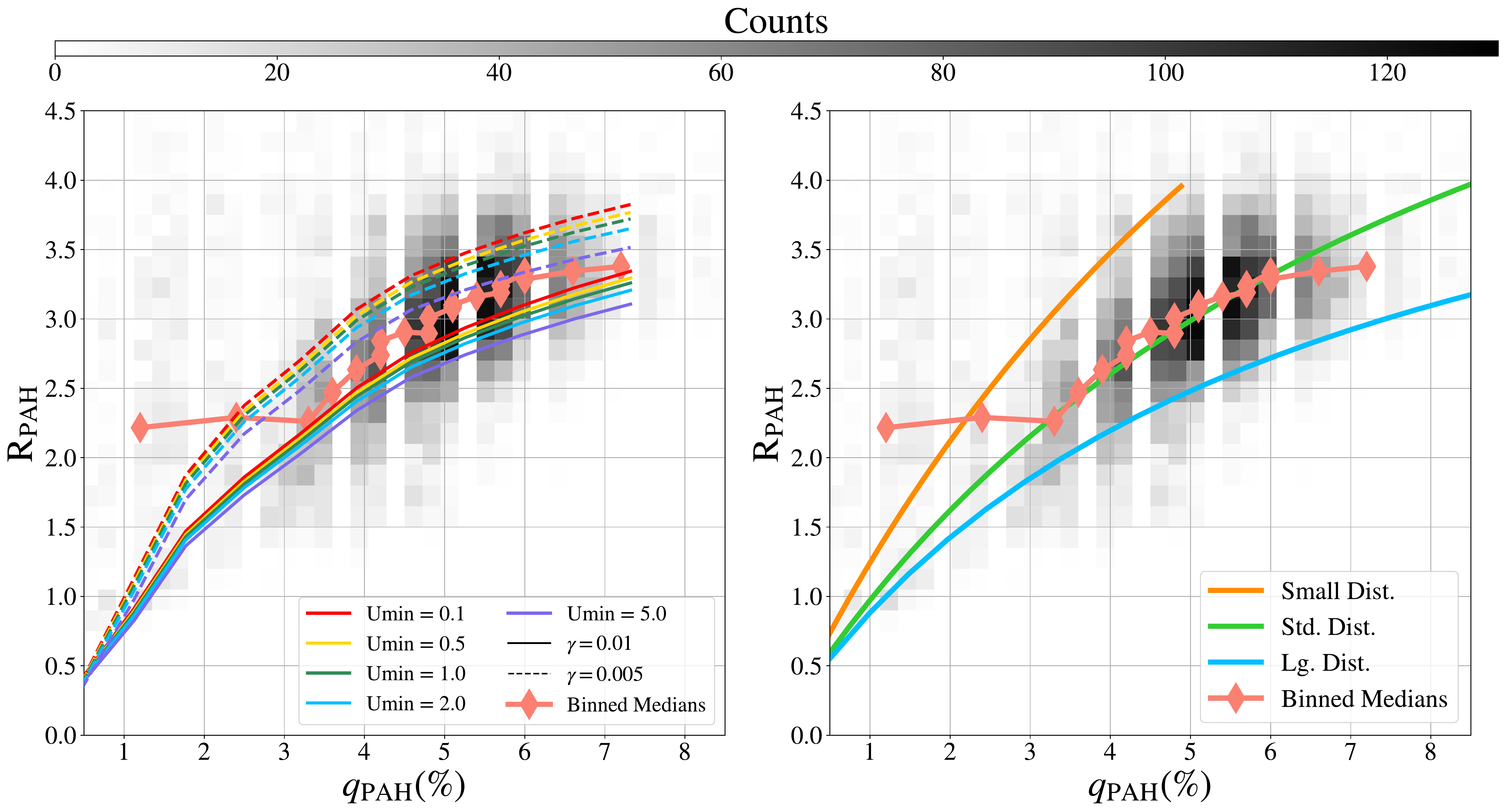}
    \caption{\rpah plotted against the modeled \qpah values for the five galaxies with \qpah maps (gray histogram, and pink diamonds representing the binned data).  Overplotted are predictions for \rpah as a function of \qpah from the \citet{Draine&Li2007} dust models (left) and \citet{Hensley2023} (right).  In the \citet{Draine&Li2007} models, we vary $U_{\rm min}$, the minimum value of the interstellar radiation field from 0.1 to 5.0, and use two values of $\gamma$, the fraction of dust heated by a radiation field greater than $U_{\rm min}$:  $\gamma = 0.01$ (solid lines) and $\gamma = 0.005$ (dashed lines).  The \citet{Hensley2023} models were measured using the small, standard, and large size distribution and standard ionization distribution, and varying the relative emission from PAHs and graphite, in a radiation field $U=1.0$.}
    \label{fig:Rpahqpahall}
\end{figure*}

The results of this analysis are shown in Figure~\ref{fig:Rpahqpahall}, along with the measurements from the sample with \qpah estimates, displayed as a gray histogram.  The estimates produced by the \citet{Draine&Li2007} models are shown on the left, where solid and dashed lines show the model values of \rpah for the range of \qpah available.  Different colors represent different values of $U_{\rm min}$, the minimum interstellar radiation field heating the dust and the solid and dashed lines represent two values of $\gamma$, the fraction of dust heated by a radiation field greater than $U_{\rm min}$, 0.01 (solid lines) and 0.005 (dashed lines).  The data almost completely fall within the ranges predicted by the models.  These models show a directly proportional relationship between \rpah and \qpah, with systematic shifts driven by changes in the radiation field heating the dust.

Three models of \rpah as a function of \qpah from \citet{Hensley2023} are overplotted on our sub-sample in the right panel.  The \cite{Hensley2023} models were produced using three PAH size distributions: the small, standard, and large size distributions described in \citet{Draine2021}, a $\log_{10}$U of 1.0, and $\gamma=0$. We note that in this comparison, the radiation field is assumed to have a fixed spectral shape, given by the \citet{MMP1983} Solar neighborhood radiation field. Changes in the radiation field hardness can also shift the \rpah-\qpah relationship.  As shown in Appendix~\ref{app:rpah_alt}, changing the PAH ionization fraction does not noticeably shift the measurement of \rpah, as long as U and the size distribution are held constant, so changes to the ionization fraction are not included here.  This seems to be due to the fact that charged PAHs emit more in the F770W band, while neutral PAHs emit more in the F1130W band, so changing the charge distribution only shifts the relative strengths of the two features, but not the total emission in these two bands.  In these models, \qpah is varied by changing the $B_1$ and $B_2$ parameters in the PAH distribution \citep[see][equation 18]{Hensley2023}.  All other \citet{Hensley2023} model parameters are kept at those recommended in the paper.   For all three size distributions, \rpah increases with \qpah, although the smallest PAHs produce the largest values of \rpah for a given \qpah. 

To summarize, the basic theoretical expectation from dust models is that \rpah and \qpah show a clear, characteristic correspondence and that the radiation field intensity perturbs this only moderately over the range $U\sim0.1-10$.

\begin{figure*}[t]
    \centering
    \includegraphics[width=\textwidth,]{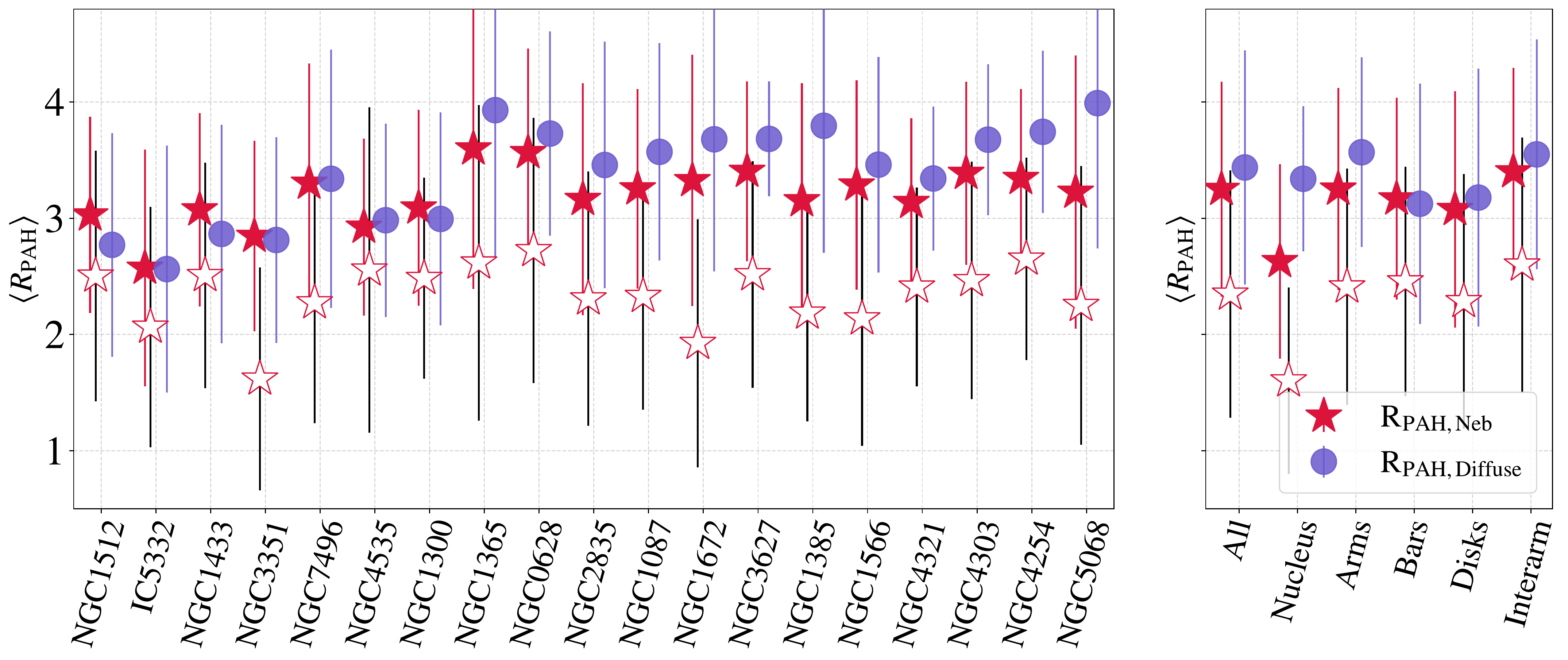}
    \caption{Median \rpah values for each galaxy within (red stars, \rpahneb) and outside (blue circles, \rpahdiff) of the nebular regions classified by \citet{Groves2023}. Vertical bars on each point represent $1 \sigma$ scatter.  Empty red stars are the \halpha-weighted averages of \rpahneb, with black error bars representing the \halpha-weighted scatter. Galaxies are ordered by increasing specific star formation rate.}
    \label{fig:NEBHII}
\end{figure*}

Additionally, while we see a relatively linear relationship between \rpah and \qpah  in both the models and the data at low \qpah (with the exception of data from NGC~1365, which is driving the observed flattening in \rpah at low \qpah as it contains the bulk of our low \qpah pixels), the slope of this relationship decreases slightly in the models at an \rpah value of $\sim$3.5 above \qpah of 5\%.  Examinations of the \citet{Draine&Li2007} models suggest this flattening effect is caused by the smaller relative contribution of non--PAH continuum emission to the F2100W band at high \qpah.  Although there are no PAH emission features in this band, the \citet{Draine2021} models, which split emission up into carbonaceous grains (PAHs and graphite) and AstroDust, show continuum emission from small grains, including PAHs, contributing  up to $\sim$80\% of the F2100W flux at $q_{\rm{PAH}} \geq 4.5$\%.  Despite this, as there are no longer-wavelength photometric dust continuum tracers at similar sensitivities and resolution, \rpah is the best possible tracer of PAH abundance at high angular resolution in most conditions expected to be found in typical galaxy ISMs.  Although this indicates that \rpah is less sensitive to changes in \qpah at $q_{\rm{PAH}} \geq 4.5$\%, the linear trend below \qpah of 4.5 shows that \rpah can provide a robust tracer of regions with low \qpah, where it is likely PAH destruction will lower \qpah.  


\section{Trends in $R_\mathrm{PAH}$}
\label{sec:Analysis}
With these results verifying \rpah traces \qpah in ISM conditions expected of our sample, we can now investigate how \rpah changes as a function of galaxy properties.  By comparing \rpah to the relative fractions of ionized, molecular, and atomic gas, as well as the ISM metallicity and galaxy environment, we can establish where within the ISM we expect the PAH fraction to vary.  Additionally, by isolating \rpah from the nebular regions identified in \citet{Groves2023}, we can determine how the PAH fraction in the vicinity of \hii regions (\rpahneb) differs from that in the diffuse gas (\rpahdiff).  As our definition of ``diffuse'' gas is based only on the nebular regions identified using the on average 70~pc resolution \halpha maps, it will include both diffuse gas and molecular clouds.  Throughout this paper we will use \rpahdiff to represent all \rpah measurements in regions not identified as nebular regions, which will be dominated by diffuse gas but will include other non--ionized gas as well.

\subsection{Decreases in R$_{\rm PAH}$ measured in HII Regions}

\subsubsection{Differences in Nebular Region vs Diffuse Gas \rpah Averages}
\label{sec:gal_ave_destruction}

The most notable trend in \rpah is the steep decrease seen within the nebular regions, suggestive of the destruction of PAHs within \hii regions.  This can be seen in Figure~\ref{fig:rpah_4303_big}, as lower \rpah values (darker purple color) within the nebular region contours. Similar trends in subsets of the PHANGS-JWST galaxies were identified by \citet{Chastenet2023_RPAH} and \citet{Egorov2023}.  We quantify this difference using the nebular catalogs of \citet{Groves2023} to isolate nebular regions and compare to diffuse gas.  We include all nebular regions identified in \citet{Groves2023} within the coverage of our MIRI observations.  It is worth noting that the definition of a nebular region is limited by the resolution of the MUSE maps, and the nebular catalog boundary generally extends well beyond the limit of the actual \hii regions \citep[see][for further discussion]{Barnes2022}.  Future work comparing \rpah to the higher resolution Hubble Space Telescope \halpha data could better address the effects of spatial resolution on isolating the nebular regions, but for the purposes of this work we treat the nebular regions as a conservative cut that isolates \hii regions from the surrounding neutral ISM.  In order to adjust for the fact that the nebular regions extend outside of the boundary of the actual \hii regions (i.e. include some contribution from the diffuse ISM), we also investigate the I(\halpha) weighted average \rpahneb values.  By weighting \rpahneb by the \halpha surface brightness, the pixels clearly inside the \hii regions play a larger role setting the average \rpahneb value, while those sampling mainly diffuse neutral gas will have less of an impact. 

Across the full sample of galaxies, the \halpha-surface brightness weighted average of \rpah within the nebular regions is $2.31$ with a $ \pm 1\sigma$ scatter of $\pm 0.69$ while the average of \rpah outside of the nebular regions (not weighted by \halpha) is $3.43$ with a $\pm 1\sigma$ scatter of $\pm  0.71$, showing a 1.6 times the standard deviation of \rpahneb increase in \rpah outside of nebular regions.  The difference in average values measured between the diffuse gas and nebular regions implies a lower PAH fraction within \hii regions, suggestive of PAH destruction.

\begin{figure*}[t]
    \centering
    \includegraphics[width=\textwidth,]{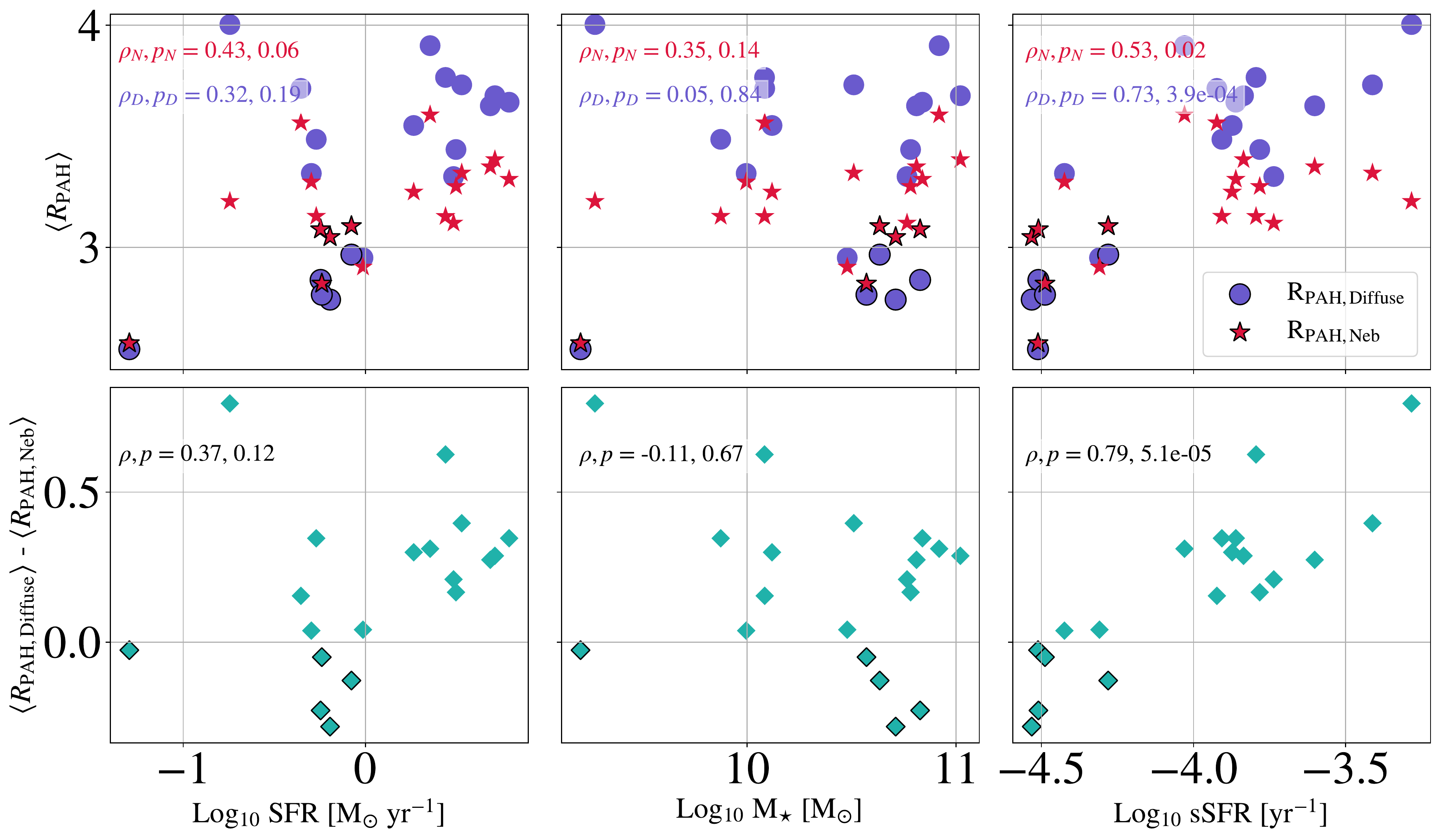}
    \caption{\textit{Top Row:} Average \rpah values plotted as a function of SFR (left), stellar mass (middle) and sSFR (right).  Red stars are the average \rpah within the nebular regions and blue circles are the average \rpah in the diffuse gas, for each galaxy.  SFR, M$_{\star}$, and sSFR were all computed using the same method as in Figure~\ref{fig:NEBHII}.  Points outlined in black represent galaxies where \avrpahneb is greater than \avrpahdiff.  \textit{Bottom Row:} The differences between the average \rpah in the nebular and diffuse gas, plotted as a function of SFR, M$_{\star}$, and sSFR.  Diamonds outlined in black represent galaxies where 
    \avrpahneb is greater than \avrpahdiff.}
    \label{fig:deltaRPAH}
\end{figure*}

The difference in \rpah between nebular and diffuse regions is further illustrated in Figure~\ref{fig:NEBHII}, which displays the median values of \rpah within the nebular regions, \avrpahneb, and outside of the nebular regions, \avrpahdiff, for each galaxy (left panel), as well as in the full sample and individual environments (right panel).  Error bars on each point represent the $1 \sigma$ standard deviation of the values of all pixels included in those regions.  The empty red stars are the \halpha-weighted \avrpahneb averages, which are consistently lower than the non-\halpha-weighted median \avrpahneb values.  This emphasizes that the highest surface brightness \hii regions, which will often have the highest U values, have the lowest values of \rpah.  We note that changing the radiation field intensity can also contribute to changing the mapping between \rpah and \qpah, though we argue (see Appendix~\ref{app:radiation}) that this is a minor effect for the range of radiation field intensities we expect.  The low \rpah values in the regions with high \halpha surface brightness and radiation field intensities was also explored by \citet{Egorov2023}, who found anti-correlations between \rpah and ionization parameter and \hii region surface brightness brightness.  

By examining the right panel of Figure~\ref{fig:NEBHII}, we find that across the various galactic environments (arms, bars, nuclei, disks, and interarm regions), the difference between \avrpahneb and \avrpahdiff is largest for the nuclei and spiral arms, that together host the bulk of the star formation.  In the more quiescent disks, bars, and inter-arm regions, the differences between \rpah in the nebular and diffuse gas shrinks. In these more quiescent environments, it is also expected that the nebular regions would be smaller and therefore more difficult to resolve in the MUSE data.  This could lead to some miss-identified pixels, which would drive the diffuse and nebular \rpah measurements closer together.  Further discussion of this point is presented in Section~\ref{sec:1kpc}.

In the left panel of Figure \ref{fig:NEBHII}, the galaxies are sorted by the specific star formation rate (sSFR), where sSFR was computed only including data within the area covered by the JWST footprints, determined using the maps from \citet{Belfiore2023} (see Section~\ref{sec:data_prods} for details). Using this order, we see that while both the weighted and unweighted values of \avrpahneb are relatively constant across the sample, the \avrpahdiff increases with sSFR.  This is further demonstrated in Figure~\ref{fig:deltaRPAH}, whose top row shows  \avrpahneb and \avrpahdiff plotted as a function of SFR, M$_{\star}$, and sSFR. The bottom row plots the difference between the average \rpah in the diffuse and nebular regions, as a function of these same variables.
Spearman's rank correlation coefficient ($\rho$) and p-value ($p$) are listed in each panel of Figure~\ref{fig:deltaRPAH}, showing that the most significant trends we observe are between \avrpahdiff and log$_{10}$sSFR  with $\rho_D = 0.73$, and 
\avrpahdiff$-$\avrpahneb
and log$_{10}$sSFR  with $\rho = 0.79$.  All of the other panels have correlations that would not be considered significant based on their $p$-values.

While the majority of the galaxies in this sample have \avrpahdiff $>$ \avrpahneb, a subset with low sSFR show the opposite trend: NGC~1512, NGC~1433, NGC~3351, IC~5332.  In addition, NGC~7496 and NGC~1300 have nearly equivalent \rpahdiff and \rpahneb.  Of these galaxies, four share a similar structure, with a strong bar that extends to a bright outer ring (NGC~1512, NGC~1433, NGC~1300, and NGC~3351). Within these JWST images, the stellar bar takes up a large fraction of the field of view, and lacks any significant SF.  By examining the \halpha data, we find that the bars in these galaxies appear ``dark'', with little ongoing star formation. This phenomenon is widely observed in barred galaxies, and described as ``star formation deserts'' \citep{James2018,Neumann2020}. Because there are few nebular regions within these bars or the areas surrounding them, they make a substantial contribution to the \avrpahdiff for these sources.  Across the maps of these four galaxies, we find that these environments have very low fluxes in all MIRI bands, likely indicative that the dominant interstellar radiation field heating the dust is from an older stellar population or a low gas surface density.   
These low-SFR bars could therefore be driving down our measurements of \avrpahdiff through softer radiation fields that have fewer UV photons to excite the PAHs across these star--formation deserts, even if \qpah has not changed. 

The other two galaxies that show lower \avrpahdiff compared to \avrpahneb are IC~5332, the lowest metallicity galaxy of our sample, and NGC~7496, which contains both a bar and an AGN.  While the difference in IC~5332 \textit{could} be due to its lower metallicity, NGC~5068, which has a similar metallicity, has one of the largest positive differences between \avrpahdiff and \avrpahneb, making it unlikely metallicity alone is driving this shift.  Alternatively, IC~5332 and NGC~7496 both have low sSFR, while NGC~5068 has the highest sSFR of our sample, suggesting that these differences are more driven by changes in sSFR than other ISM conditions.  This effect will be further investigated in Section~\ref{sec:1kpc}.

\subsubsection{Pixel--Based Measurements}
\begin{figure*}
    \centering
    \includegraphics[width=0.9\textwidth,]{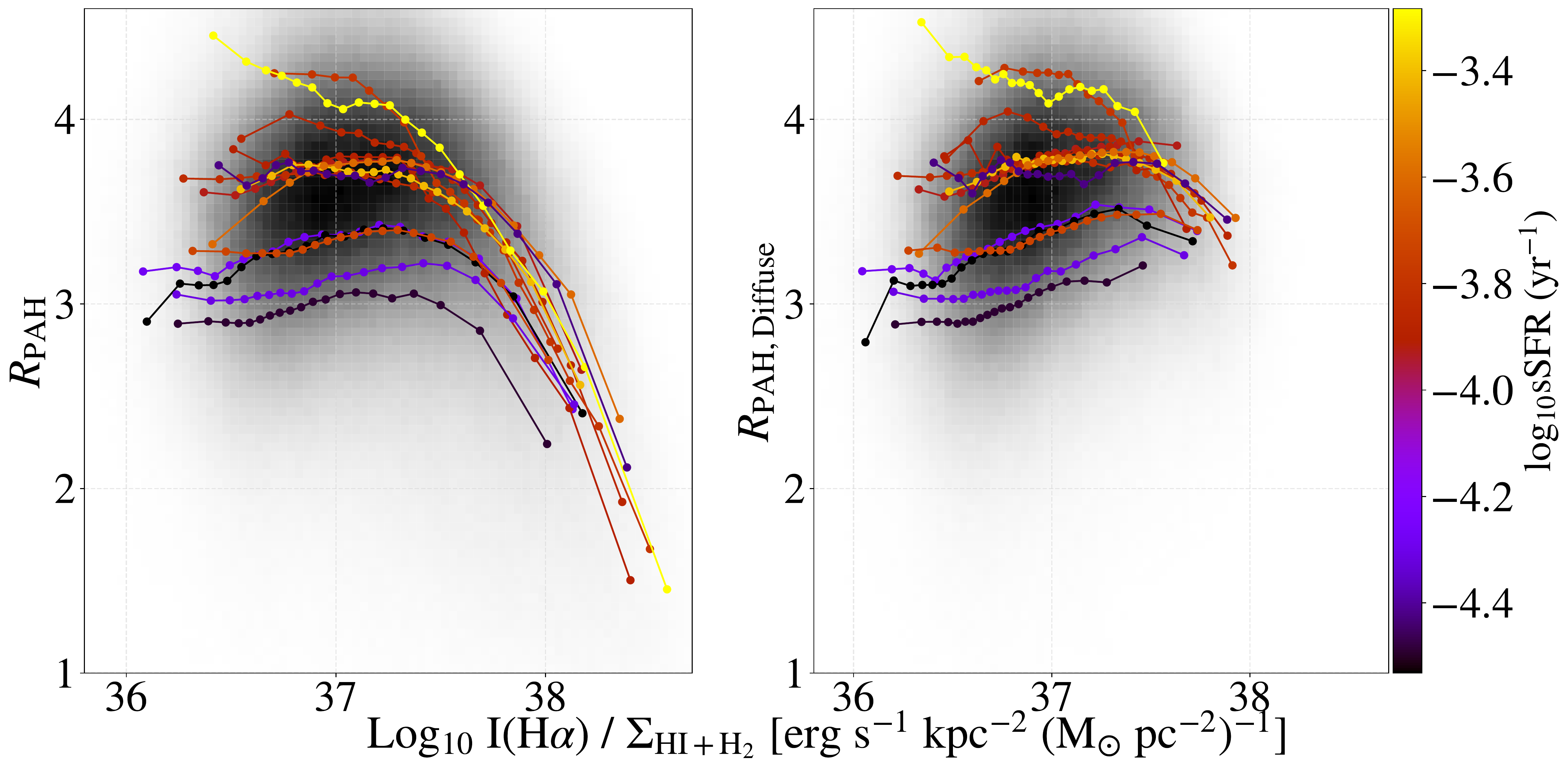}
    \caption{\rpah vs \hagasratio for the full sample with HI data.  Left panel shows all data, right panel shows only \rpahdiff.  Lines represented binned medians for each galaxy, color-coded by sSFR.}
    \label{fig:gas_all}
\end{figure*}

In addition to examining the galaxy-averaged nebular vs.\ diffuse \rpah trends presented above, we can trace changes in \rpah using our pixel--based measurements and compare to a tracer of the relative amount of ionized gas in each pixel: \newhagasratio,  following work by \citet{Chastenet2019,Chastenet2023_RPAH}.  This ratio uses the intensity of the \halpha line as a proxy for the amount of ionized gas, and normalizes by the total surface density of neutral (HI) and molecular (H$_2$) gas.  Figure~\ref{fig:gas_all} shows the behavior of \rpah, both in nebular and diffuse regions, as a function of $\log_{10} (\newhagasratio)$.
Each colored line represents the binned medians for an individual galaxy, color-coded by increasing log$_{10}$sSFR value inside the JWST footprint.  Individual galaxy pixel-based histograms are available in Appendix~\ref{sec:full_samp_hists}.     Each galaxy shows a decreasing trend above $\log_{10} (\newhagasratio)\sim$ 37.5 erg s$^{-1}$ kpc$^{-2}$ (M$_{\odot}$ pc$^{-2}$)$^{-1}$. This is similar to what was seen for the smaller sample presented in \citet{Chastenet2023_RPAH}.  Regions with $\log_{10}(\newhagasratio) \geq$ 37.5 erg s$^{-1}$ kpc$^{-2}$ (M$_{\odot}$ pc$^{-2}$)$^{-1}$ are primarily regions identified as nebular regions.  In the right panel of Figure~\ref{fig:gas_all}, we can see that \rpahdiff does not have a strong dependence on \newhagasratio.  Here we see very few pixels above $\log_{10}(\newhagasratio) = $ 37.5 erg s$^{-1}$ kpc$^{-2}$ (M$_{\odot}$ pc$^{-2}$)$^{-1}$, and no decreasing trend, validating the expectation that the decreasing trend is being driven by \rpah changes within the nebular regions. Figure~\ref{fig:gas_all} also shows that the binned median \rpahdiff\ at a given $\newhagasratio$ increases with average sSFR, following the trend seen for the galaxy-average \rpahdiff.

\subsubsection{1~kpc Scales}
\label{sec:1kpc}

In Figure~\ref{fig:NEBHII}, the average \rpah for each individual galaxy is plotted, and trends in these quantities are shown in Figure~\ref{fig:deltaRPAH}. The most significant trend we find is the correlation between \avrpahdiff$-$\avrpahneb and log$_{10}$sSFR, as shown by the Spearman's rank correlation coefficient: $\rho = 0.79, p = 5.1\times10^{-5}$.  To further examine this trend and determine whether sSFR remains the dominant driver of changes to the average \rpah on smaller scales, we examined the \avrpahdiff and \avrpahneb in 1~kpc regions.  These 1~kpc regions allow us to break down the galaxies by environment and test whether the correlations between the average \rpah and I(\halpha) or log$_{10}$SFR become tighter than log$_{10}$sSFR when smaller scales are considered.  1~kpc regions are chosen because this size is large enough to contain both significant numbers of pixels identified as nebular and diffuse, while also allowing us to obtain a meaningful measurement of sSFR, which would not be possible in isolated diffuse gas pixels.  

\begin{figure*}
\centering
  \centering  \includegraphics[width=.95\linewidth]{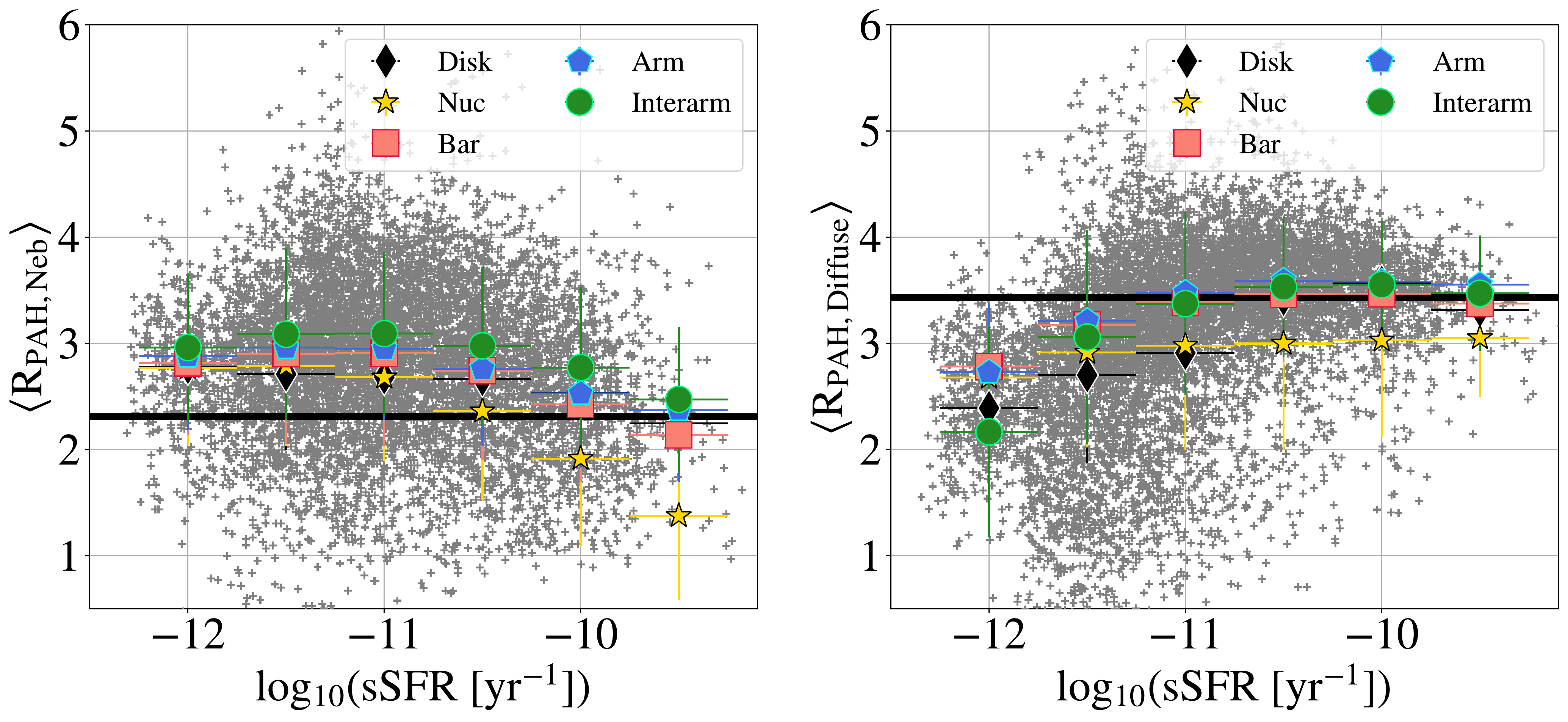}
\caption{\textit{Left:} \avrpahneb, measured in 1~kpc regions, plotted as a function of log$_{10}$sSFR.  Binned medians are shown over the full distribution, and are color-coded by environment of origin, with black diamonds being disks, red squares being bars, yellow stars being nuclei, green circles being interarm, and blue pentagons being arms.  The black line represents the H$\alpha$ weighted median value of \avrpahneb found across the full sample.  \textit{Right: }Same, but for \avrpahdiff, with the black line representing the median value of \avrpahdiff found across the full sample without H$\alpha$ weighting.}
\label{fig:1kpcRpah}
\end{figure*}

Examining the trends in sSFR, I(\halpha), and SFR with both \avrpahdiff and \avrpahneb measured in 1~kpc regions, we find that sSFR remains the best predictor of changes in \avrpahdiff.  The relationship between \avrpahdiff and log$_{10}$sSFR is shown in the right panel of Figure~\ref{fig:1kpcRpah}, with \avrpahneb in the left panel.  In this Figure, we can see  across the range of log$_{10}$sSFR, \avrpahneb decreases slightly, while \avrpahdiff increases and then remains constant. The fact that the increasing trend in \avrpahdiff and log$_{10}$sSFR is still found at 1~kpc scales implies that higher sSFR increases \rpahdiff.  This is likley due to the fact that at low sSFR the radiation field will be more dominated by softer radiation from an older stellar population that less effectively excites the PAHs \citet{Draine2021}, lowering our observations of \rpahdiff in regions with low sSFR.

\begin{figure*}
\centering
  \centering  \includegraphics[width=.95\linewidth]{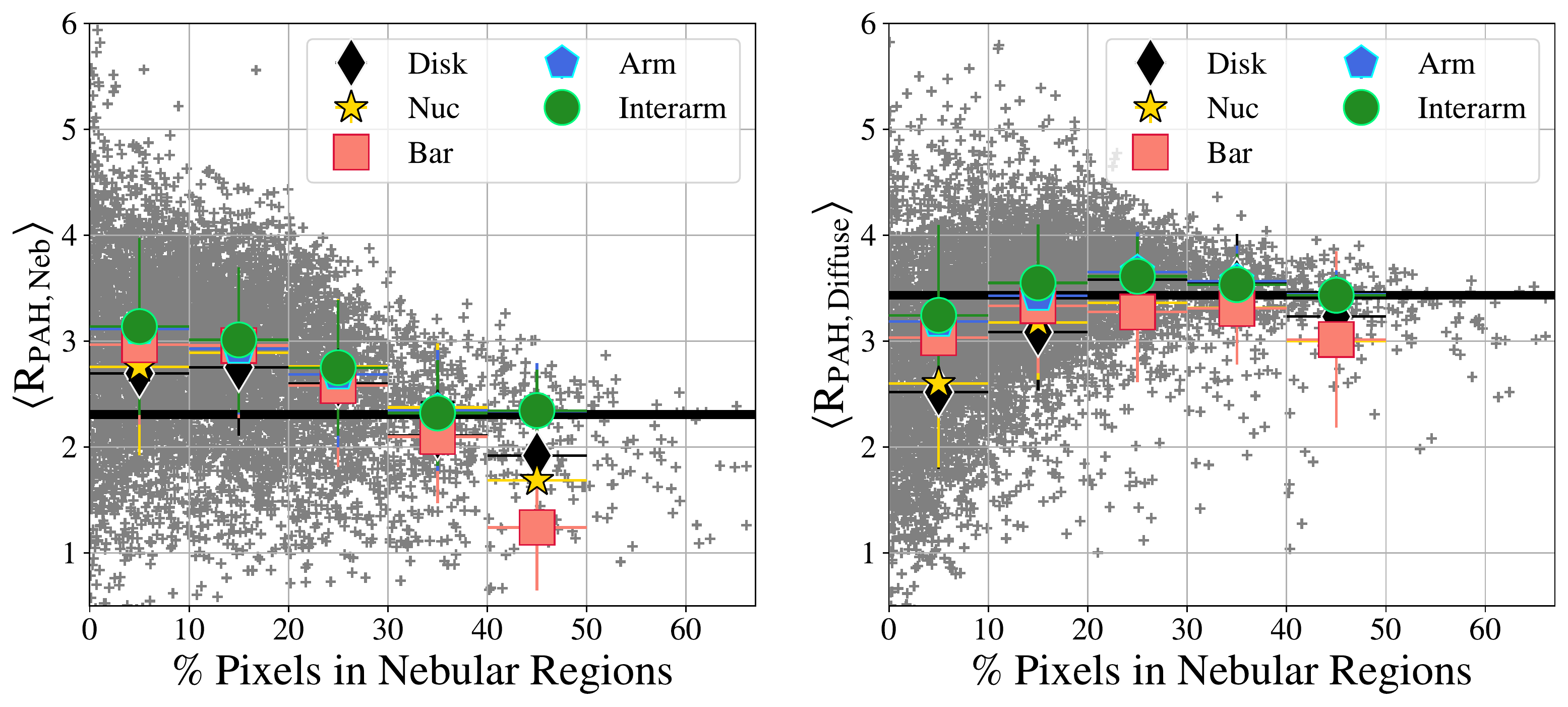}
\caption{\textit{Left:} \rpahneb, measured in 1~kpc regions, plotted as a function of the fraction of the pixels in the region in a nebular region.  Symbols are the same as in Figure~\ref{fig:1kpcRpah}. \textit{Right: }Same, but for \rpahdiff.}
\label{fig:1kpcfracneb}
\end{figure*}

Another possible factor to consider while investigating the trends observed in \avrpahdiff$-$\avrpahneb is the limited resolution of the MUSE maps. 
Due to this limited resolution, the nebular regions defined in \citet{Groves2023} contain some diffuse gas. This is caused by the inability to perfectly resolve the smallest \hii regions with MUSE.  This effect was confirmed in \citet{Barnes2022}, which found \hii regions measured by HST \halpha were smaller than those measured by MUSE, with typical radii $\lesssim 10$~pc.  If we assume the main factor changing \rpahneb across our sample is destruction of PAHs within \hii regions, we might expect that \rpahneb and \rpahdiff will have two distinct and relatively constant values representing these two phases of the ISM.  If smaller nebular regions and the edges of larger nebular regions are contaminated by diffuse gas,  we would expect \avrpahneb to shift upwards towards \avrpahdiff in nebular regions where this diffuse gas contamination is more significant.  We can test this with our 1~kpc measurements by determining what percentage of the pixels in each region are identified as within a nebular region.  1~kpc regions that cover the centers of large, well defined \hii regions will have the majority of their pixels identified as part of the nebular region catalog, while 1~kpc regions containing smaller, unresolved nebular regions or situated at the edges of a larger nebular regions will have low percentages of pixels identified as part of the nebular catalog.  If these regions with lower percentages of pixels in nebular regions have more contamination from diffuse gas, we would expect \avrpahneb to increase toward the \avrpahdiff value of 3.43.   We test this prediction in Figure~\ref{fig:1kpcfracneb}, where the left panel shows the \avrpahneb measured in the 1~kpc regions as a function of the \% of pixels identified as within a nebular region and the right panel shows the trends in \avrpahdiff.  The black line in the left panel represents the \halpha weighted value of \avrpahneb found above, while the black line in the right pane represents the non-H$\alpha$-weighted \avrpahdiff.  In the 1~kpc regions with near 30\% of their pixels in nebular regions, we see both \avrpahdiff and \avrpahneb flatten towards these black lines.  This implies that around the larger nebular regions, where the MUSE data can better distinguish between nebular and diffuse gas, \rpah becomes remarkably constant in both the nebular regions (at a value of 2.31 with $\pm1\sigma$ scatter of 0.78) and in the diffuse gas (at a value of $3.43$ with $\pm1\sigma$ scatter of 0.98).

\subsection{Further Investigations of R$_{PAH}$ Outside of HII Regions}

As shown above, the most obvious factor driving changes in \rpah is likely the destruction of PAHs within \hii regions.  Outside of \hii regions, \rpah is relatively constant at a value of $\sim 3.43$, indicating either a constant \qpah in the diffuse gas or possibly that we are in a high \qpah regime, where the mapping between \rpah and \qpah begins to flatten (see Section~\ref{sec:rpah_qpah} for details).  Comparing to the \citet{Hensley2023} models, we can determine what possible values of \qpah correspond to the average value of \rpah we measure in the diffuse gas.  A range of possible \qpah values for different single-$U$ radiation field strengths and PAH size distributions are listed in Table \ref{tab:HD2023}. We find that our average \rpahdiff of 3.43 corresponds to a \qpah of 5.5\% using the \citet{Hensley2023} dust models assuming the \citet{Draine2021} standard grain distribution, 8.1\% assuming the large grain distribution, or 3.91\% using the small grain distribution, in a radiation field with $\log_{10}$U=0.0 and $\gamma=0$.  We note that fixing $\gamma=0$ (i.e. using single values of $U$ rather than a distribution) will produce the widest range of inferred \qpah values for a given \rpah, since there is no contribution to the MIR emission from radiation fields that extends up to high intensities. As can be seen in \citet{Draine&Li2007} Figure 18, contributions from the power-law component can be important at 21 \micron.  Therefore, Table~\ref{tab:HD2023} represents the widest range of \qpah measurements that could yield \rpah$=3.43$.  In the remainder of this Section, we will examine any secondary trends in \rpahdiff that we can measure with the full suite of ancillary datasets available.  

\begin{deluxetable}{lccc}
    \tablecaption{Predicted \qpah\ from \citet{Hensley2023} for \rpah=3.43}
    \tablehead{
    \colhead{Model} & \colhead{log$_{10}$U$= -1 $} & \colhead{log$_{10}$U$= 0$ } & \colhead{log$_{10}$U$= 1$} }
    \startdata   
        \textbf{Sm} & 3.20\% &  3.91\% &   4.88\%   \\
        \hline
         \textbf{Std} & 4.55 \% &  5.55\%  & 9.14\%   \\
         \hline
         \textbf{Lg}  & 6.70\% & 8.12\% &  13.1\%    \\
         \hline
    \enddata
    \tablecomments{Predictions of \qpah from the \citet{Hensley2023} dust models with single radiation field intensities, to match \rpahdiff~=~3.43 based on the averages from the 1~kpc regions measured in the full sample.  Sm, Std, and Lg correspond to the Small, Standard, and Large PAH size distributions from \citet{Draine2021}.} 
    \label{tab:HD2023}
\end{deluxetable}

\subsubsection{\rpahdiff and Galaxy Environment}

\begin{figure}
    \centering
    \includegraphics[width=0.49\textwidth,]{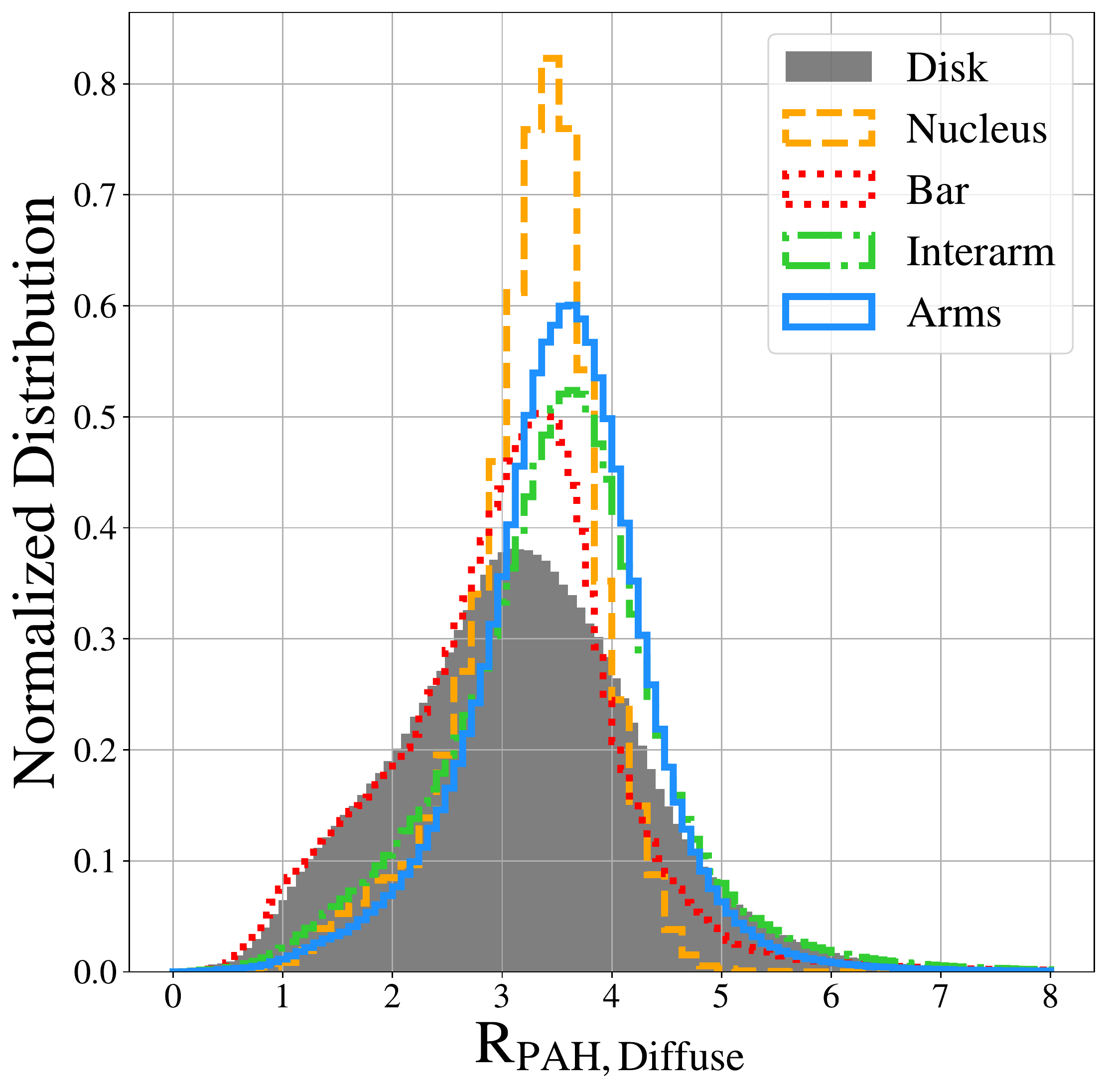}
    \caption{Histograms showing the distribution of \rpah for different galactic environments, as defined by \citet{Querejeta2021}.  Nebular regions have been excluded to remove the effects of PAH destruction in \hii regions.  Histograms showing individual galaxies can be found in Appendix~\ref{sec:full_samp_hists}}
    \label{fig:env_hist}
\end{figure}
Using the environment maps from \citet{Querejeta2021} which allow us to sort pixels by the structure they fall in, we examine how \rpahdiff varies as a function of galactic environment. Histograms showing the pixel-based distribution of \rpahdiff in galaxy nuclei, bars, spiral arms, interarm regions, and disks without significant arms are shown in Figure~\ref{fig:env_hist}.  Examining these histograms, we find that \rpahdiff tends to be highest in the spiral arms, and lowest in bars and disks without spiral arms.  Galaxy nuclei have the narrowest distributions of \rpahdiff, while bars and extended disks have the broadest.  Overall though, all five environments show similar distributions of \rpahdiff, suggesting that galactic environment does not drive changes to \rpahdiff.   Histograms of the distribution of \rpahdiff in isolated environments for each galaxy in our sample can be found in Appendix~\ref{sec:full_samp_hists}

\subsubsection{\avrpahdiff and Integrated Gas Properties}
\begin{figure*}
    \centering
    \includegraphics[width=0.9\textwidth,]{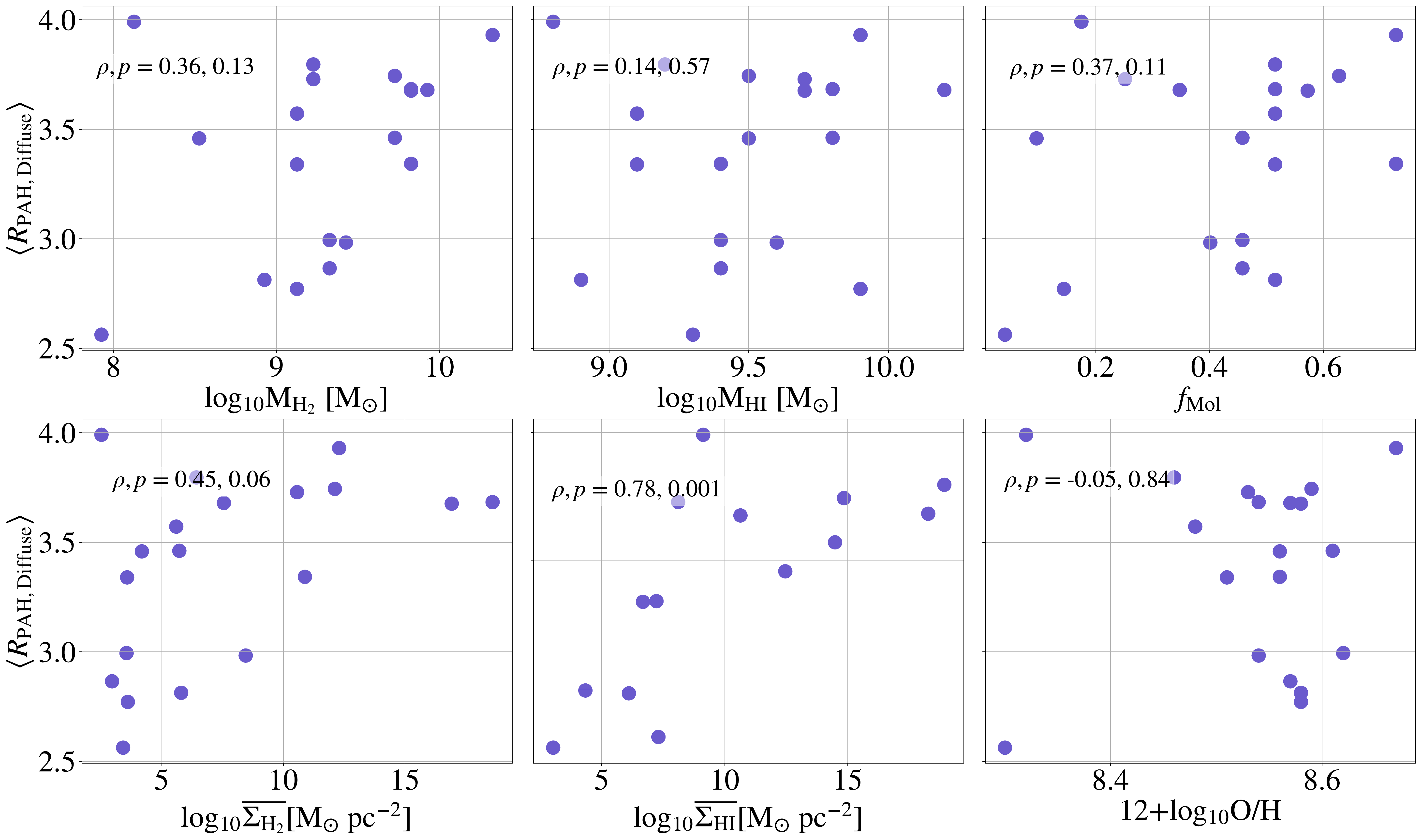}
    \caption{\textit{Top row:}  \avrpahdiff as a function of global gas mass, both molecular and atomic and global molecular gas fraction ($f_{\rm{mol}}$) from \citet{Lee2023}.  \textit{Bottom row:} \avrpahdiff as a function of average atomic ($\overline{\Sigma_{\rm{HI}}}$) and molecular gas ($\overline{\Sigma_{\rm{H}_2}}$) surface density determined in the area covered by the JWST data and 12+log$_{10}$O/H estimated at $R_{\rm{eff}}$ from \citet{Groves2023}.}
    \label{fig:diff_integrated}
\end{figure*}
Similar to the analysis in Section~\ref{sec:gal_ave_destruction}, here we examine how \avrpahdiff for each galaxy varies as a function of global galaxy gas properties.  Figure~\ref{fig:diff_integrated} shows \avrpahdiff measurements over the full JWST maps as a function of global M$_{\rm{H_2}}$, M$_{\rm{HI}}$,\fmol, and $12+$log$_{10}$O/H from \citet{Lee2023}, as well as the average $\overline{\Sigma_{\rm{H}_2}}$ and $\overline{\Sigma_{\rm{HI}}}$ within the area covered by the JWST data.  Spearman's $\rho$ and $p$ values are listed in each panel.  While we find no significant trends in average \rpahdiff as a function of global galaxy properties, we do see some indication of increases in \rpahdiff with higher M$_{\rm{H_2}}$ and $\overline{\Sigma_{\rm{HI}}}$.  It is worth noting that while M$_{\rm{H_2}}$ is well correlated with the SFR, as one would expect based on the Kennicutt-Schmidt relationship, we find none of these quantities is correlated with sSFR, suggesting any correlations found between \avrpahdiff and M$_{\rm{H_2}}$ or $\overline{\Sigma_{\rm{HI}}}$ is not driven by the correlation between \avrpahdiff and sSFR, but a separate ISM process.

\subsubsection{\rpahdiff as a function of Metallicity and H$_2$ gas fraction on Pixel Scales}

\begin{figure*}
\centering
  \centering
  \includegraphics[width=.45\linewidth]{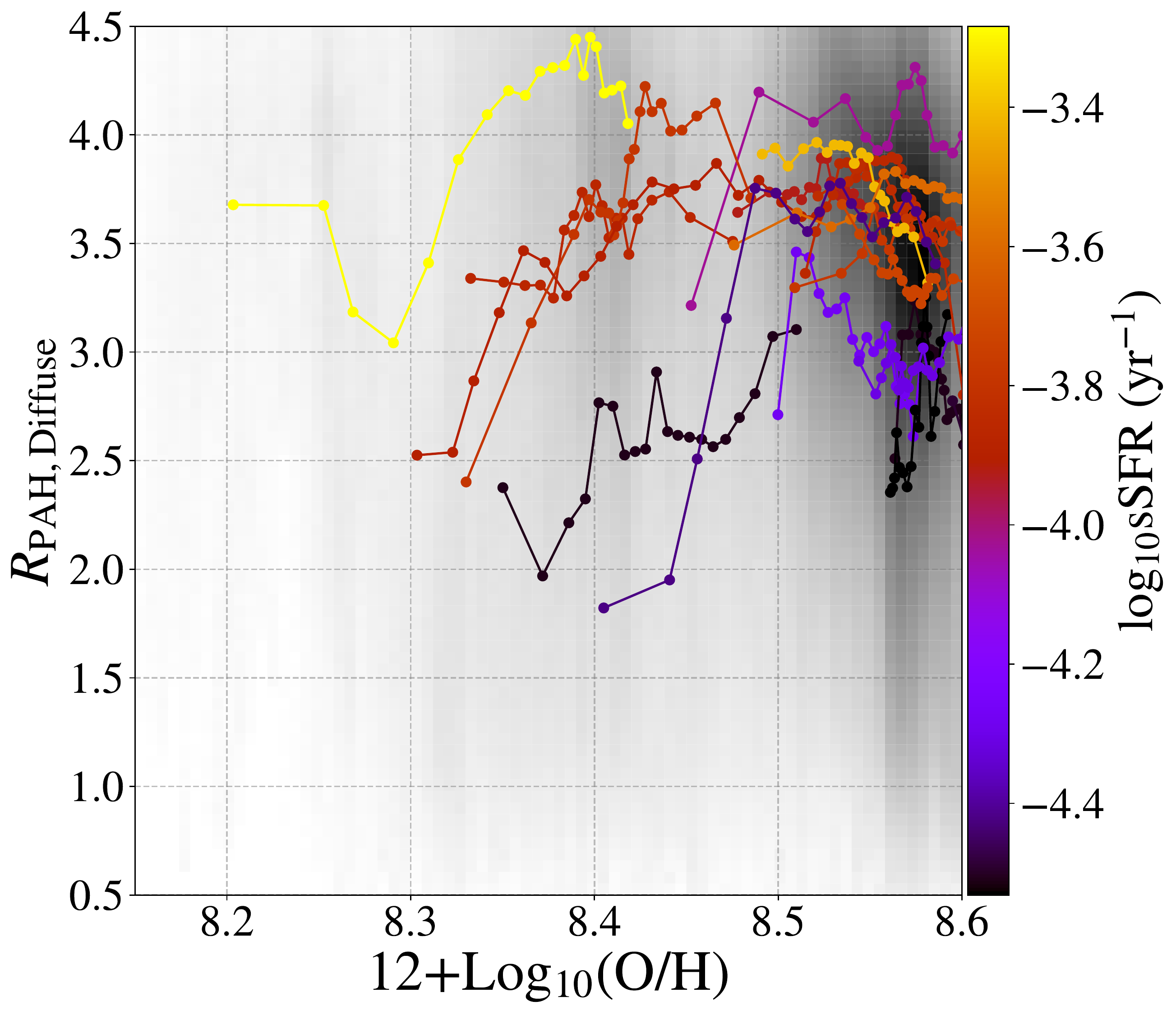}
\includegraphics[width=.45\linewidth]{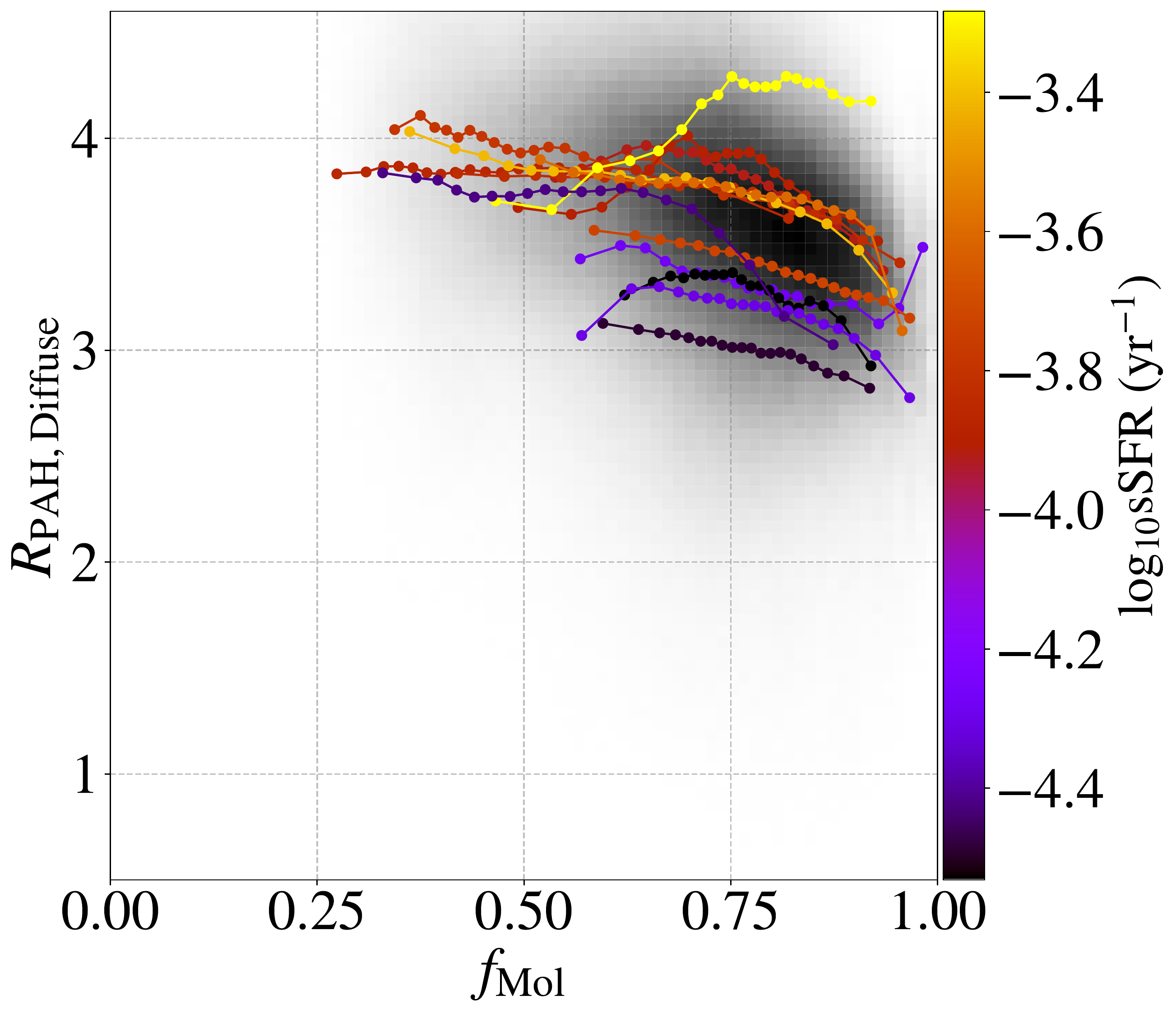}

\caption{The per pixel values of \rpahdiff plotted as a function of metallicity (right) and \fmol (left).  Metallicities were determined using metallicity maps of \citet{Williams2022}.  Binned medians are shown for each galaxy, color-coded by sSFR.  Histograms for individual galaxies are displayed in Appendix~\ref{sec:full_samp_hists}.}
\label{fig:fmol&metals}
\end{figure*}

Studies of PAH emission from the Small and Large Magellanic Clouds \citep{Chastenet2019, Paradis2023} as well as in nearby galaxies \citep[see e.g.][]{Engelbracht2005,Calzetti2007, Draine2007,  Gordon2008, Li2020} have shown that PAH abundance tends to decrease in low-metallicity environments.  We test this trend in our sample by plotting \rpahdiff as a function of 12+$\log_{10}$(O/H), using the maps of \citep{Williams2022}.  The pixel-by-pixel results of this analysis are shown in the left panel of Figure~\ref{fig:fmol&metals}.

The galaxies in our sample are all near solar metallicity and span a fairly small range ($\sim0.3$ dex) in 12+$\log_{10}$(O/H) and show no obvious trends in \rpahdiff as a function of metallicity.  A small sub-sample (IC~5332, NGC~1385, NGC~2835, NGC~5068, and NGC~7496) reach $12+\log_{10}$(O/H)$ < 8.5$.  In this sub-sample, we observe a trend with decreasing \rpah with decreasing $12+\log_{10}$(O/H).  However, due to the small amount of data below this metallicity, it is difficult to diagnose exactly what value of $12+\log_{10}$(O/H) this decreasing trend begins, especially when we consider that the galaxy that reaches the lowest $12+\log_{10}$(O/H) (NGC~5068) also has the highest sSFR, and the highest average value of \rpahdiff.

It may also be the case that the PAH fraction could decrease in regions of dense molecular gas, where PAHs may coagulate into larger dust grains \citep[e.g.][]{MivilleDeschenes2002,Kohler2015}.  Dense molecular environments may also have higher values of $A_V$, which could attenuate the UV field. This could lower \rpah at a given PAH fraction as a consequence of the softer radiation field \citep{Draine2021}. Attenuation effects have been suggested to explain high spatial resolution observations of MIR-to-FIR colors in MW molecular clouds \citep{Flagey2009}, however, at the $10-50$ pc scales we probe, the fraction of dust contributing from highly shielded regions is likely to be small. To test the possible changes in \rpah as a function of molecular gas abundance, the right panel of Figure~\ref{fig:fmol&metals} shows \rpahdiff plotted as a function of \fmol, the molecular gas fraction, defined as \fmol$ = \Sigma_{\rm H_2} / (\Sigma_{\rm H_2} + \Sigma_{\rm HI})$.  We find a fairly flat relationship between \rpahdiff and \fmol\ below \fmol=0.8, followed by a decreasing trend in \rpahdiff.

\section{Discussion}
\label{sec:Discussion}
Prior to the launch of JWST, \rpah-equivalent measurements (i.e.\ requiring only MIR observations) at $<50$pc resolution were restricted to the galaxies of the Local Group (D $\lesssim$ 1 Mpc) because of the resolution of {\em Spitzer} and WISE ($\sim6''$ and $\sim12''$ respectively at $\sim20$ \micron). For the Magellanic Clouds, existing FIR measurements also reach 10 pc spatial resolution, enabling highly spatially resolved studies of \qpah across the LMC and SMC \citep{Sandstrom2010,Paradis2011,Chastenet2019}.  The high resolution measurements of \rpah from PHANGS-JWST give new insights into the behavior of the PAH fraction at $\sim$Solar metallicity.  In this Section, we will highlight the primary results of this paper and how these results fit into the broader picture of PAH studies. 

\subsection{\rpah as a Tracer of \qpah}
\label{ss:rp_qp}
The fraction of dust in the form of PAHs with $<10^3$ C atoms \citep[\qpah,][]{Draine&Li2007} can be constrained using MIR through FIR observations. From the FIR, the peak of the thermal dust SED provides information on the radiation field intensity.  The shape of the FIR SED can also provide constraints on the distribution of radiation fields heating the dust \citep{Dale2014}.  With this information on the radiation field, the addition of MIR measurements that sample PAH emission yield a constraint on \qpah, since the primarily stochastic emission from PAHs is proportional to the radiation field intensity. The inferred PAH surface density compared with the dust mass surface density constrained by the FIR peak, yields \qpah.

The resolution of existing FIR datasets strongly limits the sample of galaxies where \qpah can be resolved at $10-50$ pc resolution to the Local Group.  For this reason, it is critical to explore MIR-only indicators of PAH fraction to extend observations beyond the nearest galaxies in order to make progress in understanding the life-cycle of PAHs. In this paper we have explored the MIR-only diagnostic \rpah $=$(F770W$_{\rm ss}$+F1130W)/F2100W as an indicator of \qpah. We have compared \rpah and \qpah at low resolution for several galaxies with FIR observations that were also covered by PHANGS-JWST and found good correlation. We also explored the relationship between the predicted \rpah to \qpah in the \citet{Draine&Li2007} dust models and their recent updates \citep{Draine2021,Hensley2023}. In general, at dust-mass weighted average radiation field intensities $\lesssim10^2$, which is typical for regions in our galaxies at $10-50$ pc resolution, \rpah is well correlated with \qpah, yielding a good diagnostic for changes in the PAH fraction (see also Appendix~\ref{app:radiation}). It is worth noting that the mapping between \rpah and \qpah is non-linear and at high values of \qpah $>5$\% (super-Milky Way \qpah values), the slope between them becomes flatter, resulting in only small changes in \rpah as \qpah increases.  On the other hand, decreases in \rpah are a good diagnostic for revealing locations where the PAH fraction drops, presuming the dust-mass weighted average radiation field is not $U\gg10^2$.

There are several questions about the \rpah-\qpah relationship that should be addressed in future studies. First, we have focused on the \citet{Draine&Li2007} model, which makes specific assumptions about the dust grain size distribution.  Other dust models make different assumptions about grain size distributions and dust optical properties which can change the interpretation of \rpah, particularly if there is a variable amount of ``very small grains'' (VSGs) that can contribute to the MIR continuum at $\sim$20 \micron.  Future work with DustEM and the THEMIS model \citep{Compiegne2011, Jones2013, Ysard2024} could explore \rpah as a tracer of small hydrocarbon grain abundance in that model framework, which includes variable VSG abundance.  

In addition, in our FIR modeling comparison and exploration of the model predictions for \rpah-\qpah mapping, we have assumed a radiation field distribution (specifically, a delta function plus power-law model). This choice directly impacts the interpretation of the MIR emission, since the power-law component (with a distribution of radiation fields extending from $U_{\rm min}$ to $U=10^7$ and a slope fixed to $\alpha=2$) contributes MIR continuum emission at $\sim20$ \micron. In models that assume a single radiation field, that MIR emission is often attributed to changes in the grain size distribution \citep[specifically, an increase in very small grains, see Figure III.15 of][for a very clear visualization of the differences between these types of models]{Galliano2022}.  The applicability of a delta function plus power-law distribution has been explored in a wide variety of contexts for unresolved galaxies or $\sim$kpc scale measurements \citep{dale2005,Aniano2012,Galliano2018}. At a very high spatial resolution, in theory one might expect to reach a scale where a single radiation field may be applicable, but in general on $10-50$ pc scales it seems likely that a distribution of radiation field intensities is present. Little is known about the detailed small scale distribution of radiation field intensities throughout the ISM.  Future work investigating this will be critical to improve understanding of the \rpah-\qpah mapping and for reconciling differences between models that interpret the MIR continuum as changes in small grain abundance, PAH fraction, and radiation field intensity and distribution.

\subsection{The Destruction of PAHs in \hii Regions}
\label{ss:PAH_hii}
\hii regions crisply stand out in our \rpah maps as regions with low PAH fraction. This observation agrees well with previous studies in MW \hii regions \citep[e.g.][]{Cesarsky1996,Kassis2006,Povich2007},  with extragalactic observations that have the spatial resolution to isolate \hii regions \citep[e.g.][]{Chastenet2019}, and with early results on JWST observations of our sample \citep{Chastenet2023_RPAH,Egorov2023}.  Additionally, this follows from the trends between sSFR (measured by 70\micron\ surface brightness) and \qpah measured at lower spatial resolution described in \citep{Aniano2020}.  When we separate ``nebular'' regions using the \citet{Groves2023} catalog, we find that these boundaries encompass nearly all of the low \rpah regions in our maps. We also find that the \rpah weighted by H$\alpha$ shows an even more dramatic decrease in the nebular regions. To confirm this result is valid even if some \hii regions were not included in the nebular catalog of \citet{Groves2023}, we also examine trends in \rpah as a function of \newhagasratio, a pixel-based tracer of the amount of ionized gas, and find that \rpah falls dramatically at a value of \newhagasratio$=37.5$erg~s$^{-1}$~kpc$^{-2}$~(M$_\odot$pc$^{-2}$)$^{-1}$, confirming that \rpah decreases in regions dominated by ionized gas.

The good correspondence of the presence of \hii regions with the decrease in \rpah across a sample of \hii regions with a wide range of ages and luminosities suggests that PAH destruction occurs quickly compared to the \hii region lifetime. Indeed, observations of PAH emission in MW \hii regions suggest PAH survival times on the order of $\sim$few thousand years \citep{Kassis2006,Compiegne2007} inside the ionization front. Recent JWST observations of the Orion Bar photodissociation region (PDR) have emphasized the sharp decrease in PAH emission near the ionization front \citep{Chown2023,Peeters2023}, corroborating a quick destruction timescale in the photoionized gas. 

The exact mechanism of PAH destruction in \hii regions is not fully understood. Laboratory studies of PAHs have suggested that fragmentation of PAH molecules is likely to occur when they are irradiated by photons with energies between 8--40~eV, although there is some uncertainty as to what photon energies will primarily fragment PAHs and which energies will lead only to photoionization of the PAHs \citep{Jochims1996, Zhen2015}. Theoretical calculations suggest that in photoionized gas with \hii region-like densities, thermal sputtering would require long timescales \citep[$\sim10^7$ years;][]{Micelotta2010}, which would be difficult to reconcile with the strong correspondence between \hii region locations and dips in \rpah across our large sample of regions. Observations of MW \hii regions provide evidence for extreme UV (EUV) photons playing an important role in PAH destruction \citep{Povich2007}. Comparison of optical line ratios from MUSE to the regions of low \rpah in the PHANGS-JWST targets also show \rpah is correlated with tracers of the hardness and intensity of the UV field \citep{Egorov2023}. More detailed investigation of the correlation of \rpah and PAH properties judged from spectroscopy with tracers of the ionized gas conditions and radiation field should provide key insights into the destruction pathways. 

Because of the resolution of the MUSE and JWST data, our nebular regions always include contributions from both the \hii region, surrounding PDR, and depending on the \hii region size and galaxy distance and inclination, a substantial contribution from the surrounding ``diffuse'' gas. This means that the low \rpah signal associated with \hii regions is diluted, particularly for small \hii regions (see Section~\ref{sec:1kpc} for more discussion). Future work comparing the \rpah maps to high resolution H$\alpha$ from HST observations or Paschen-$\alpha$ from JWST should better resolve the boundary of the \hii region enabling less contaminated measurements of the decrease in \rpah.  We do see, however, that in cases where \hii regions are large or the line of sight is dominated by ionized gas (see Figure~\ref{fig:gas_all}) the decrease in \rpah is more stark. This is likely to be a combination of a better match between the MUSE nebular region and the actual \hii region size, and the fact that large \hii regions may be big enough to fully puncture through the galaxy disks, giving a line of sight that has little foreground or background contamination from the surrounding neutral ISM.  It is worth noting that in the case of \qpah$\sim0$, \rpah does not asymptote to 0, but something more like \rpah$\sim0.5$ in the \citet{Draine&Li2007} and \citet{Hensley2023} models. The nature of the MIR continuum in situations where there are truly no PAHs is not well characterized since few lines of sight with those characteristics have been studied in the tracers needed to constrain the grain population in the MW.  Observations with \textit{Spitzer} and SOFIA of Galactic \hii regions NGC~7023 \citep{Croiset2016} and the Eagle Nebula \citep{Flagey2011} show little to no PAH emission inside the photodissociation regions (PDRs), but these observations were not the main comparison source for the \citet{Draine&Li2007} and \citet{Hensley2023} models.  This means that at very low \qpah, the \qpah-\rpah mapping may be more uncertain.  Regardless, it is clear that we see the lowest \rpah values in lines of sight that are most dominated by ionized gas. Our observations are consistent with a picture where ionized gas is essentially free of PAHs, as would be expected based on observations of MW \hii regions and estimates of the PAH destruction timescale.

The observation that \hii regions are holes in the distribution of PAHs in the ISM provides a potential explanation for the observed trends of kpc-scale \qpah and \rpah with SFR and sSFR \citep[e.g.][]{Aniano2020}.  As the SFR surface density increases, a larger fraction of the ISM in that kpc is comprised of \hii regions, leading to a lower average PAH fraction. Also, because the kpc-scale averages are luminosity-weighted, and the surroundings of \hii regions have some of the highest MIR luminosities, low \qpah material will contribute more to setting the kpc-average \rpah. An interesting consequence of the SFR or sSFR trends in \rpah may be to exaggerate the observed dependence of \qpah on metallicity. Because low metallicity galaxies tend to have low stellar mass and therefore have higher sSFR \citep[and in particular many of the low metallicity galaxies previously studied in the MIR are starbursts with very high sSFR;][]{Wu2006,Engelbracht2008}, some part of the observed metallicity trend may in fact be a sSFR trend.  Future JWST observations should be able to clearly disentangle these two effects by resolving the \rpah maps of these sources and isolating the \hii regions from the surrounding neutral ISM.

\subsection{PAH Fraction as a Function of Metallicity}\label{sec:metal}

One of the strongest observed environmental trends in PAH fraction is its dependence on metallicity \citep{Engelbracht2005,Madden2006,Draine2007,RemyRuyer2015,Chastenet2019,Shivaei2024}. Galaxy integrated \qpah measurements suggest a decrease in the PAH fraction at $12+\log$(O/H)$\sim$8.1-8.3 \citep{Draine2007,Aniano2020,Galliano2021}. This decline has been attributed to enhanced PAH destruction \citep{Madden2006,Gordon2008} or impeded PAH formation \citep{Sandstrom2010} at low metallicity, among other possibilities. Observations of the Magellanic Clouds have revealed that at 10 pc resolution, where \hii regions and diffuse gas can be separated, the diffuse neutral gas PAH fraction shows a sharp decrease between the metallicity of the LMC and SMC, and that the diffuse neutral gas in the LMC has a nearly MW \qpah value of $\sim5$\% \citep{Chastenet2019}. 

Understanding the origin of this metallicity trend is critical for using PAH emission to trace SF and for predicting attenuation curves for galaxies across a range of metallicities and redshifts. It also holds key information about the PAH life cycle---a subject that remains poorly understood. As discussed in \citet{Aniano2020}, the galaxy-integrated \qpah measurements as a function of metallicity do not provide a clear distinction between a model where the PAH fraction decreases gradually with metallicity or where there is a sharp decrease at some threshold metallicity.

Although our sample does not span a large metallicity range, nor does it reach the $12+\log$(O/H)$\sim$8.1-8.3 metallicities where the PAH fraction becomes substantially lower than MW, our observations of a nearly constant \rpahdiff throughout our sample suggest that metallicity trends are steep, rather than gradual. We see no clear metallicity trends in \rpahdiff over $\sim0.3$ dex in $12+\log$(O/H) and in fact one of our lowest metallicity galaxies has one of the highest \rpahdiff values.  The lack of any systematic metallicity trend  may lend some support to the scenario where there is a sharp drop in \qpah at some metallicity. This agrees well with the LMC results \citep{Chastenet2019} showing a MW-like \qpah in the diffuse neutral gas of that galaxy, despite its low metallicity.  An important future direction for understanding PAH fraction as a function of metallicity is to separate \rpah in \hii regions and diffuse gas in lower metallicity targets. If the metallicity dependence is sharp, we may expect to see \rpahdiff in galaxies persist at high values until some threshold metallicity.

\subsection{Behavior of the PAH Fraction Outside \hii Regions}
\label{ss:PAH_diff}
Outside of the \hii regions identified with the nebular catalogs of \citet{Groves2023}, we find that \rpah is fairly constant, with an average value of $\sim$3.4.  Based on the dust emission models of \citet{Hensley2023}, this corresponds to a \qpah of 5.55\%, assuming the ``standard'' PAH size distribution and a single radiation field with $U=1$.  The \qpah calibration in \citet{Hensley2023}, places the MW diffuse ISM at \qpah$=5.9$\%, very similar to what we infer in the diffuse ISM of galaxies in our sample. 
If we instead translate \rpah to \qpah using a distribution of radiation fields with $U_{\rm min}=1$ and $\gamma=0.005$ in the \citet{Draine&Li2007} model, we find that \rpah$=3.4$ translates to \qpah$\sim4.5-5$ (see Figure~\ref{fig:Rpahqpahall}). Within the \citet{Draine&Li2007} model, the MW diffuse ISM \qpah$=4.55$\%, again very similar to what we infer for our galaxies. These comparisons suggest that we are seeing \qpah values in the diffuse neutral gas of our galaxies that is very similar to what is found in the Milky Way.  In addition, the consistency of \rpah outside of nebular regions suggests that the diffuse ISM of these galaxies has a fairly uniform \qpah.   The small degree of environmental variation at $\sim$Solar metallicity in PAH fraction also helps explain observations showing a tight, nearly linear correlation of PAH emission with gas column \citep{Chown2021,Leroy2023b}.  This is similar to what has been found in the Milky Way, where gas column and dust emission are tightly correlated in the high latitude cirrus \citep{Boulanger1996, Plank2011, Lenz2017}.

The most significant driver of changes to the average value of \rpahdiff across our sample is the specific star formation rate (sSFR), a correlation that we tested both over the full JWST footprint (Figure~\ref{fig:deltaRPAH}) and within 1~kpc regions (Figure~\ref{fig:1kpcRpah}).  The positive correlation between \rpahdiff and sSFR could be explained if regions with higher sSFR have an overall harder radiation field in the diffuse ISM compared to lower sSFR regions, which would increase \rpah measured in the diffuse gas at a given \qpah.  Future efforts to model the interstellar radiation field using information from the stellar populations observed by MUSE should provide important insights into these trends.  In addition, studies of \rpah in the vicinity of star clusters, where the radiation field can be predicted with stellar population modeling, show trends consistent with \rpah responding to changes in radiation field hardness \citep[][Dale et al.\ in prep]{Dale2023}. Lastly, work with PHANGS-JWST exploring PAH band ratios has shown evidence of the dependence of PAH emission feature strengths on the hardness of the radiation field \citep{Chastenet2023PAHratios,Baron2024}. Accounting for the response of PAH emission to changes in radiation field spectrum will be a critical next step in interpreting \rpah variations.

In addition to the correlation between \rpahdiff and sSFR, we see a slight decreasing trend in \rpahdiff and \fmol.  This decrease could be explained by PAHs coagulating or sticking to larger dust grains in the dense gas. 
\citet{Ysard2013} suggest that coagulation of dust grains will occur above a density threshold of a few~10$^3$~H~cm$^{-3}$. Above this threshold, smaller grains may form fluffy aggregates, producing a different dust population than in the lower density diffuse ISM and lowering \qpah.  This aggregation could potentially cause some of the decrease in \rpahdiff we observe at high \fmol, where we expect the gas to be the densest.  Additionally or alternatively, in regions of dense molecular gas, $A_V$ is expected to be highest, blocking UV light that would excite PAHs from reaching these regions.  This high attenuation of UV light could also be lowering \rpah even if \qpah remains constant, it softens the radiation field, leading to lower \rpah at a given \qpah.  Determining whether the main cause of this decreasing trend is coagulation or UV attenuation requires further spectral information not available with this dataset.  This trend appears consistent across our full sample with the exception of NGC~1300 and NGC~5068, which show slight increases in \rpahdiff at high \fmol.   

It should also be noted that our isolation of \rpah from diffuse gas is limited by our ability to identify \hii regions with MUSE.  Some of the decrease observed in \rpahdiff at high \fmol\ could be caused by embedded star-forming regions undetected by MUSE \citep[as found in a small set of cases in][]{Hassani2023}.  These embedded regions would decrease \rpahdiff through PAH destruction, as observed in the lower \rpahneb measurements above.  The magnitude of this effect is likely to be limited by the short duration of this highly embedded phase of star formation \citep{Kim2023} and the overall excellent correspondence between 21 \micron\ sources and H$\alpha$ \citep[with 92\% of F2100W bright sources falling within an identified nebular region][]{Hassani2023}.

\section{Summary}
\label{sec:Conclusions}
This work represents an in-depth analysis of the JWST photometry based indicator of the fraction of dust in the form of PAHs: \rpah~=~(F770W$_{\rm{ss}}$+F1130W)/F2100W.  Using MIRI photometeric maps of 19 nearby galaxies from the PHANGS sample, we are able to assess the behavior of \rpah in a wide variety of ISM environments.  Through this initial analysis, we provide the necessary verification that \rpah as defined in this work is a robust tracer of the PAH fraction.  We determine this through comparisons between \rpah and \qpah, the PAH fraction as defined by the \citet{Draine&Li2007} dust models.  Model predictions of \rpah increase with \qpah up to a \qpah of $\gtrsim$5$-6$\% using the dust emission models of both \citet{Hensley2023} and \citet{Draine&Li2007}, assuming reasonable radiation field properties (both single radiation fields and distributions of radiation field intensity) for $10-50$ pc scales in galaxies. Above \qpah$\sim5-6$\%, the modeled relationship between \rpah and \qpah continues to increase with a slightly flatter slope in both the \citet{Hensley2023} and \citet{Draine&Li2007} models, for a range of PAH size or charge distributions and interstellar radiation fields.  We also compare observed values of \rpah and \qpah at low resolution for galaxies with both JWST and FIR observations and find good correlation. Based on this behavior, we argue that \rpah is a good indicator of PAH fraction in most typical ISM conditions observable at $\geq 10$pc scales.   Across our sample, we find \avrpahdiff=3.43, which closely corresponds to the \rpah values predicted for the Milky Way of \qpah$\sim5.9$\% listed in \citet{Hensley2023} models or \qpah$\sim 4.6$ listed in the \citet{Draine&Li2007} models.

We use \rpah to assess the PAH fraction across the disks of the 19 PHANGS galaxies, and look for correlations with environment through comparisons to optical, millimeter, and sub-mm observations available in the PHANGS dataset.  We find that the dominant change in \rpah is a steep decrease observed in \hii regions, as noted both by lower average values of \rpah within nebular regions and a sharp decrease in \rpah above a log$_{10}$(\hagasratio) of $\sim$37.5~erg s$^{-1}$ kpc$^{-2}$ (M$_{\odot}$ pc$^{-2}$)$^{-1}$.  The exact destruction method of PAHs within \hii regions is not well known, but is likely caused by the UV radiation from young stars. The good correspondence between the low \rpah values and the nebular regions, which span a wide range of ages and luminosities, suggest that PAH destruction occurs quickly compared to the \hii region lifetime.

We also find that the difference between the average \rpah measured in nebular regions and the diffuse gas is correlated with the sSFR.  This trend seems to be driven by the correlation between \avrpahdiff and sSFR, which was observed in 1~kpc regions and when examining the integrated measurements from the full area mapped by JWST.  This could indicate that in galaxies with the highest sSFR, a harder radiation field is irradiating the PAHs in the diffuse ISM, producing a higher value of \rpah for a given \qpah.

We further test what may affect the PAH abundance outside \hii regions by examining how \rpahdiff  changes as a function of galactic environment (nuclei, arms, bars, etc.), molecular gas fraction, and metallicity.  We find little variation of \rpah with metallicity, although this is likely caused by the limited range of $12+\log$(O/H) covered by our sample.  Within the metallicity range included in our sample, there is some indication of a slight decrease in the binned median values of \rpahdiff below $12+\log_{10}$(O/H)$\sim$8.4, but as only 5 galaxies in our sample have Z lower than this threshold it is difficult to ascertain the veracity of this trend.  We find a slight decrease in \rpahdiff with the molecular gas fraction $f_{\rm{Mol}}$, which could be caused by the PAHs forming into larger grains and decreasing the PAH fraction in denser gas or attenuation of UV photons that would excite PAH features limiting the PAH emission this ISM phase.

This paper lays the groundwork for using MIRI photometry to trace the PAH fraction at 10--50~pc scales in nearby galaxies.  JWST has opened a new window for studying the distribution of PAHs at high sensitivity and high spatial resolution.  Further observations of nearby galaxies including dwarfs, starbursts, and early type galaxies with JWST will continue to expand our understanding of how the PAH fraction both impacts and is impacted by its local environment.

\section*{}
This work has been carried out as part of the PHANGS collaboration. This work is based on observations made with the NASA/ESA/CSA {\it JWST}. The data were obtained from the Mikulski Archive for Space Telescopes at the Space Telescope Science Institute, which is operated by the Association of Universities for Research in Astronomy, Inc., under NASA contract NAS 5-03127 for {\it JWST}. These observations are associated with program 2107. The specific observations analyzed can be accessed via \dataset[DOI: 10.17909/ew88-jt15]{https://doi.org/10.17909/ew88-jt15}.

JS, KS acknowledge funding from JWST-GO-2107.006-A. RSK and SCOG acknowledge funding from the European Research Council via the Synergy Grant ``ECOGAL'' (project ID 855130), from the German Excellence Strategy via the Heidelberg Cluster of Excellence (EXC 2181 - 390900948) ``STRUCTURES'', and from the German Ministry for Economic Affairs and Climate Action in project ``MAINN'' (funding ID 50OO2206). AKL, DP, and RC gratefully acknowledge support by grants 1653300 and 2205628 from the National Science Foundation, by award JWST-GO-02107.009-A, JWST-GO-03707.001-A, JWST-GO-04256.001-A,  and by a Humboldt Research Award from the Alexander von Humboldt Foundation. MC and LR gratefully acknowledge funding from the DFG through an Emmy Noether Research Group (grant number CH2137/1-1). COOL Research DAO is a Decentralized Autonomous Organization supporting research in astrophysics aimed at uncovering our cosmic origins. OE and KK acknowledge funding from the Deutsche Forschungsgemeinschaft (DFG, German Research Foundation) in the form of an Emmy Noether Research Group (grant number KR4598/2-1, PI Kreckel) and the European Research Council’s starting grant ERC StG-101077573 (“ISM-METALS"). MB acknowledges support by the ANID BASAL project FB210003 and by the French government through the France 2030 investment plan managed by the National Research Agency (ANR), as part of the Initiative of Excellence of Université Côte d’Azur under reference number ANR-15-IDEX-01.
JC acknowledges funding from the Belgian Science Policy Office (BELSPO) through the PRODEX project “JWST/MIRI Science exploitation” (C4000142239).
EWK acknowledges support from the Smithsonian Institution as a Submillimeter Array (SMA) Fellow and the Natural Sciences and Engineering Research Council of Canada.
ER acknowledges the support of the Natural Sciences and Engineering Research Council of Canada (NSERC), funding reference number RGPIN-2022-03499.

This paper makes use of the following ALMA data: ADS/JAO.ALMA\#2012.1.00650.S, 
ADS/JAO.ALMA\#2013.1.01161.S,
ADS/JAO.ALMA\#2015.1.00925.S,
ADS/JAO.ALMA\#2017.1.00392.S,
ADS/JAO.ALMA\#2017.1.00886.L. 
ALMA is a partnership of ESO (representing its member states),
NSF (USA), and NINS (Japan), together with NRC (Canada),NSC and ASIAA (Taiwan), and KASI (Republic of Korea), in cooperation with the Republic of Chile. The Joint ALMA Observatory is operated by ESO, AUI/NRAO, and NAOJ. The National Radio Astronomy Observatory is a facility of the National Science Foundation operated under cooperative agreement by Associated Universities, Inc.

%

\vspace{5mm}
\facilities{JWST (MIRI), MUSE, ALMA, VLA, MeerKAT} 


\software{astropy \citep{astropy:2022, astropy:2018, astropy:2013}
{\tt jwst} \citep{bushouse_howard_2023_8157276},
{\tt matplotlib} \citep{Hunter:2007},
{\tt numpy} \citep{harris2020array},
{\tt photutils} \citep{larry_bradley_2022_6825092},
SAOImageDS9 \citep{2003JoyeMandel},
{\tt scipy} \citep{2020SciPy-NMeth},
{\tt pjpipe} \citep{Williams2022}
         }



\appendix
\section{Histograms for individual galaxies}
\label{sec:full_samp_hists}

\begin{figure*}[h!]
    \centering
    \includegraphics[width=\textwidth,]{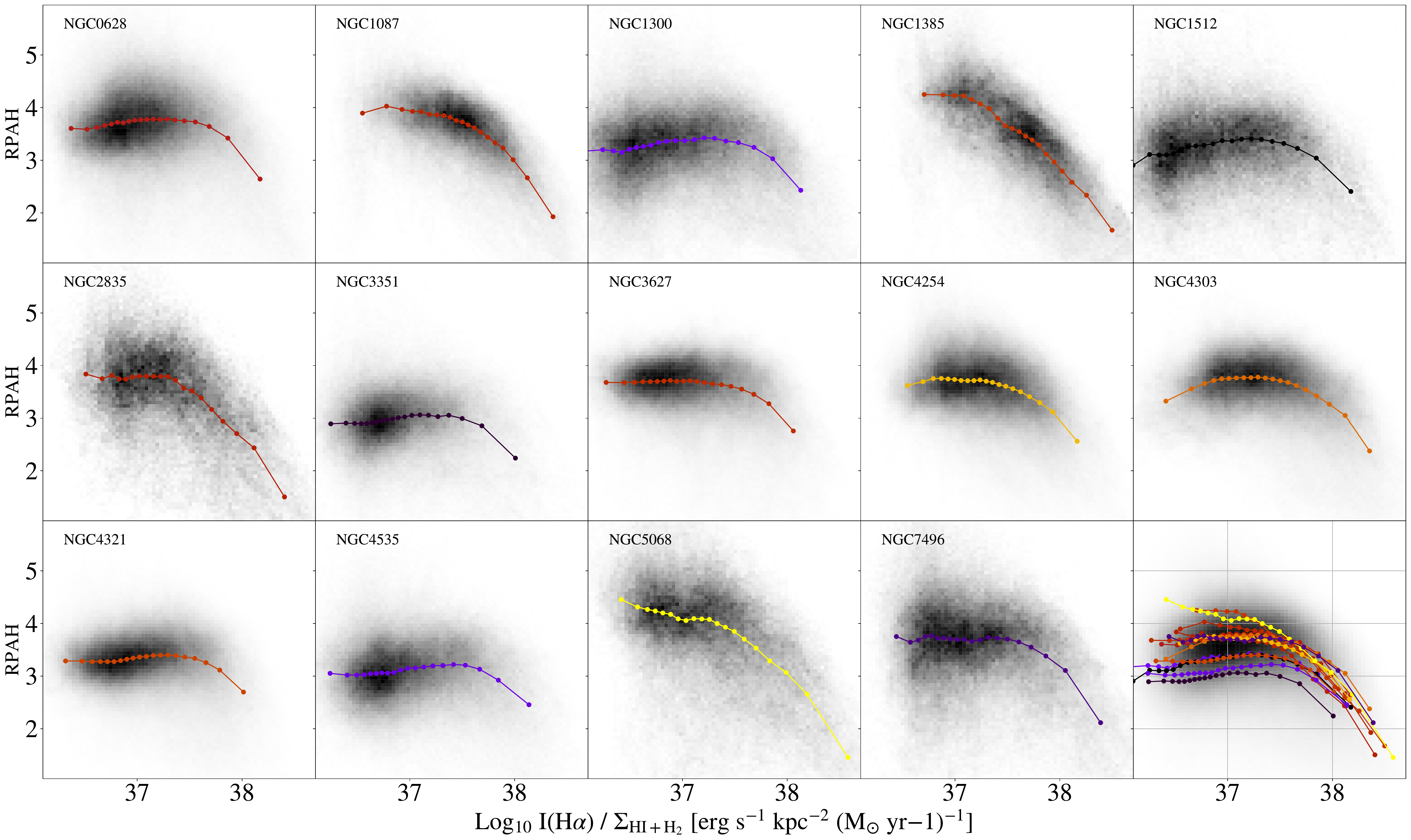}
    \caption{\rpah plotted as a function of \hagasratio for each galaxy with an HI map available.  Binned medians are displayed in each panel.  The final panel combines the binned medians from each galaxy for comparison.  Medians are color--coded by increasing log$_{10}$sSFR, using the same colors as in the main text.}
    \label{fig:gas}
\end{figure*}

\begin{figure*}[h!]
    \centering
    \includegraphics[width=\textwidth,]{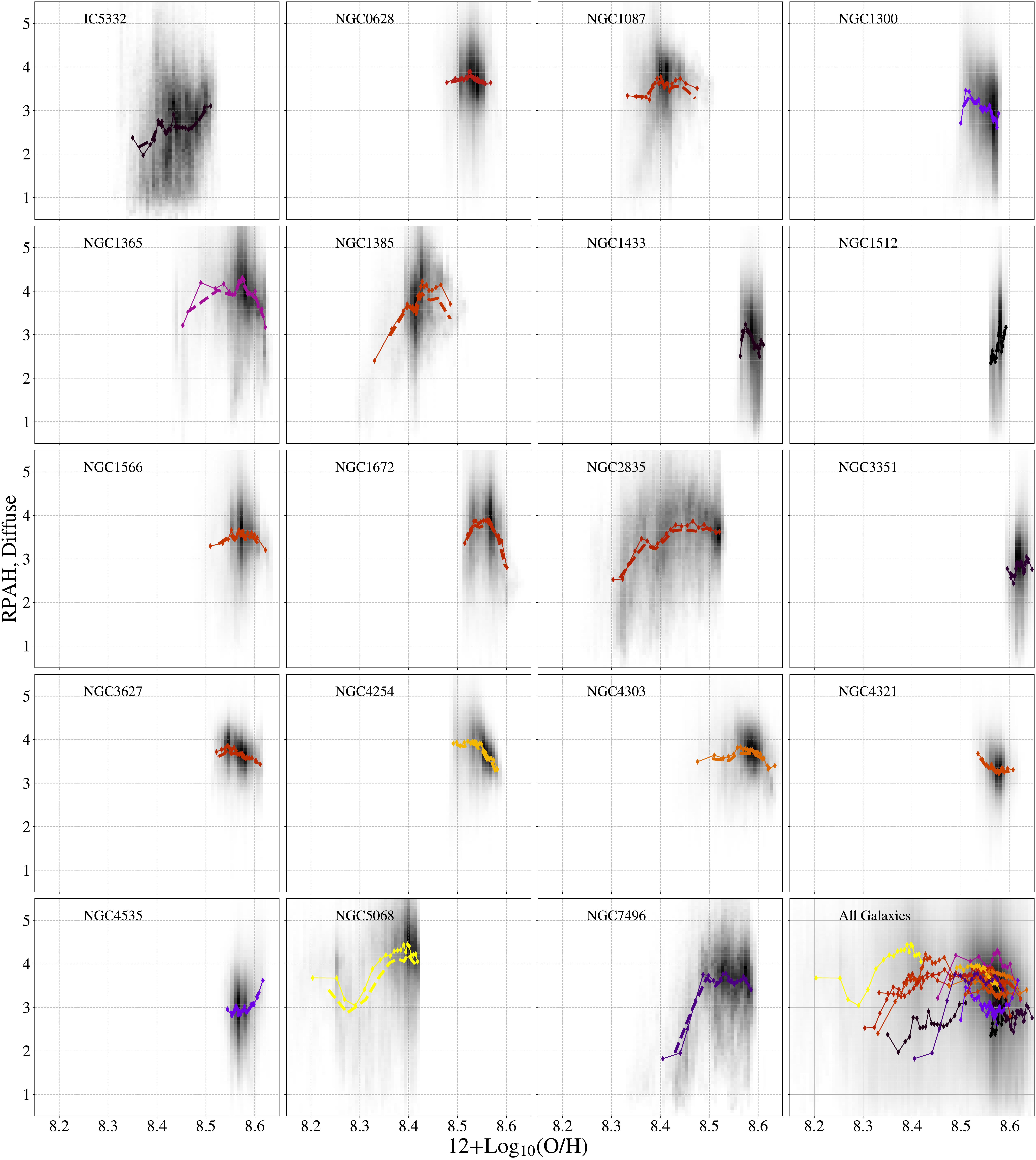}
    \caption{The per pixel values of \rpahdiff plotted as a function of metallicity.  Metallicities were determined using metallicity maps of \citet{Williams2022}.  Binned medians are shown in each panel and plotted together in the final panel, color-coded by sSFR.  Dashed lines represent the binned medians for all the measurements of \rpah.}
    \label{fig:metals}
\end{figure*}

\begin{figure*}
    \centering
    \includegraphics[width=\textwidth,]{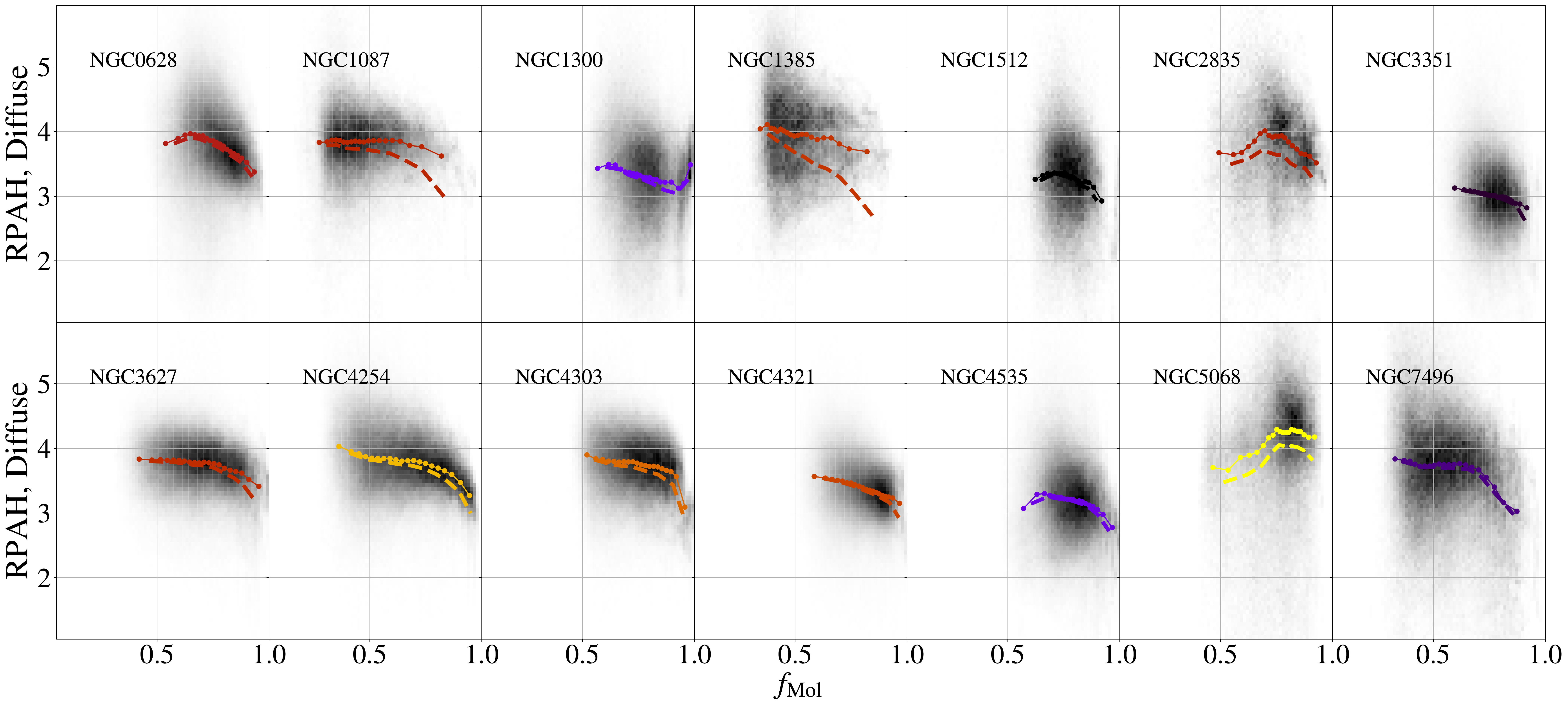}
    \caption{\rpahdiff plotted as a function of the molecular gas fraction, \fmol.  \fmol\ was determined using the \coto-derived $\Sigma_{\rm{H}_2}$ and HI-derived $\Sigma_{\rm{HI}}$. Binned medians are color-coded using the same sSFR schematic described in Figures~\ref{fig:gas} and \ref{fig:metals}.  Dashed lines represent the binned medians for all the measurements of \rpah.}
    \label{fig:fmol}
\end{figure*}

\begin{figure*}
    \centering
    \includegraphics[width=\textwidth,]{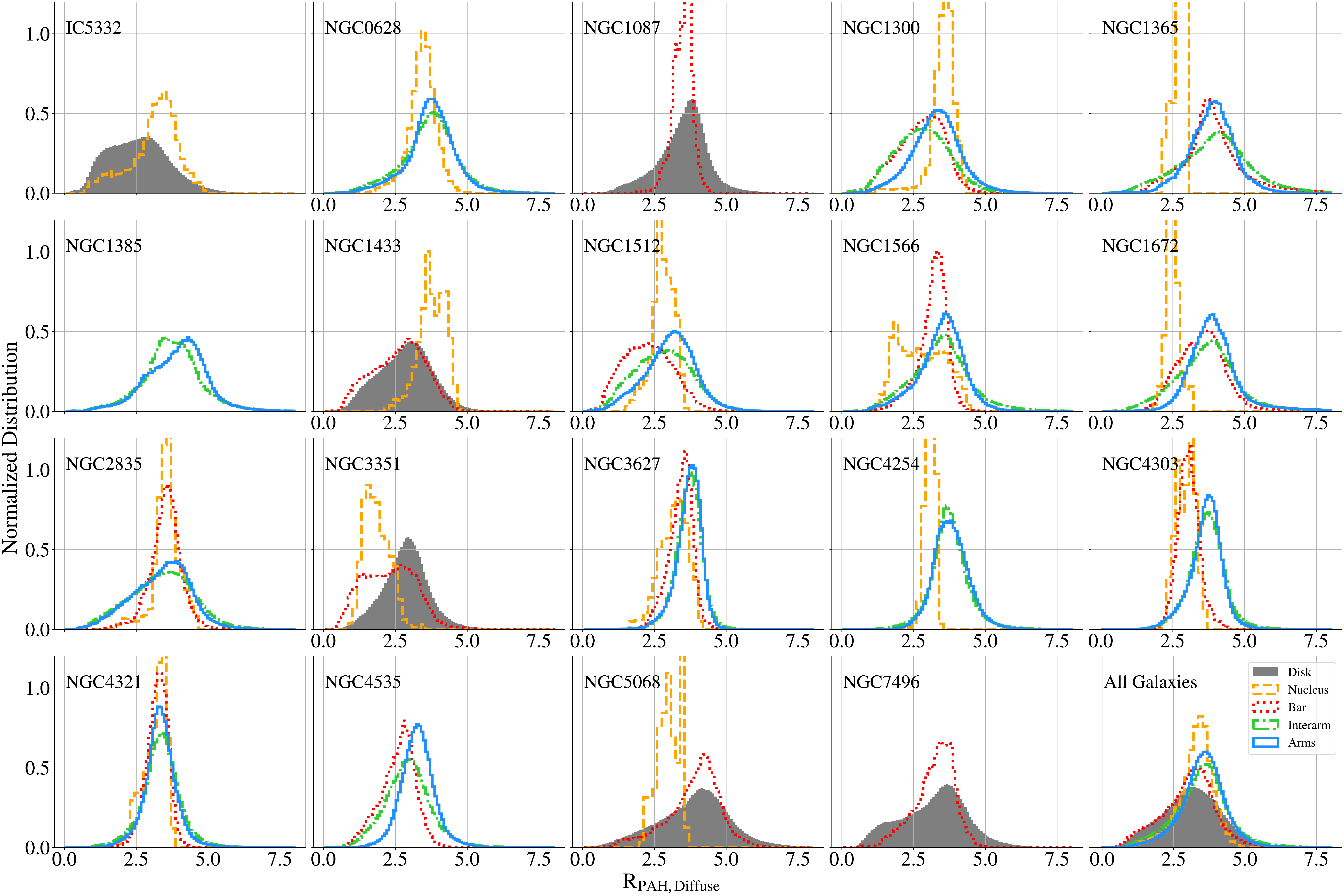}
    \caption{Histograms of \rpahdiff from isolated galactic environments as defined by \citet{Querejeta2021} for isolated galaxies.  Gold histograms represent measurements from the nucleus, purple are spiral arms, green are interarm, red are bars, and black are disks without spiral arms.}
\end{figure*}

\section{\rpah Alternatives}
\label{app:rpah_alt}
In some cases, only one of the PAH-dominated filters that we use in \rpah may be available. In this appendix we evaluate the relationship between \rpah\ and F770W/F2100W and F1130W/F2100W to understand how well, empirically, these more restricted measurements can be expected to capture the full behavior seen in \rpah\ in the main text.  Although this paper uses the full (F770W+F1130W)/F2100W as a charge-agnostic PAH fraction tracer, we find that either F770W or F1130W individually could be used as a good proxy for PAH emission, as shown by the strong correlations between \rpah and F770W/F2100W and F1130W/F2100W.  

\begin{figure*}[ht!]
    \centering
    \includegraphics[width=\textwidth]{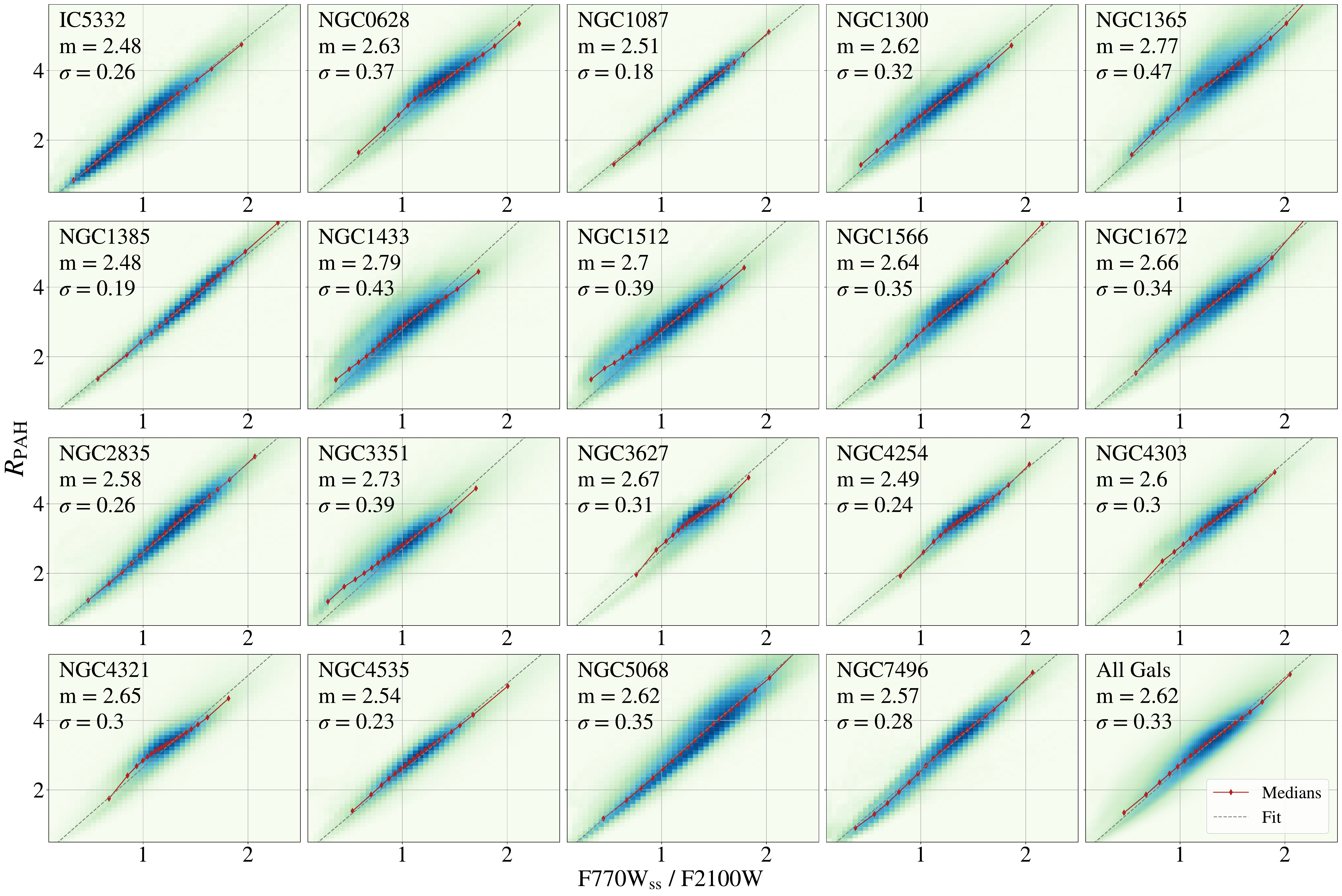}
    \caption{The relationship between F770W$_{\rm{ss}}$/F2100W and \rpah for each galaxy in our sample, as well as the full sample in the last panel.  Blue histograms show pixel distributions, red lines show binned medians, and the gray dashed lines show the best fit. }
    \label{fig:f770ssf2100_rpah}
\end{figure*}

Figure~\ref{fig:f770ssf2100_rpah} shows the relationship between \rpah and $F770W_{\rm ss}$/$F2100W$.  $F770W_{\rm ss}$ is the F770W flux after starlight has been removed using the method described in Section~\ref{sec:starsub}.  While the expected linear relationship is clear, we can see there are some galaxy-to-galaxy variations in the slope between \rpah\ and $F770W_{\rm ss}$/$F2100W$.  Across the full sample, we find: \rpah $ = 2.57 \times (F770W_{\rm ss}/F2100W)$, with a galaxy to galaxy 1$\sigma$ scatter of 0.09 in the range of predicted slope values. 

\begin{figure*}[ht!]
    \centering
    \includegraphics[width=\textwidth]{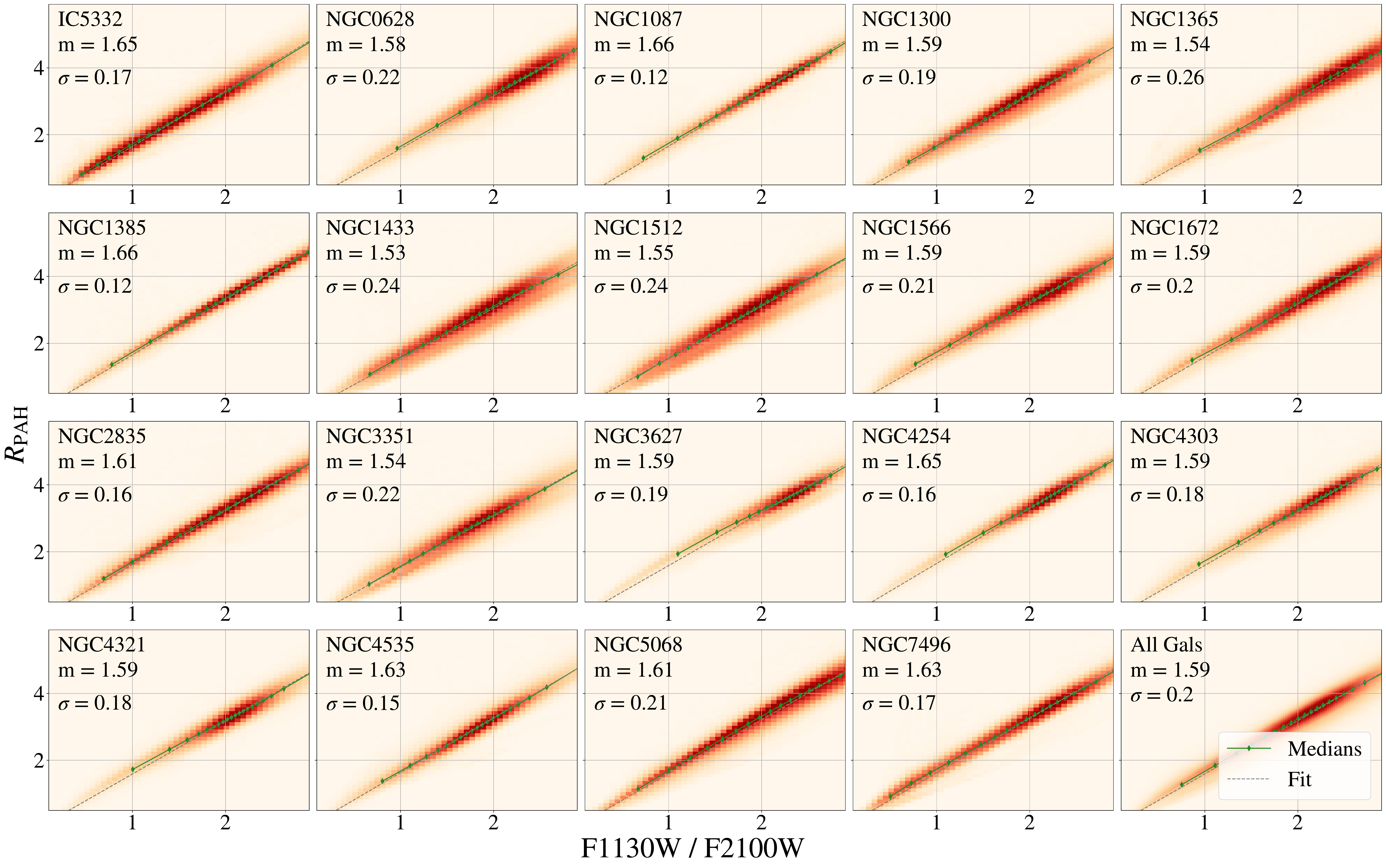}
    \caption{The relationship between F1130W/F2100W and \rpah for each galaxy in our sample, as well as the full sample in the last panel.  The histograms represent all pixels, green lines represent binned medians, and gray lines represent the best fit of the form \rpah = $m \times \frac{F1130W}{F2100W}$. }
    \label{fig:f1130f2100_rpah}
\end{figure*}

Figure~\ref{fig:f1130f2100_rpah} shows the relationship between  $F1130W/F2100W$ and \rpah.  Again, a reliable linear relationship is found between these two ratios, suggesting that $F1130W/F2100W$ can provide constraints on the PAH fraction as well.  For all pixels across our sample, we find \rpah $=1.59\times (F1130W/F2100W)$ with a galaxy to galaxy 1$\sigma$ scatter in the slope values of 0.04.  While we recommend using \rpah as an indicator of the PAH fraction as it has less potential to be effected by the average charge of the dust grains \citep[see e.g.][]{Draine&Li2007, Draine2021}, we find that both $F770W_{ss}$/$F2100W$ and $F1130W/F2100W$ provide viable alternatives.  

\section{Dust Model Predictions of \rpah}
\label{app:models}
\begin{figure}[ht!]
    \centering
    \includegraphics[width=0.5\textwidth]{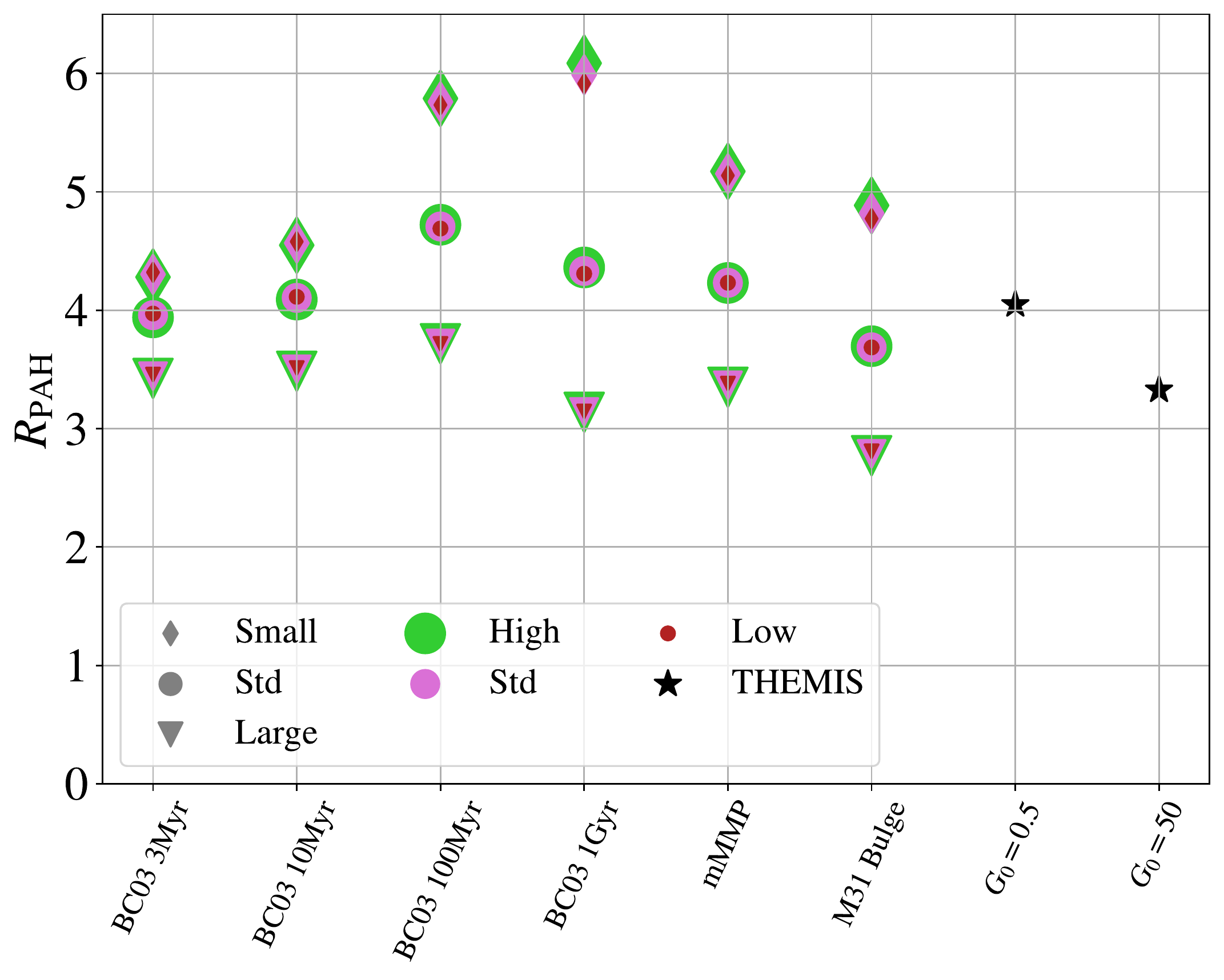}
    \caption{Predictions of \rpah for a range of available \citet{Draine2021} and THEMIS dust models.  For the \citet{Draine2021} models we vary the size, charge, and ISRF heating the grains.  For the THEMIS models we vary $G_0$ of the radiation field heating the grains.  All models produce \rpah between 3 and 6, matching values found in our data.}
    \label{fig:D21}
\end{figure}

In addition to using the \citet{Draine&Li2007} and \citet{Hensley2023} dust models to determine the extent to which \rpah traces \qpah in Section~\ref{sec:rpah_qpah}, we also examined how changing the size and charge distribution of PAHs and the ISRF altered the value of \rpah produced by the \citet{Draine2021} dust models.  These models provide nine PAH populations: three different size distributions (small, standard, and large) and three different ionization distributions (low, standard, and high).  Each model includes the predicted PAH emission spectrum as well as the predicted spectrum of a population of larger dust grains (astrodust) for a range of interstellar radiation fields (ISRFs).  We show the predicted values of \rpah determined for each of the size and charge distributions included in the \citet{Draine2021} models  in Figure~\ref{fig:D21} for a range of possible ISRFs at $U=1$.  In addition, we show the predicted \rpah values for two of The Heterogeneous dust Evolution Model for Interstellar Solids \citep[THEMIS;][]{Jones2013} for two values of the incident far--ultraviolet radiation field, $G_0$.  THEMIS \rpah predictions were computed using the standard distribution and summing the emission from the carbonaceous and silicate grains.  All models produce \rpah between 4 and 6, matching our observations for areas with higher PAH fractions.  This is expected for the \citet{Draine2021} models, which all assume a Milky Way \qpah of 4.5\%. 

The overlap between points of different colors shows that changing the ionization state of the PAH population does not strongly influence measurements of \rpah, which is expected as the 11.3~\micron\ and 7.7~\micron\ features trace neutral and ionized PAHs, respectively.  The ISRF heating the PAHs has some effect on the predicted value of \rpah.  The range of ISRFs provided by the \citet{Draine2021} models are shown on the x-axis in order of relative radiation field hardness (harder radiation fields on the left, softer on the right).  The first five ISRFs included are from the \citet[][BC03]{Bruzual2003} stellar population synthesis models for stellar populations of ages 3~Myr, 10~Myr, 100~Myr, 300~Myr, and 1~Gyr.  The next ISRF, labeled mMMP, is a modified \citet{MMP1983} ISRF which represents the average ISRF in the Milky Way, and the final \citet{Draine2021} model shows PAHs illuminated by a ISRF matching the old stellar population found in the bulge of M31 (M31 Bulge).  The next two models show the predicted values from the THEMIS dust model with $G_0$ of 0.5 and 50.  

\section{Flux weighting vs pixel weighting}
In order to represent the large amount of data included in this analysis, many of the figures in this paper displayed the binned averages of the datasets.  It is worthwhile to note that there are multiple ways to determine the placement of these averages.  For simplicity and uniformity, we have chosen to split the data up into bins of equal numbers of pixels and take the average value of all pixels in in said bin.  This is equivalent to giving each pixel in the bin an equal weight.

\begin{figure*}[ht!]
    \centering
    \includegraphics[width=\textwidth]{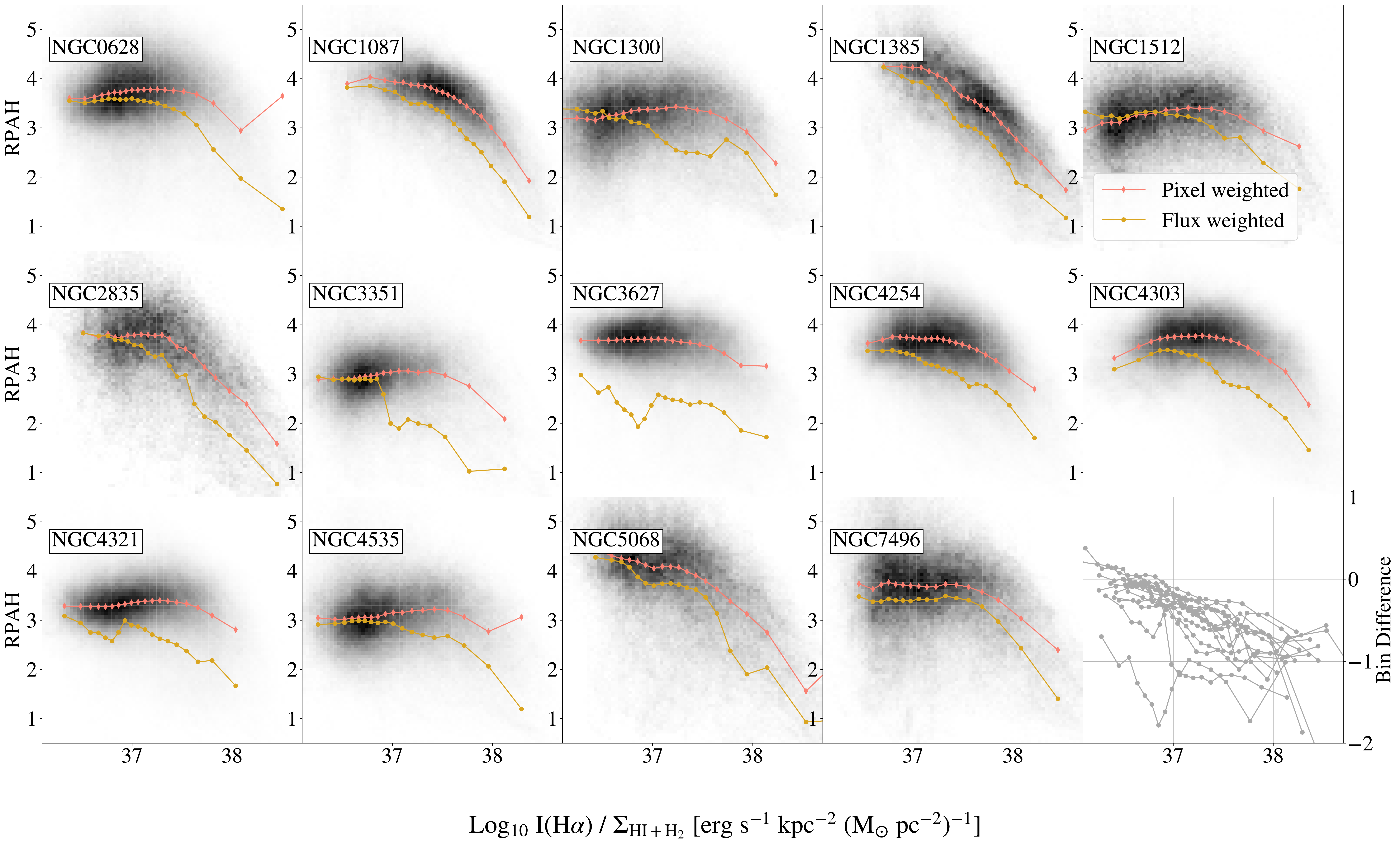}
    \caption{\rpah plotted as a function of \newhagasratio used to demonstrate the difference in using a pixel-weighted average vs a flux weighted average.  The gray histograms show the full pixel distributions, while the salmon lines show the pixe-weighted binned medians and the yellow lines show the flux--weighted binned medians.  The final panel shows the difference between the flux and pixel weighted binned medians.  We can see that these values are primarily negative, driven by the brightest F2100W values dominating the flux--weighted averages.}
    \label{fig:bin_ex}
\end{figure*}

For quantities like \rpah, which are ratios of other measurements, another method for placing the running averages would be summing the values of F770W+F1130W for each pixel in the bin and dividing by the summation of the F2100W values.  This would be similar to taking a flux-weighted average, as the pixels with the highest fluxes will have the greatest impact on the measured value.  Figure~\ref{fig:bin_ex} uses these two methods for determining the binned medians to recreate the relationship shown in Figure~\ref{fig:gas}.  Pink lines show the pixel-weighted medians, while yellow lines show the flux-weighted medians.  The final panel displays the differences between the flux and pixel-weighting.  

We can see that in most cases the flux-weighted average values are lower than the pixel-weighted values.  This suggests that the brightest pixels that dominate the flux-weighted average have lower \rpah measurements, likely driven by larger F2100W values.  This is similar to what was found in \citet{Pathak2023}, where they show that within areas of active star formation, the F2100W distribution has a relatively higher peak than the F770W and F1130W.  These few bright pixels in F2100W dominate the F2100W total fluxes, lowering the flux-weighted average \rpah measurements in comparison to the pixel-weighted averages.  This explains the differences in the average \rpah values shown in Figure~\ref{fig:bin_ex}.  A more detailed analysis of the MIR flux distribution is presented in \citet{Pathak2023}.

\section{Results from Previous 10 pc Far-IR SED modeling}
\label{app:radiation}
Although this work provides an unprecedented view of the PAH fraction in galaxies outside the Local Group, previous observations of the Large Magellanic Cloud (LMC) with \textit{Spitzer} and \textit{Herschel} provide similar spatial resolution infrared data.  This allows for full FIR spectral energy distribution (SED) modeling.  As these models are fit to both MIR and FIR data, they are able to include the emission from larger dust grains which gives constraints on the radiation field heating the dust.  Using these models as a guide, we can establish what range of radiation field intensities we might expect to be heating the dust at the 10-50~pc scales we measure within our sample. 

\begin{figure*}[ht!]
    \centering
    \includegraphics[width=0.49\textwidth]{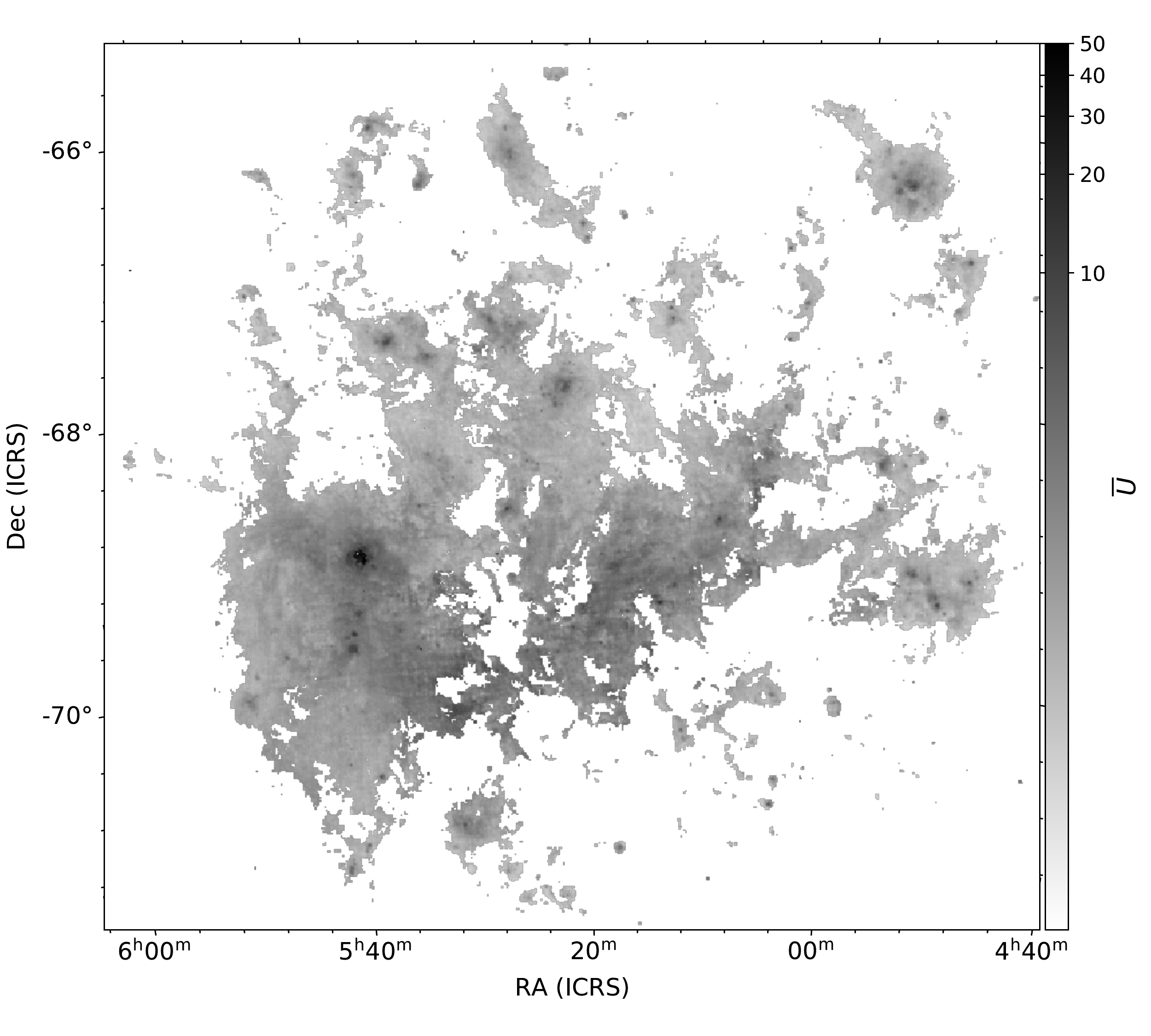}
    \caption{A map of the modeled $\overline{U}$ for the LMC.  A log scale is used to better display the range of $\overline{U}$ found across the LMC.  The star--forming region 30 Doradus has the highest values of $\overline{U}$, reaching $\overline{U}\gtrsim30$.}
    \label{fig:LMC_logu}
\end{figure*}

\begin{figure*}[ht!]
    \centering
    \includegraphics[width=0.49\textwidth]{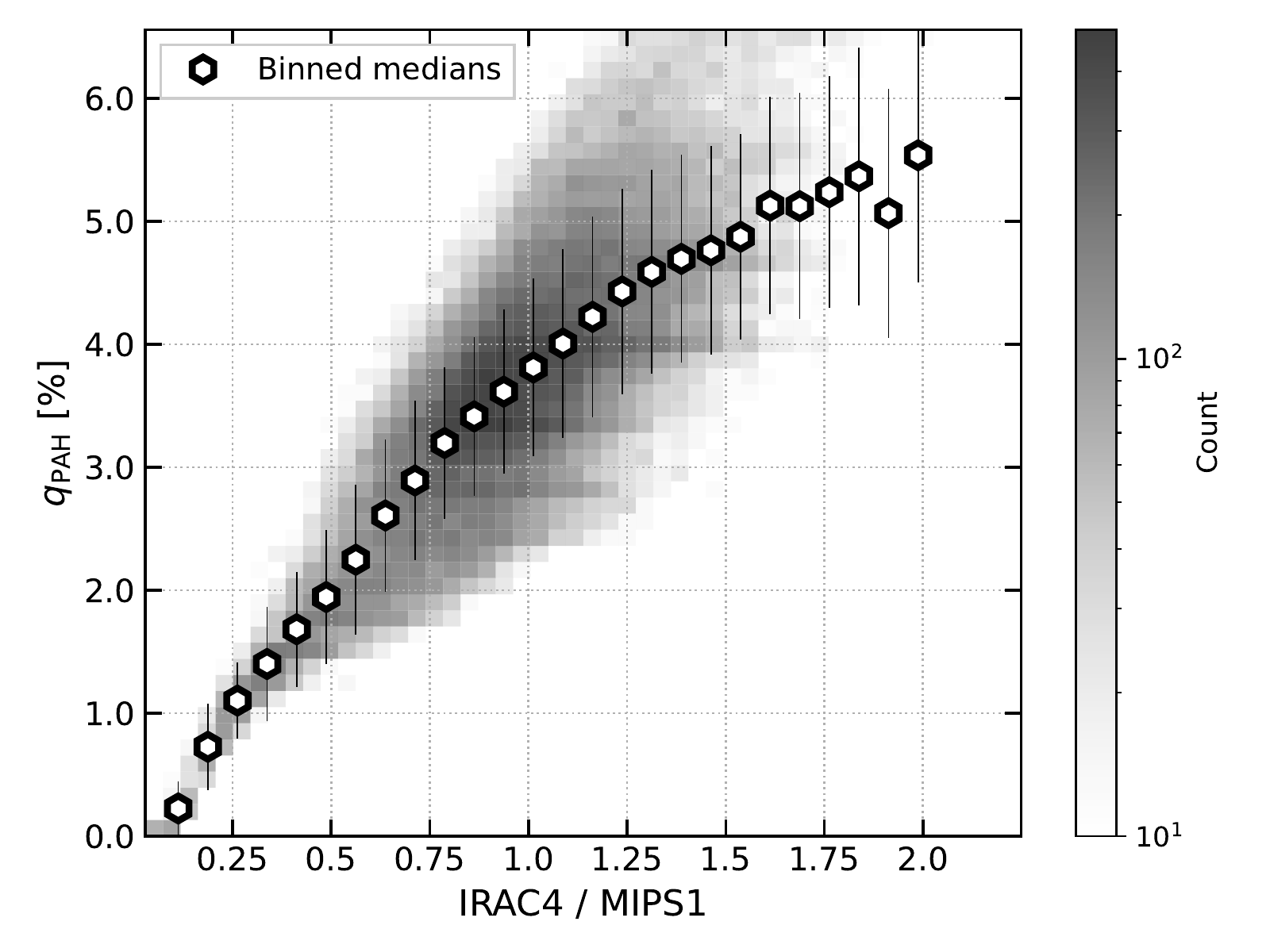}
    \includegraphics[width=0.49\textwidth]{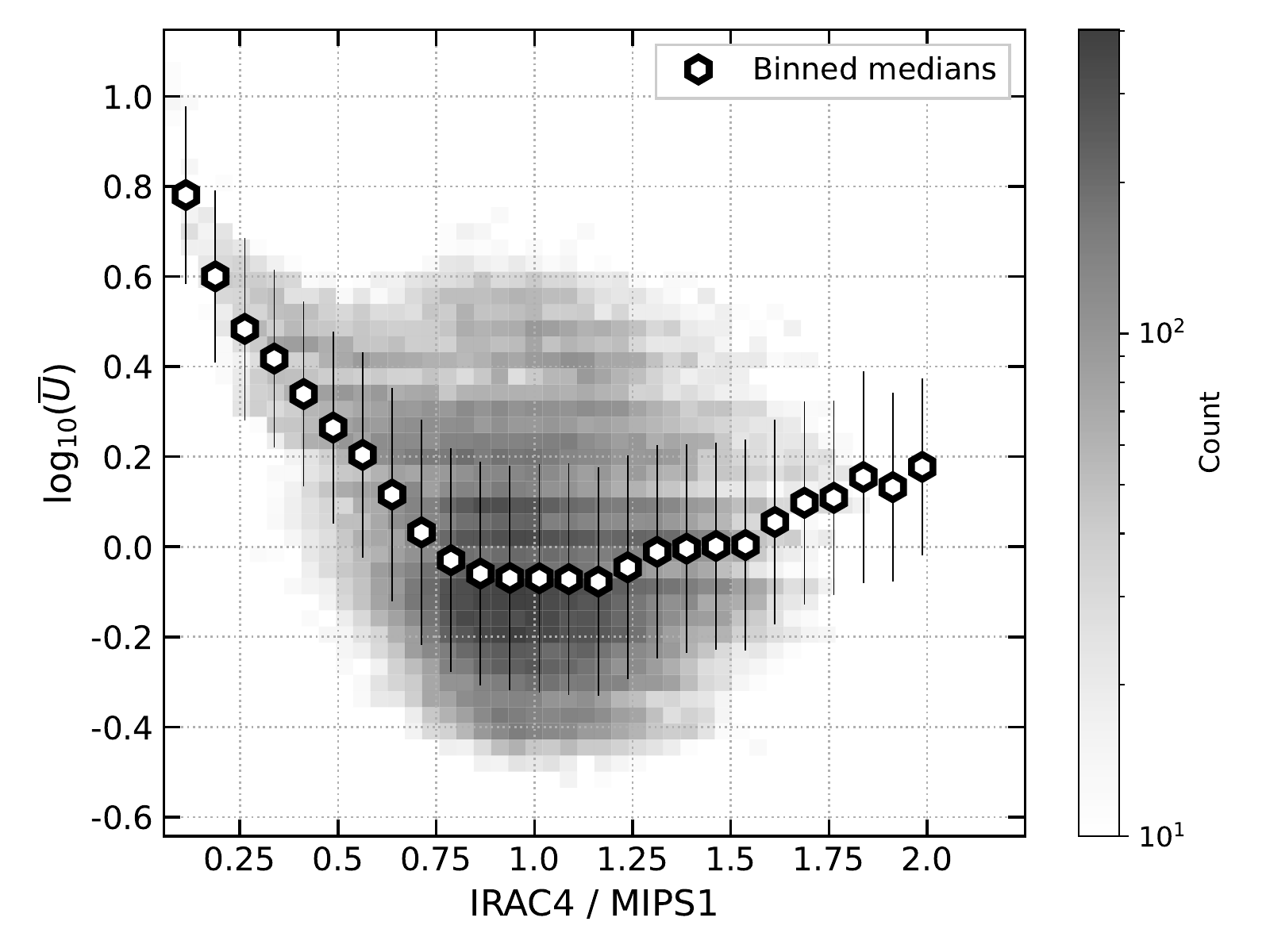}

    \caption{\textit{Left:} Modeled values of \qpah plotted as a function of IRAC4/MIPS1 fluxes. Here, the relationship between IRAC4/MIPS1 is clearly linear, suggesting that changes in this flux ratio are driven by changes in \qpah, not changes in $U$, for the range of radiation fields measured at these 10~pc scales. \textit{Right:} Modeled values of $\log_{10}\overline{U}$ plotted as a function of IRAC4/MIPS1 fluxes. We can see a decreasing trend between $\overline{U}$ and the IRAC4/MIPS1 at low IRAC4/MIPS1 or low \qpah.  Despite this trend at low IRAC4/MIPS1, \qpah is still well correlated with IRAC4/MIPS1 at similar values.}
    \label{fig:LMC_hists}
\end{figure*}

Of particular interest is investigating the range of ${\overline{U}}$, the average radiation field in units of the \citet{MMP1983} solar radiation field, that is heating the dust.  Using the FIR SED models produced in \citet{Chastenet2019}, we present a map of $\overline{U}$ for the LMC in Figure~\ref{fig:LMC_logu}.  Across this map, we find that $\overline{U}$ spans a range of $\sim 0-10$ on 10~pc scales, with the exception of the starburst region 30 Doradus (where on scales $\ll10$pc the radiation field intensity is very high), and $\overline{U}$ reaches $\sim50$, the highest value of $\overline{U}$ measured in this map. This suggests that for the physical resolution we obtain using JWST for the galaxies presented in this work ($\sim10-50$ pc), we can expect $\overline{U}$ to fall within typically fall within the range: $\sim 0-10$, with some extreme star forming regions potentially reaching a maximum $\overline{U}$ of $\sim50$.  

This is relevant to our proposed use of \rpah as a tracer of \qpah because very high values of $U$ \citep[$1\times10^2$ to $1\times10^4$, see][Figure 10]{Draine2021}, will push the peak of the black body emission produced by dust grains in equlibrium with the radiation field into the F2100W band, and at the highest values of U ($1\times10^6$ to $1\times10^7$) the peak can even shift into the F1130W or F770W bands and most grains are no longer stochastically heated.  Large contributions of emission at F2100W from hot small grains in equilibrium with the radiation field would drive down \rpah regardless of any changes to \qpah.  Based on the modeled values of ${\overline{U}}$ found within the LMC at similar spatial resolution, it is unlikely that $U$ reaches the necessary values for this to happen within our sample.  For this reason, we maintain that the decrease in \rpah within the nebular regions is driven by PAH destruction, and not the increased intensity of the radiation field heating the dust.

Furthermore, by using these models of the IR SED of the LMC, we can test how the \textit{Spitzer}--based photometric \qpah tracer, IRAC4/MIP1, varies with changing ${\overline{U}}$.  This is shown in the left panel of Figure~\ref{fig:LMC_logu}.  While there is a decreasing trend in IRAC4/MIPS1 with increasing $\log_{10}{\overline{U}}$, at low IRAC4/MIPS1, we can see that at these same values IRAC4/MIPS1 is still well correlated with \qpah.  This suggests that the relationship observed between IRAC4/MIPS1 and $\log_{10}{\overline{U}}$ at these $\log_{10}{\overline{U}}$ values is caused by a change to \qpah, not a change to how the dust grains emit light.  If we instead plot the trend between IRAC4/MIPS1 and \qpah, as shown in the right panel of this figure, we see a clear positive correlation.  This suggests that at the 10~pc scales measured in these maps of the LMC, changes to \qpah drive changes in IRAC4/MIPS1, while changes to $\overline{U}$ have only a small impact.

We note that the conclusions described above also depend on the assumption of a distribution of radiation fields heating the dust. In the delta-function plus power-law radiation field model, a very small amount of dust heated by high radiation field intensities can contribute significantly to the MIR continuum emission at F2100W \citep[see Figure 18 from][]{Draine&Li2007}. In models that make different assumptions about the radiation field distribution (e.g.\ a single radiation field intensity), the mapping between \rpah\ and \qpah\ can differ significantly. Future studies that use different dust and radiation field models beyond \citet{Draine&Li2007} with the delta function plus power-law model may draw different conclusions about the role radiation field plays in interpreting \rpah.


\bibliography{main}{}
\bibliographystyle{aasjournal}



\end{document}